\documentclass[a4paper,11pt]{article}
\usepackage{pos}
\usepackage{multirow}
\usepackage{makecell}
\usepackage{amsmath,amsthm,mathrsfs,bbm}
\usepackage{latexsym,amscd,amsbsy,amsfonts,dsfont}
\usepackage{graphicx,tikz,pgfplots,subcaption,float}
\usepackage{xcolor}
\usepackage{parskip}
\usepackage[font=small,labelfont=bf]{caption}
\numberwithin{equation}{section}
\usetikzlibrary{shapes.geometric}
\usetikzlibrary{arrows.meta}
\usetikzlibrary{backgrounds}
\title{Black holes in the classical and quantum world}
\newcommand{\beq}{\begin{equation}}
\newcommand{\eeq}{\end{equation}}
\newcommand{\beqa}{\begin{eqnarray}}
\newcommand{\eeqa}{\end{eqnarray}}
\newcommand{\bea}{\begin{eqnarray}}
\newcommand{\eea}{\end{eqnarray}}

\newcommand{\ie}{{i.e.,\ }}
\newcommand{\eg}{{e.g.,\ }}

\newcommand{\lp}{\left(}
\newcommand{\rp}{\right)}

\newcommand{\mc}[1]{\mathcal{#1}}

\author*[a,b]{Roberto Emparan}
\author[c]{Elvis Barakovic}
\author[d]{Roukaya Dekhil}
\author[e]{Filip Rescic}

\affiliation[a]{Instituci\'o Catalana de Recerca i Estudis
Avan\c cats (ICREA),\\ Passeig Llu\'{\i}s Companys 23, E-08010 Barcelona, Spain}
\affiliation[b]{Departament de F{\'\i}sica Qu\`antica i Astrof\'{\i}sica, Institut de
Ci\`encies del Cosmos, Universitat de
Barcelona, Mart\'{\i} i Franqu\`es 1, E-08028 Barcelona, Spain}
\affiliation[c]{Faculty of Natural Sciences and Mathematics, Department of Mathematics,\\ University of Tuzla, Ul. Urfeta Vejzagića br. 4, 75000 Tuzla, Bosnia and Herzegovina}
\affiliation[d]{Arnold Sommerfeld Center for Theoretical Physics, Ludwig-Maximilians-Universit\"at München \\Theresienstrasse 37, 80333 M\"unchen, Germany}
\affiliation[e]{Faculty of Physics, University of Rijeka,\\  Ul. Radmile Matejčić 2, 51000 Rijeka, Croatia}

\emailAdd{emparan@ub.edu}
\emailAdd{elvis.barakovic@untz.ba}
\emailAdd{Roukaya.Dekhil@physik.uni-muenchen.de}
\emailAdd{filip.rescic@uniri.hr}

\abstract{These are the lecture notes for an introductory course on black holes and some aspects of their interaction with the classical and quantum world.
The focus is on phenomena of ``fundamental physics'' in the immediate surroundings of the black hole (classical and quantum fields, with little astrophysics). We aim more at qualitative, intuitive understanding than at quantitative rigor or detail. Accordingly, we only assume previous exposure to a conventional introduction to the elements of General Relativity and a glancing acquaintance with the Schwarzschild solution, but not more. We use many figures for illustrations and provide a set of carefully guided exercises. 
\\
Topics: \\
(1) The black hole as a tale of light and darkness.\\
(2) The black hole that vibrates.\\
(3) The black hole that rotates.\\
(4) The black hole that evaporates.\\
(A) Guided problems.
}

\FullConference{Second Training School of COST Action CA18108\\
3-10 September 2022\\
Belgrade, Serbia}
\pgfplotsset{compat=1.18}

\tableofcontents

\begin{document}
\maketitle
\section*{Introduction}
\addcontentsline{toc}{section}{Introduction}

These are the lecture notes for a short course on black holes delivered at the Second Training School of COST Action CA18108
``Quantum gravity phenomenology in the multi-messenger approach''.

We begin in the classical realm, explaining the defining feature of a black hole---the event horizon---and reviewing several central results of classical black hole theory: Penrose's singularity theorem, Hawking's black hole area theorem, and the no-hair theorem. Then we study how black holes react when they are disturbed (\eg by infalling particles or by the presence of classical fields), how they settle down to quiescence following their formation in \eg a merger, and the consequences that these phenomena have for gravitational wave observations.  We will also see that the black hole can be set into rotation, and, as it does so, qualitatively new phenomena appear with new opportunities for astronomical observation. Afterwards, we discuss what happens when the black hole is surrounded by quantum fields (as it always is). The black hole turns out to emit a very subtle quantum radiation with deep consequences, some of which remain perplexing and incompletely understood.

\paragraph{Breadth and depth.} The focus is on phenomena of ``fundamental physics'' in the immediate surroundings of the black hole. These will be classical and quantum fields, with little astrophysics. Due to time restrictions, much of great interest is not covered: for instance, we do not touch at all on any aspects of binary configurations and their inspirals, nor astrophysics of the medium around black holes such as accretion disks. Of fundamental physics, nothing more exotic than Hawking radiation is discussed, the idea being that a good grasp of the widely accepted, conventional lore in these lectures is necessary for assessing current speculations that aim at revealing new physics---even quantum gravity---from observations of the black holes in our universe.

\paragraph{Target audience.} When designing this course we have had in mind a reader who does not necessarily intend to become an expert on General Relativity or black hole theory, but who is interested in understanding the elementary concept of a black hole and the physics behind some of its main effects. This reader does not need to see the detailed computations but wants to get the gist of what the experts who perform them are doing, and why. Qualitative and intuitive explanations are then of more value than elaborate technical detail.

We will assume that the reader has already been exposed to a conventional, elementary introduction to GR---with the basic tensor calculus for understanding what the Einstein equations are, and with a glancing acquaintance with the Schwarzschild solution. No more advanced differential geometry is assumed, nor knowledge of the full structure of the Schwarzschild solution, even less so of the Kerr solution. When discussing quantum phenomena, we do not assume any more than an elementary appreciation of what a quantum field is; correspondingly, the exposition of these aspects remains at an even more qualitative level than for the classical ones. 

We have included a set of problems with detailed guidance, which should allow the interested reader to delve into aspects that are only briefly discussed in the course.


We have not attempted to provide a detailed and complete list of original references. Instead, we make a few suggestions for texts---comprehensive reviews or pedagogical textbooks, preferably both---where the reader will readily find how to go beyond each of the four lectures:
\begin{itemize}

\item For lecture 1: almost any textbook of General Relativity covers in detail the material we discuss. The reader may want to consult, at an increasingly technical level, \cite{Hartle:2021pel}, \cite{Carroll:2004st} and \cite{Wald:1984rg}.

\item For lecture 2: \cite{Berti:2009kk,Konoplya:2011qq}.

\item For lecture 3: \cite{Wiltshire:2009zza,Brito:2015oca}.

\item For lecture 4: \cite{Fabbri:2005mw,Susskind:2005js}, and, for the latest developments, \cite{Almheiri:2020cfm}.

\item As a recent comprehensive black hole textbook that covers most of what we discuss here and much more, we recommend \cite{Grumiller:2022qhx}.

\end{itemize}


\section{The black hole as a tale of light and darkness}

As you already know, Einstein's general theory of relativity reduces gravity to an effect created by the curvature of spacetime. More precisely, it posits that spacetime is a dynamical four-dimensional Lorentzian manifold, and the information about the geometry of spacetime and how it is curved is contained in the  metric $g_{ij}(x)$, which expresses the distance relations between nearby points through the line element
\beq
ds^2=g_{ij}(x)dx^idx^j, \quad i,j=0,\dots\, , 3\, .
\eeq

Physical magnitudes must not change under generic coordinate transformations (diffeomorphisms) $x^i\rightarrow x'^{i}(x^j)$. For instance, if we demand that the line element remains invariant under an infinitesimal transformation $x^i \rightarrow x^i+\varepsilon v^i(x)$ with $ \varepsilon \ll 1$, the metric coefficients must change as
\begin{align}\label{infdiff}
    g_{i j} \rightarrow g_{i j}+\varepsilon\left(\partial_i v_j+\partial_j v_i\right)\,, \qquad \partial_i \equiv \frac{\partial}{\partial x^i}\,.
\end{align}

The metric is the fundamental dynamical variable in GR. By considering second-order variations of the metric one can construct another invariant object, the Riemann tensor,\footnote{A tensor $T$ is a locally defined object that is invariant under coordinate transformations. Its components in a basis of vectors $\partial_i$ and one-forms $dx^i$ (a.k.a.\ covariant and contravariant vectors),
\beq
T=T^{ij\dots}_{\quad kl\dots}\partial_i\otimes\partial_j\otimes\cdots \otimes dx^k\otimes dx^l\otimes \cdots 
\eeq
transform under a diffeomorphism in such a way that $T$ remains unchanged.} which characterizes the curvature of the geometry. Simpler tensors can be obtained from it, such as the Ricci tensor $R_{ij}$ and the curvature scalar (Ricci scalar), which contain partial but convenient information about the curvature averaged over all directions.

\subsection{Matter and geometry, or why you are not attractive (and who is)}

The Einstein equations describe the dynamical manner in which the geometry and the matter couple to each other. They relate the components of the Ricci tensor $R_{ij}$ and its trace, the Ricci scalar, to the matter stress-energy tensor $T_{ij}$ as
\beq \label{einstein}
R_{ij}-\frac{1}{2}g_{ij}R=\frac{8\pi G}{c^2}T_{ij}\,,
\eeq
where $c$ is the speed of light and $G$ is Newton constant. These two fundamental constants are necessary in order to meaningfully connect the two sides of this equation---geometry and matter---since they allow us to express all the material notions such as mass, energy, and pressure, in geometrical units of length. Observe that for this translation to be possible we need both $c$ and $G$, that is, we need a \emph{relativistic} theory of \emph{gravity}. To be concrete, given a mass $M$ we can always construct a quantity with dimensions of length that is proportional to it,
\begin{align}\label{LM}
    L_M=\frac{G}{c^2}M\,,
\end{align}
and this implies that we can \emph{measure mass in meters}. For instance, instead of saying that the mass of an object is 1~kg, we can equivalently say that it weighs $10^{-29}$~m. Your own weight is then $\sim 10^{-27}$~m, which is manifestly much less than your physical size. What this fact means is that \emph{you are not very attractive}---at least gravitationally speaking. You are not alone in this: the mass of the Earth is only $\sim 1$~cm, so our beloved planet is not that attractive either. 

To see how this notion works out in the equations \eqref{einstein}, imagine that you have an object of a characteristic size  $R_o$, which bends the geometry around it so much that the curvature radius is comparable to $R_o$. Schematically, \eqref{einstein} says that
\begin{align}
    \textrm{Curvature}=\frac{G}{c^2}\,\textrm{Mass density}\,,
\end{align}
so the Einstein equations determine that for such an object to exist, it must  satisfy\footnote{The Riemann and Ricci tensors are obtained taking two derivatives of the metric, so they measure an inverse square curvature radius.}
\begin{align}
    \frac1{R_o^2}=\frac{G}{c^2}\frac{M}{R_o^3}\,,
\end{align}
and therefore 
\begin{align}\label{RM}
    R_o\sim \frac{GM}{c^2}\,.
\end{align}
That is, an object so massive and compact that its physical size $R_o$ is comparable to its mass in length units \eqref{LM}, manages to create a very large spacetime curvature. These objects (and not you) are indeed attractive, so we have given them a very sexy name: black holes.\footnote{Beware the morality tale here: don't try to be too attractive, or no one will ever see you.}

In these lectures we will only consider the simplest instance of the theory, namely in the absence of matter. The vacuum Einstein equations \eqref{einstein} then simplify to
\beq 
\label{vac}
R_{ij}=0.
\eeq
Since, unlike Newton's theory, GR is a non-linear theory, it is possible to have matter-free but non-trivial gravitational fields, i.e., self-bending spacetime geometries. It is indeed mind-boggling how much richness and intricacy of physics is encoded in equations that are conceptually as simple as \eqref{vac}.

\subsection{Schwarzschild's solution}

The simplest nontrivial exact solution of \eqref{vac} was found by Karl Schwarzschild while he was in the Eastern front in December 1915, barely a month after Einstein obtained the vacuum form of his equations.\footnote{Schwarzschild's article was published in January 1916.} 

This solution describes a static, spherically symmetric spacetime that is supposed to model the empty exterior of a static spherical star, or the geometry that remains after a very massive star has collapsed on itself, a situation of strong field behavior in GR.  

\begin{figure}[t]
\centering
\includegraphics[width=\textwidth]{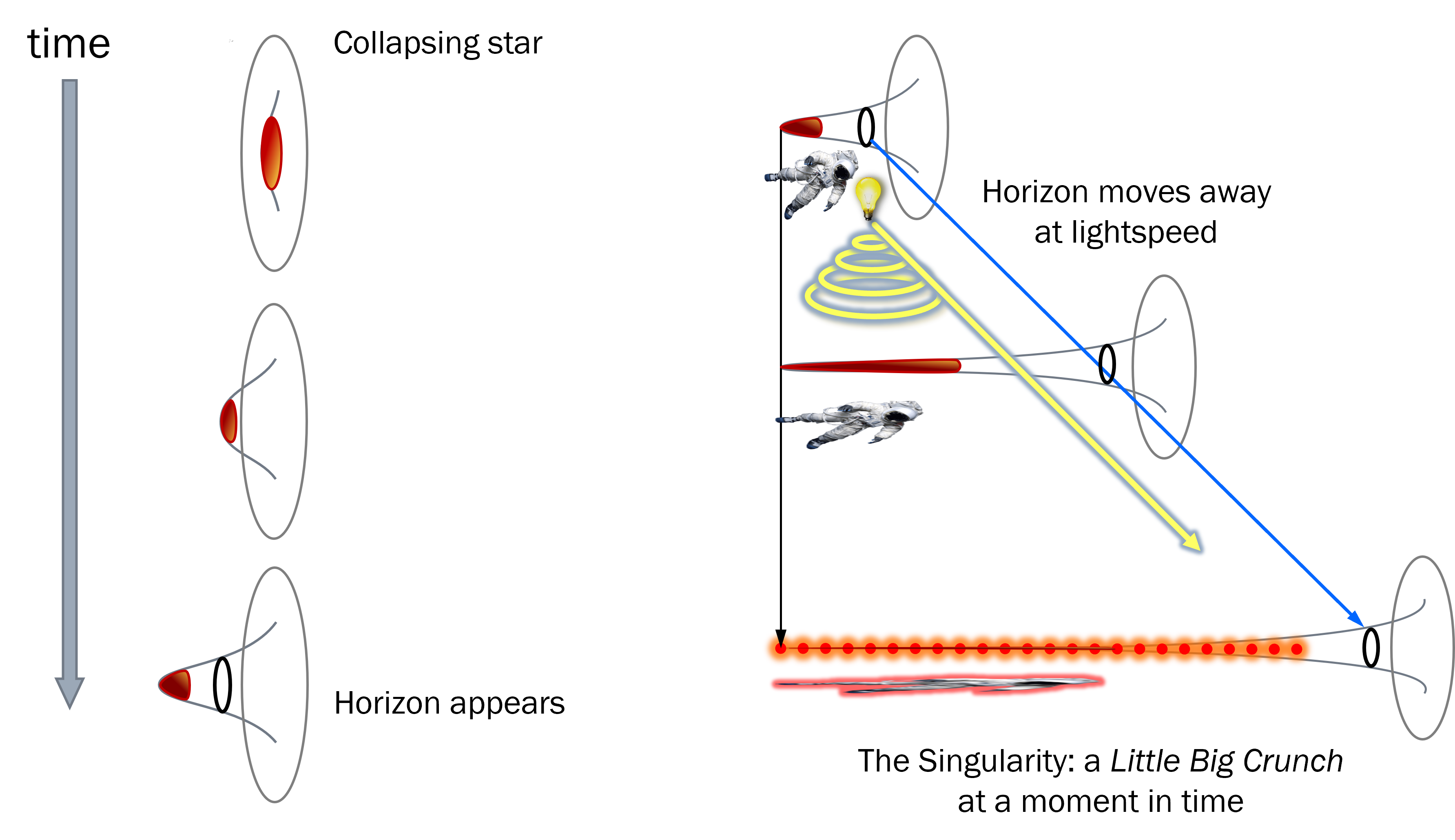}
\caption{Left: A collapsing star distorts the geometry around it giving rise to a horizon. Right: If 
you are a reckless astronaut riding along with the collapsing star, you will be increasingly stretched (in height) and compressed (in width), while the horizon recedes away from you at the speed of light, making it impossible for you to come back out again. Eventually, the distortion is so extreme that the notion of a smooth geometry of space and time loses sense. This final moment is the Singularity. The interior of the black hole can be regarded as an anisotropic Little Big Crunch, a strongly time-dependent geometry where the universe locally comes to an end. In contrast, the exterior remains calm and essentially unchanging.}
\label{fig:nofuture}
\end{figure}

Let us anticipate some striking notions that will emerge from our study of this solution. We illustrate them in figure~\ref{fig:nofuture}. Once you are inside the black hole, the horizon recedes away from you at the speed of light, so you will not be able to catch up with it to cross it back outside: \emph{there is no escape for you}.

This horizon shields the final stages of the collapse in the interior, which ends at a singularity. The singularity is a \emph{terminal instant} in the future of anything inside the black hole. It is not a fiery point that you could see if you were inside the black hole---because, simply, you cannot see the future. And, once inside the black hole, you cannot avoid the singularity any more than you can prevent Monday coming after Sunday: the singularity is ineludible for you because time passes and the future will arrive. You can think of the black hole interior as undergoing a Little Big Crunch, a local end of the universe that is invisible to anyone who remains outside the black hole. If you enter the black hole, you cannot escape this Little Big Crunch: \emph{there is no future for you}. So it goes.

We now proceed to flesh out these gloomy observations with a proper analysis of the Schwarzschild geometry. In a convenient and conventional form using spherical coordinates, it is written as
\beq \label{schw}
ds^2=-\lp 1-\frac{r_0}{r}\rp dt^2+\frac{dr^2}{1-\frac{r_0}{r}}+r^2\lp d\theta^2+\sin{\theta}^2d\phi^2\rp\,,
\eeq
where the last terms stand for the 2-sphere line element, and $r$ denotes the radial coordinate.

This metric describes a manifestly static (time-independent) and spherically symmetric geometry. It is characterized by a length parameter $r_0$, whose meaning we will presently clarify. The metric is asymptotically flat, meaning that far away from the gravitational spherical body, where $r\gg r_0$, the universe is well described by the flat Minkowski metric. 

We can learn more about $r_0$ by taking the weak-field Newtonian limit, expanding the time component of the metric around the flat background $g_{tt}\simeq -1$, so that
\beq\label{gttPhi}
g_{tt} \approx -1-\frac{2\Phi_N}{c^2}, \quad \text{for} \ \lvert\Phi_N\rvert\ll c^2\, ,
\eeq
where $\Phi_N$ is the Newtonian potential. We know what this potential is for a mass $M$, and thus we can compare it to \eqref{schw} at distances $r\gg r_0$,
\beq
\Phi_N=-\frac{GM}{r}\Rightarrow r_0=\frac{2GM}{c^2}\, .
\eeq
This gives the physical interpretation of $r_0$ and confirms what we said about \eqref{RM}. 

It can be proven that this is the unique solution to the Einstein equations that describes a spherically symmetric vacuum. Thus, it describes the empty \textit{exterior} of any spherically symmetric object in the universe. If, on the other hand, we are interested in the geometry in the interior of the star, we must go back to the general equation \eqref{einstein} and solve it with an energy-momentum tensor $T_{ij}$ that describes the stellar matter. In this way, one derives the relativistic equations of stellar structure.\footnote{Schwarzschild found the first such solution in 1916, shortly before his premature death. Years later, his son Martin went on to become a distinguished expert in the theory of stellar structure.}

A most intriguing feature of the Schwarzschild metric \eqref{schw} is the presence of an apparent singularity at $r=r_0$. Historically, this was initially dismissed by saying that, as long as the radius of a star is $R_*\gg 2GM/c^2$, we do not have to worry about it since only the exterior geometry is described by \eqref{schw}---and for a star like our Sun, the radius $R_\odot\simeq 750,000$~km is certainly much larger than $2GM_\odot/c^2 \simeq 3$~km (so even the Sun is not that attractive). It took decades until the nature of the surface $r=r_0$ began to be understood, and even longer until it was regarded as relevant to astrophysics. Now we know that it corresponds to the defining feature of a black hole: its horizon.

\subsection{The horizon}

Henceforth we set natural units $G=c=1$. We consider that there is no matter anywhere and study the Schwarzschild solution \eqref{schw}, which we now write as
\beq \label{schw2}
ds^2=-\lp 1-\frac{2M}{r}\rp dt^2+\frac{dr^2}{1-\frac{2M}{r}}+r^2d\Omega_2^2\, ,
\eeq
where we abbreviate $d\Omega_2^2$ for the line element of the unit two-sphere $S^2$. 

This metric, as a matrix with components $g_{ij}$, has a singularity at $r=2M$. A matrix is non-singular if all its components are finite and if it is invertible, that is, if its determinant is non-zero and finite. But some singularities of the matrix of metric coefficients can be artifacts of the coordinates we use---indeed, the metric above is not invertible at the poles of the $S^2$, but we know that this is simply a feature of polar coordinates, not any physical difficulty, and it can be avoided by choosing different coordinates. That is, the coefficients $g_{ij}$ change under coordinate transformations, and if we can find coordinates where the singular behavior at a point disappears, then this means that there is no physical singularity associated with that point.

Let us then take a closer look at the singularity at $r=2M$ and see if it is just an artifact of the specific coordinates we are using.

In order to do this, we study what happens to ingoing radial light rays once they reach this singularity.\footnote{Problems~\ref{prob:rindler} and \ref{prob:kruskal} discuss other approaches.} In general, light rays are characterized by a vanishing line element $ds^2=0$. After setting the angles $\theta$ and $\phi$ to a constant, we obtain a relation between the temporal and radial displacements for these light rays,
\beq
dt=-\frac{dr}{\lvert 1-\frac{2M}{r}\rvert}\,,
\eeq
where we have chosen the sign to describe ingoing trajectories. This can be easily integrated to give
\beq
\label{time_coordinate}
t=-r_* + \text{const.}\,,
\eeq
where we have introduced the radial \textit{tortoise coordinate} $r_*$ defined by
\beq \label{tort}
dr_*=\frac{dr}{\lvert 1-\frac{2M}{r}\rvert} \Rightarrow r_*=r+2M\log \lvert r-2M\rvert\,.
\eeq
Observe that
\begin{align}
    r_*\rightarrow r \rightarrow \infty &\quad \text{as} \ r\rightarrow \infty, \\
    r_*\rightarrow -\infty &\quad \text{as} \ r\rightarrow 2M\,.\label{r*hor}
\end{align}
Let us now define a new coordinate
\beq \label{efcoord}
v\equiv t+r*
\eeq
which remains constant along the ingoing light ray \eqref{time_coordinate}. We will use it instead of $t$. This is convenient: when traveling along a light ray, in order to get to $r\to 2M$, we must take $t\to+\infty$ (see \eqref{time_coordinate} and \eqref{r*hor}), so in terms of $t$ it would seem to take an infinite time to get to $r=2M$, which sounds strange. However, $t$ is just a coordinate whose meaning is clear at large distances, but much less so at smaller $r$. Near $r=2M$ it is more convenient to use $v$, which remains constant, and therefore finite, along the ingoing light ray even when it gets to $r=2M$.

Since
\beq
dt=dv-dr_*=dv-\frac{dr}{1-\frac{2M}{r}}\,,
\eeq
when we make the coordinate transformation $(t,r)\rightarrow(v,r)$ in the Schwarzschild metric \eqref{schw2} we obtain
\beq \label{efmetric}
ds^2=-\lp1-\frac{2M}{r}\rp dv^2+2dvdr+r^2d\Omega_2^2\, .
\eeq
Now the matrix of the metric coefficients 
\beq
g_{ij} = \lp \begin{array}{cccc}
    -\lp 1-\frac{2M}{r} \rp & 1 & 0 & 0\\
     1 & 0 & 0 & 0\\
     0 & 0 & r^2 & 0\\
     0 & 0 & 0 & r^2 \sin^2\theta
\end{array} \rp\,,
\eeq
does not have any element that diverges at $r=2M$, and since its determinant, det$(g_{ij})=-r^4\sin^2\theta$, is non-zero there, it is also invertible. 

We conclude that the ingoing light rays encounter $r=2M$ as a smooth place. The peculiar behavior of the metric \eqref{schw2} at $r=2M$ is a coordinate singularity and not a physical one. The coordinates $(v,r)$ that make this manifest are called \textit{ingoing Eddington-Finkelstein coordinates} (see also Problems~\ref{prob:outEF} and \ref{prob:kruskal}).

\begin{figure}
\centering
\includegraphics[width=.65\textwidth]{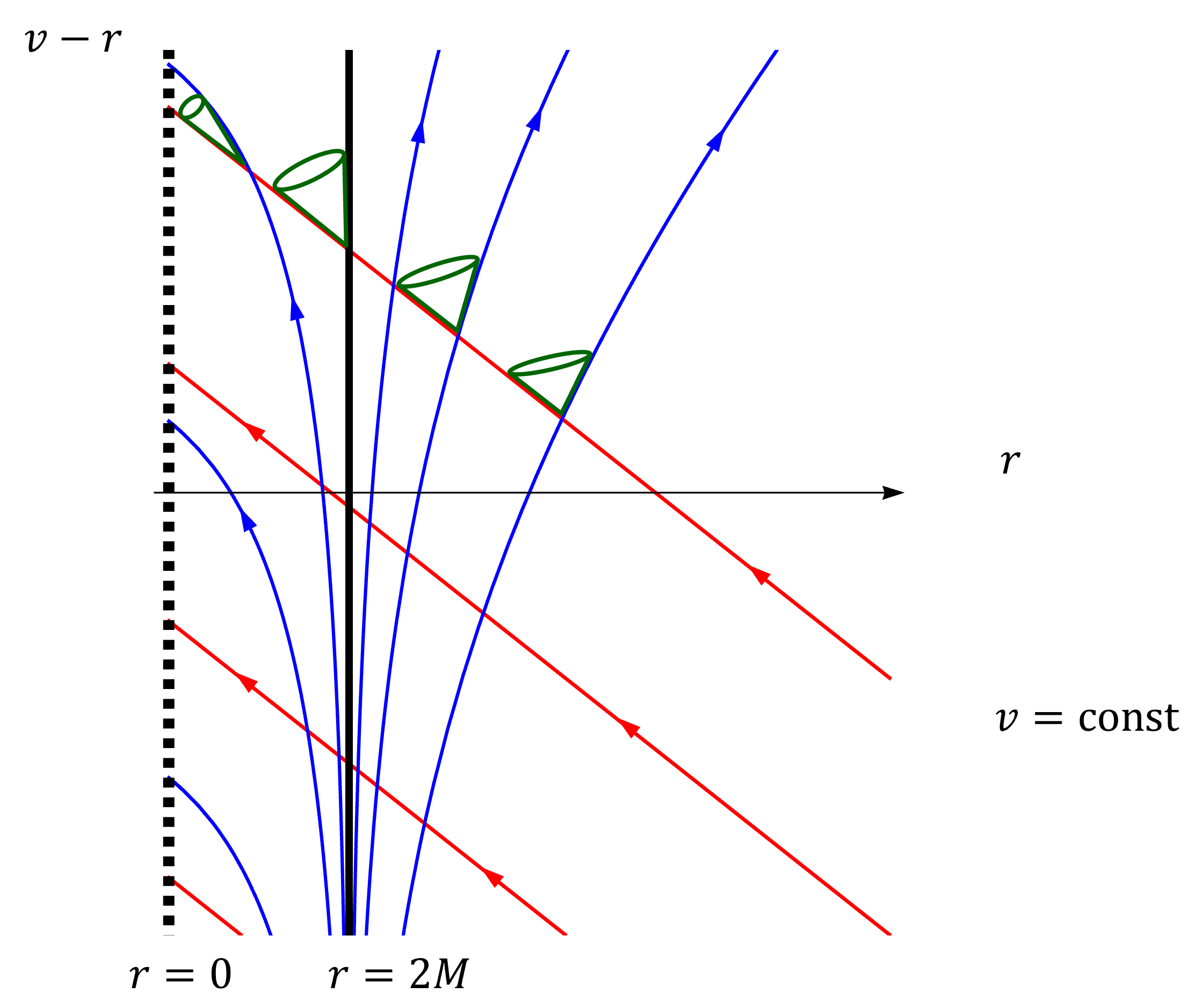}
\caption{Different types of radial light rays. Red lines: ingoing light rays with constant $v$. Thick black vertical line: light rays frozen at  $r=2M$. Blue lines:  locally outgoing light rays---for $r<2M$ they fail to escape away to infinity. In green, the behavior of light cones.}
\label{fig:lightrays}
\end{figure}

Using these coordinates and solving for $ds^2=0$ in \eqref{efmetric} we easily find three types of radial light rays:
\begin{itemize}
    \item Ingoing rays $v=\text{const.}$ 
    \item `Frozen' rays at constant radius $r=2M$.
    \item Outgoing rays $v=2r_*+\text{const.}$
\end{itemize}
We illustrate them in figure \ref{fig:lightrays}. The sphere where the light rays are frozen is what, as we will explain next, we can rightly call the \emph{horizon}.

You may now want to pause for a moment to see how this diagram encodes the qualitative properties in figure~\ref{fig:nofuture}. For instance, figure~\ref{fig:lightrays} makes it apparent that, since you always move inside a light cone, when you fall into the black hole you will get increasingly distorted while you are headed towards a singularity that you cannot see.

\subsection{The event horizon in black hole collapse and mergers}\label{subsec:collmer}

Our aim now is to provide a more pictorial description of these findings, which will lead us to an understanding of the meaning of the black hole and its event horizon.

\begin{figure}
\centering
\begin{subfigure}{0.45\linewidth}
  \centering
\includegraphics[scale=0.37]{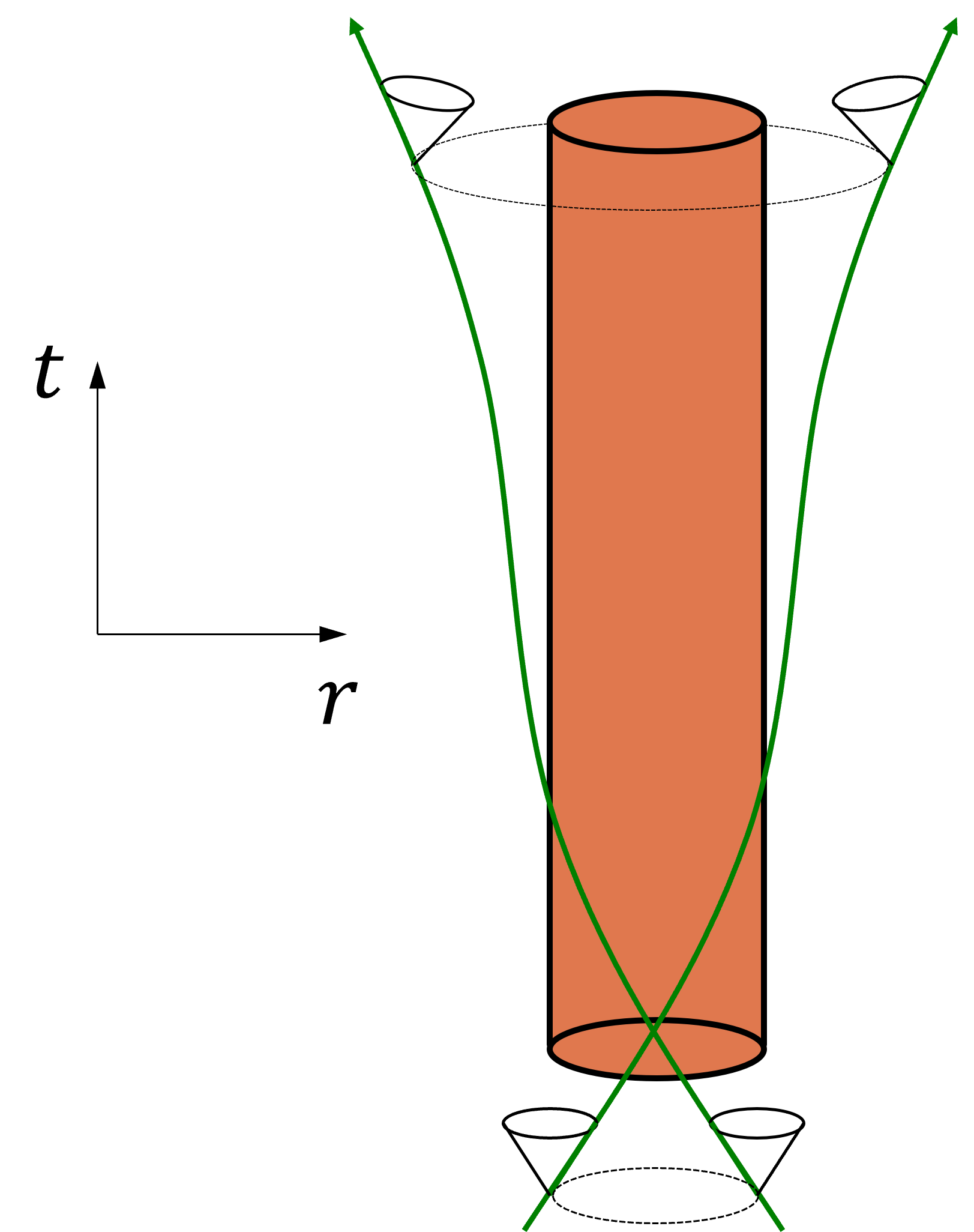}
  \caption{Light rays curve under the gravity of a static star. Outgoing lightfronts contract.}
  \label{fig:star}
\end{subfigure}\hfill
\begin{subfigure}{0.45\linewidth}
  \centering
\includegraphics[scale=0.4]{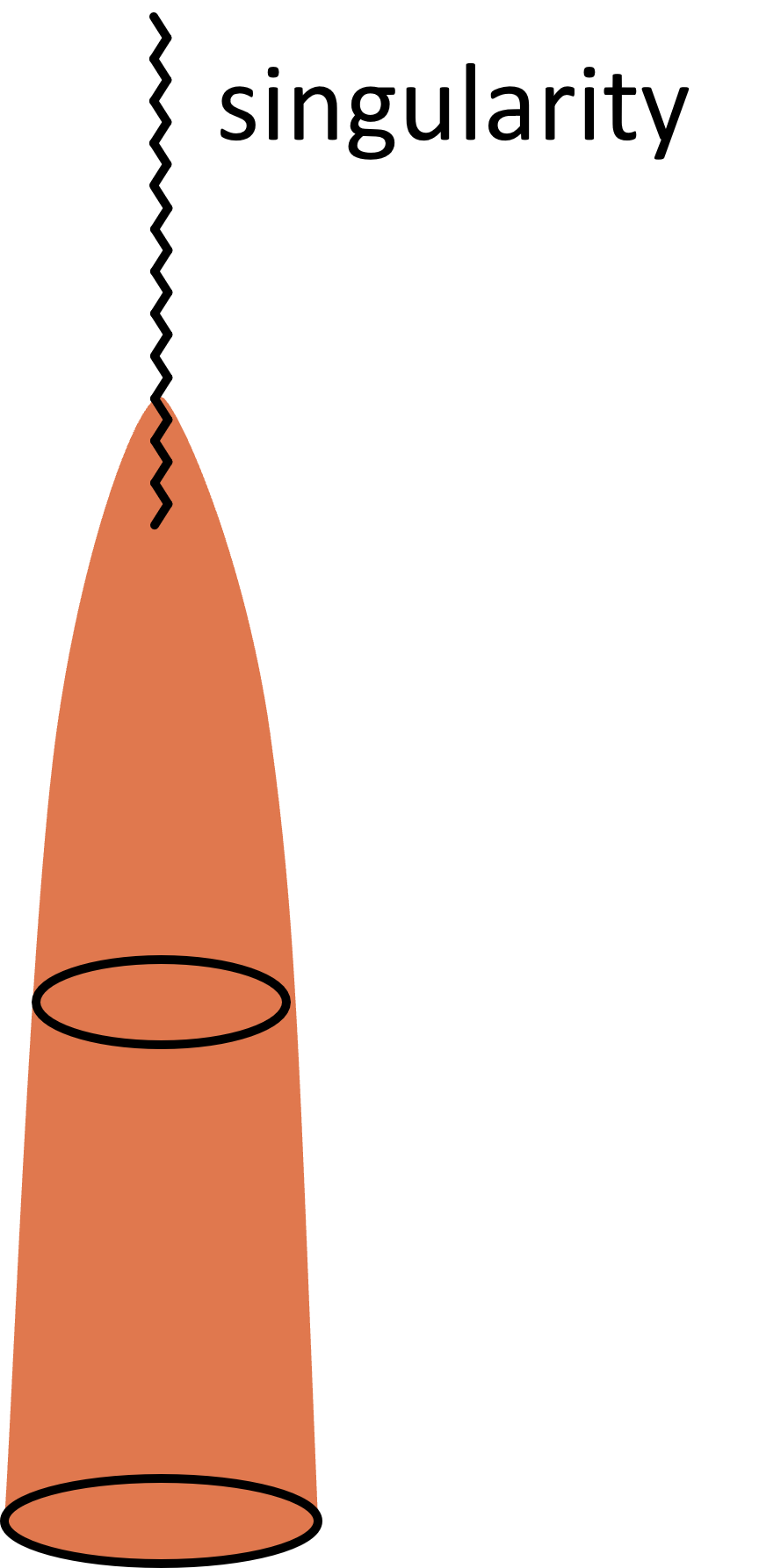}
  \caption{A star that collapses to form a singularity.}
  \label{fig:collapse}
\end{subfigure}
\caption{Stars: static and collapsing.}
\label{fig:starGR}
\end{figure}

Consider the propagation of light rays
in the spacetime of a spherically shaped star. An initial spherical lightfront that is contracting
will reach zero size inside the star, and then (since
the interior of the star may be quite hot, but the geometry there is still smooth and not that different from flat space), it will expand again, feeling only some attraction from
the star which slightly delays its expansion. This is what figure \ref{fig:star} illustrates.

\begin{figure}
\centering
\begin{subfigure}{0.45\linewidth}
  \centering
\includegraphics[scale=0.4]{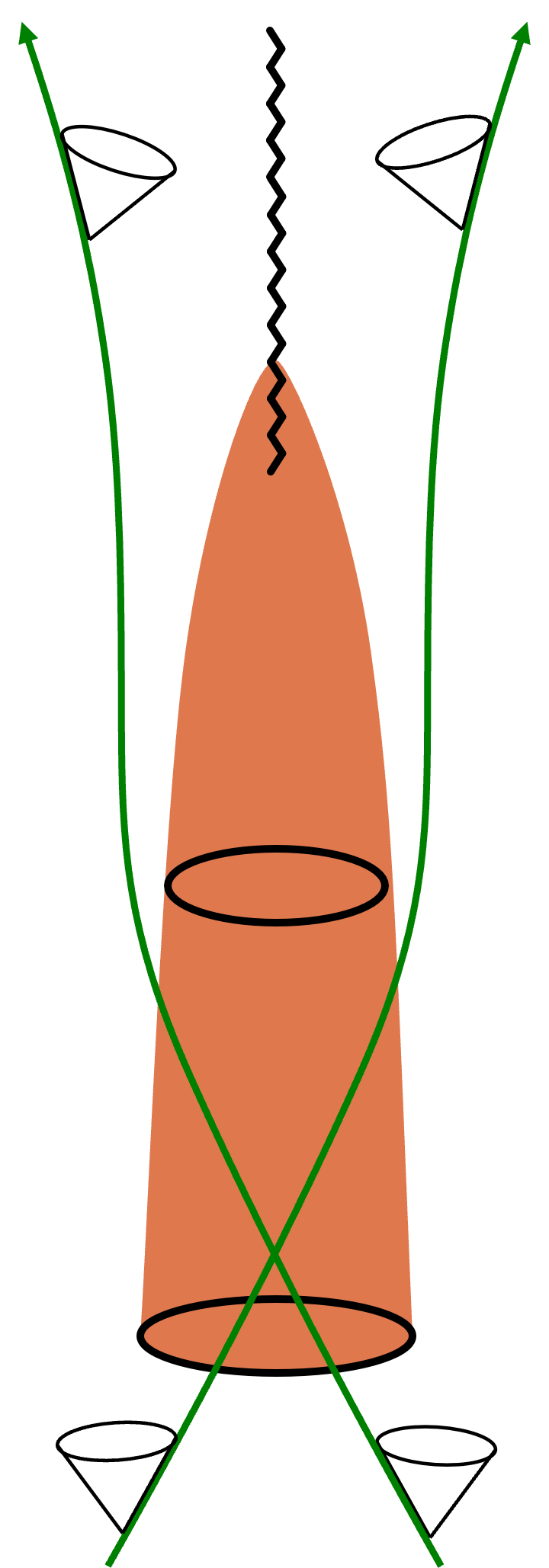}
\caption{Lightfronts that begin early
enough contract to zero, and then expand.}
  \label{fig:cones}
\end{subfigure}\hfill
\begin{subfigure}{0.45\linewidth}
  \centering
\includegraphics[scale=0.4]{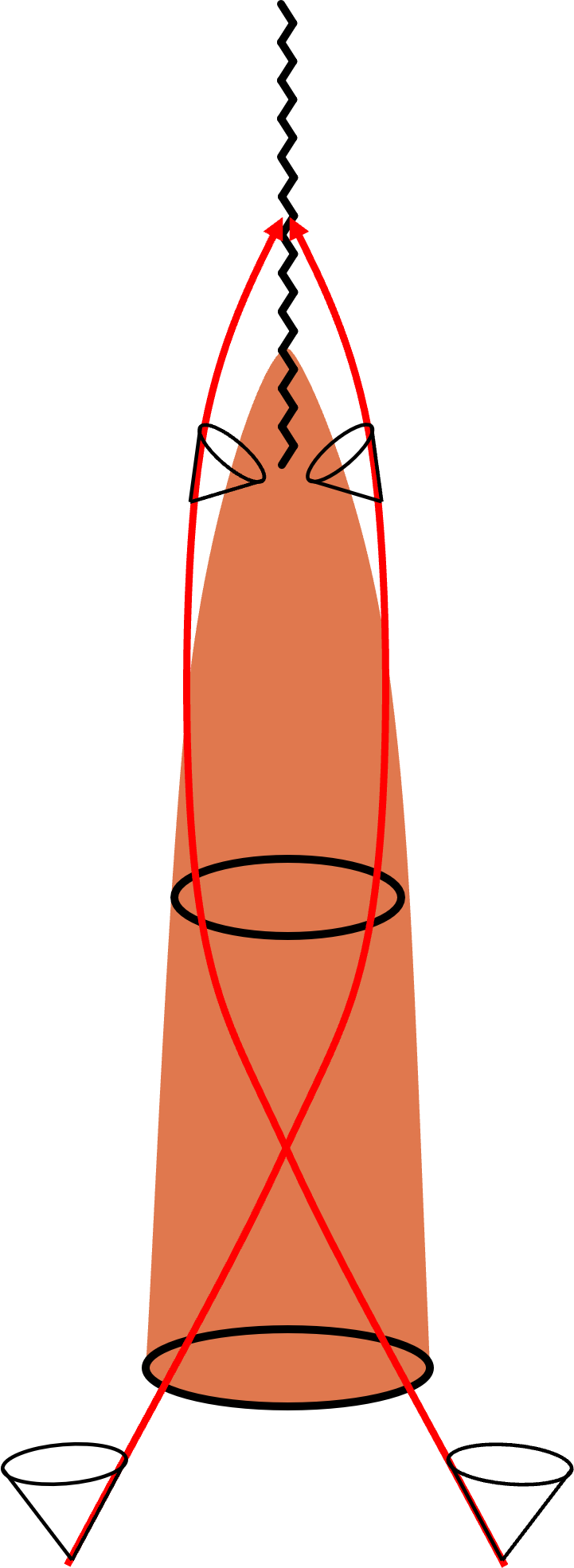}
\caption{Later lightfronts are dragged back so strongly that they collapse to the singularity and fail to
escape away.}
  \label{fig:collapse_cone}
\end{subfigure}
\caption{Fate of early and late spherical lightfronts near the collapsing star.}
\label{fig:starGR2}
\end{figure}

Now consider a star that, as in figure \ref{fig:collapse}, has collapsed to form a singularity where the geometry is not at all smooth but actually it is not even well defined. Lightfronts that begin early enough will, as before, contract to zero and then expand, as shown in figure \ref{fig:cones}. But there will also be later lightfronts that, when
they try to expand, are dragged back so strongly
that they collapse to the singularity and fail to
escape away to infinity, as in figure \ref{fig:collapse_cone}.

\begin{figure}
\centering
\begin{subfigure}[t]{0.45\linewidth}
  \centering
\includegraphics[scale=0.4]{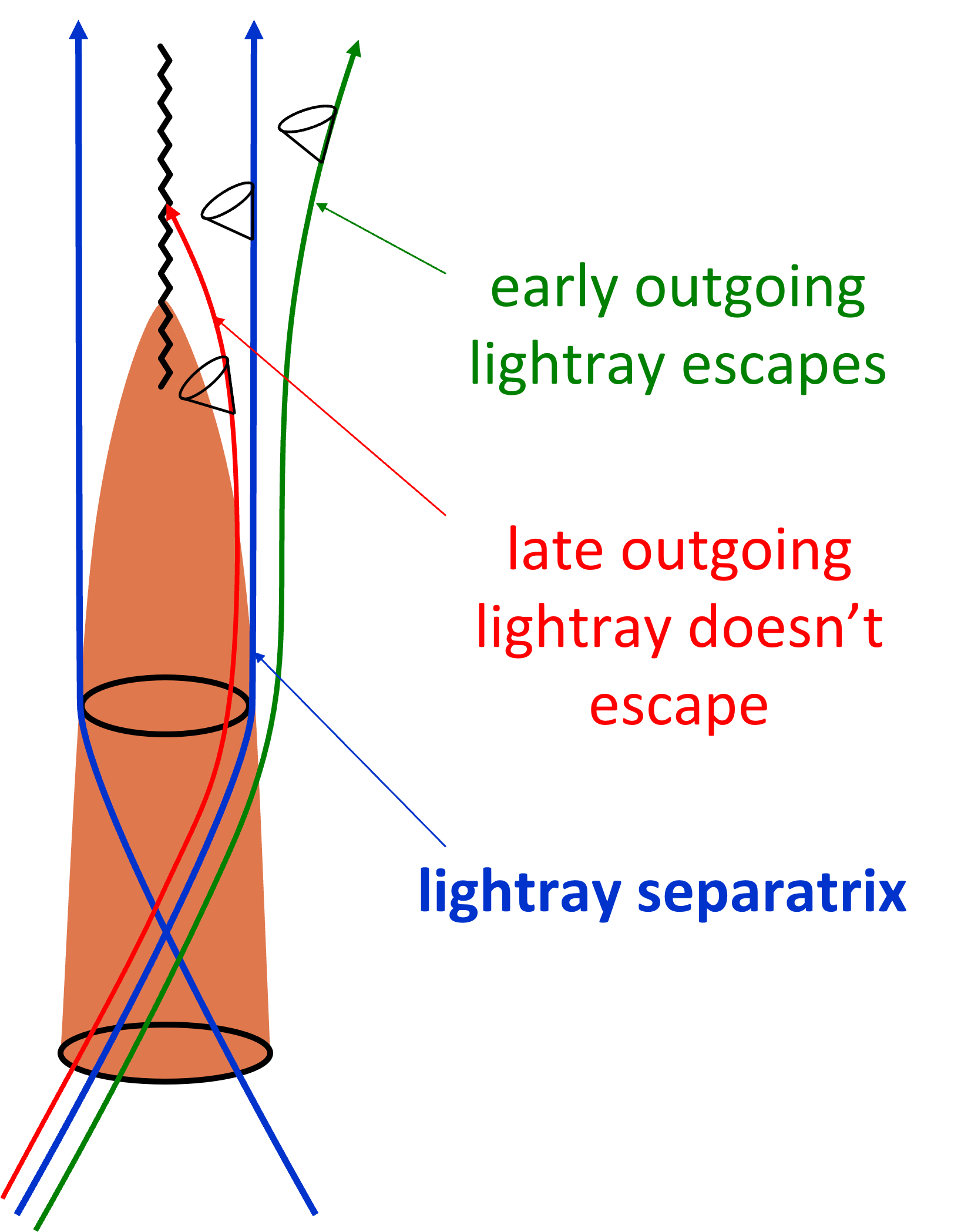}
  \caption{Three classes of light rays, according to whether they escape to infinity, fall to the singularity, or do neither of that. }
  \label{fig:collapse_lightrays_types}
\end{subfigure}\hfill
\begin{subfigure}{0.45\linewidth}
  \centering
\includegraphics[scale=0.4]{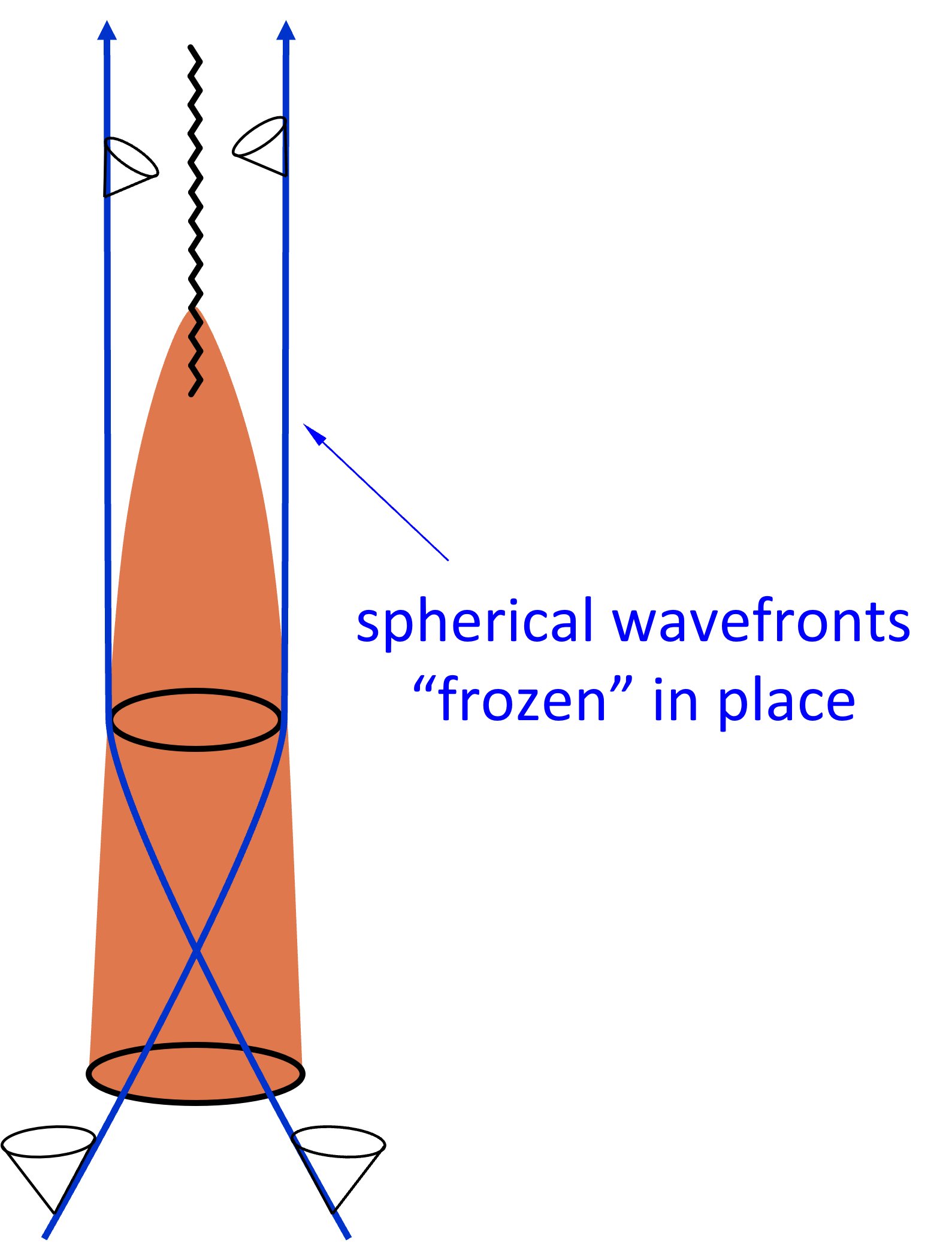}
\caption{Buridan light rays remain frozen at a fixed radius.}
  \label{fig:frozen_lightrays}
\end{subfigure}
\caption{Classes of light rays in a collapsing star. Cf.~figure~\ref{fig:lightrays}.} 
\label{fig: Light rays after collapsed star}
\end{figure}

Thus, we can identify two classes of null geodesics: those that escape and those that fall into the singularity. But clearly, there must also be a third class of null geodesics that neither escape nor fall. Like Buridan's ass, they remain frozen in place as shown in figure \ref{fig:frozen_lightrays}. This class of null geodesics is particularly interesting since they represent a causal boundary between regions in the spacetime geometry.

\begin{figure}
\centering
\begin{subfigure}[t]{0.45\linewidth}
\centering
\includegraphics[scale=0.37]{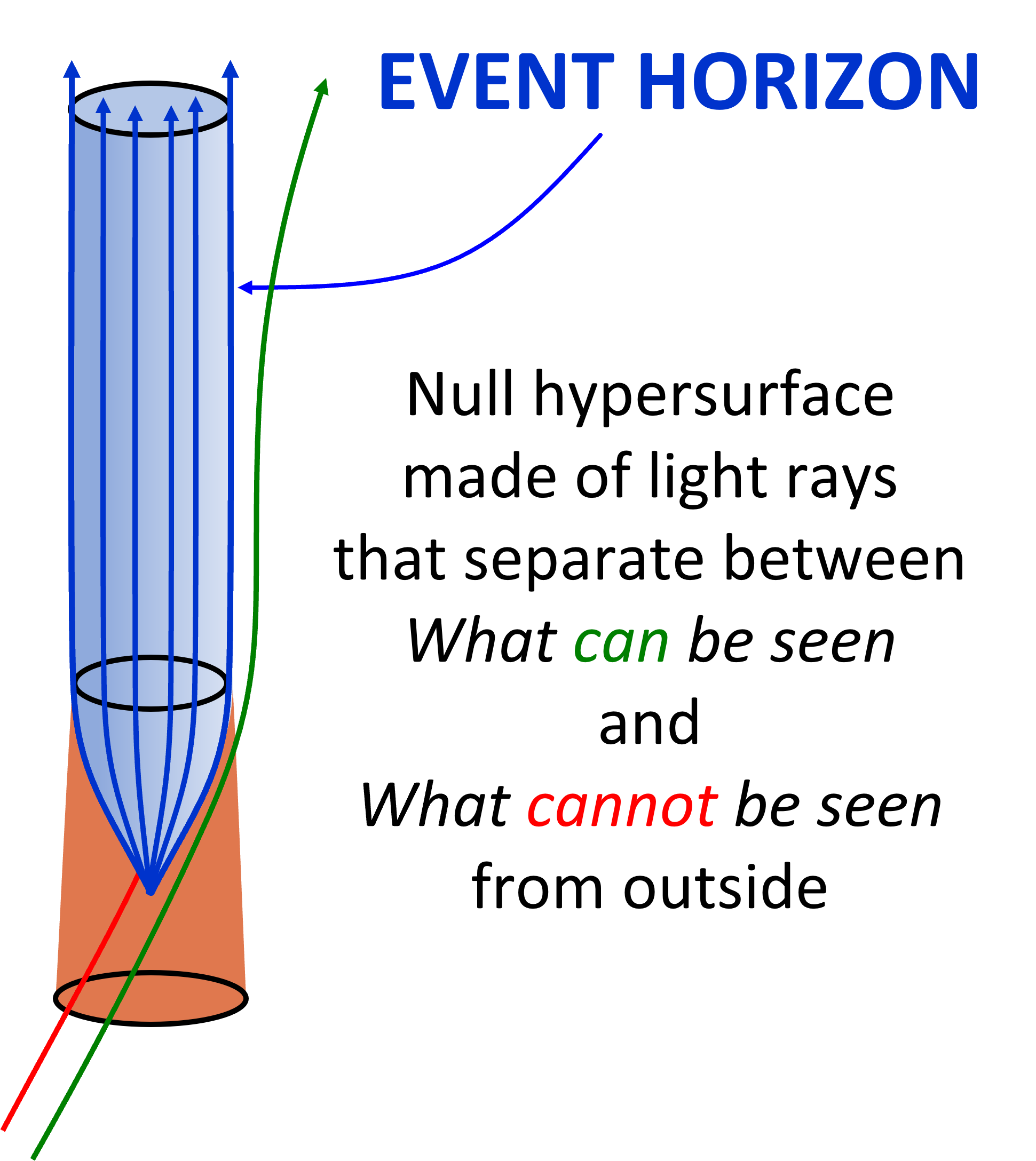}
  \caption{Event horizon.}
  \label{fig:event_horizon}
\end{subfigure}
\begin{subfigure}[t]{0.45\linewidth}
  \centering
\includegraphics[scale=0.37]{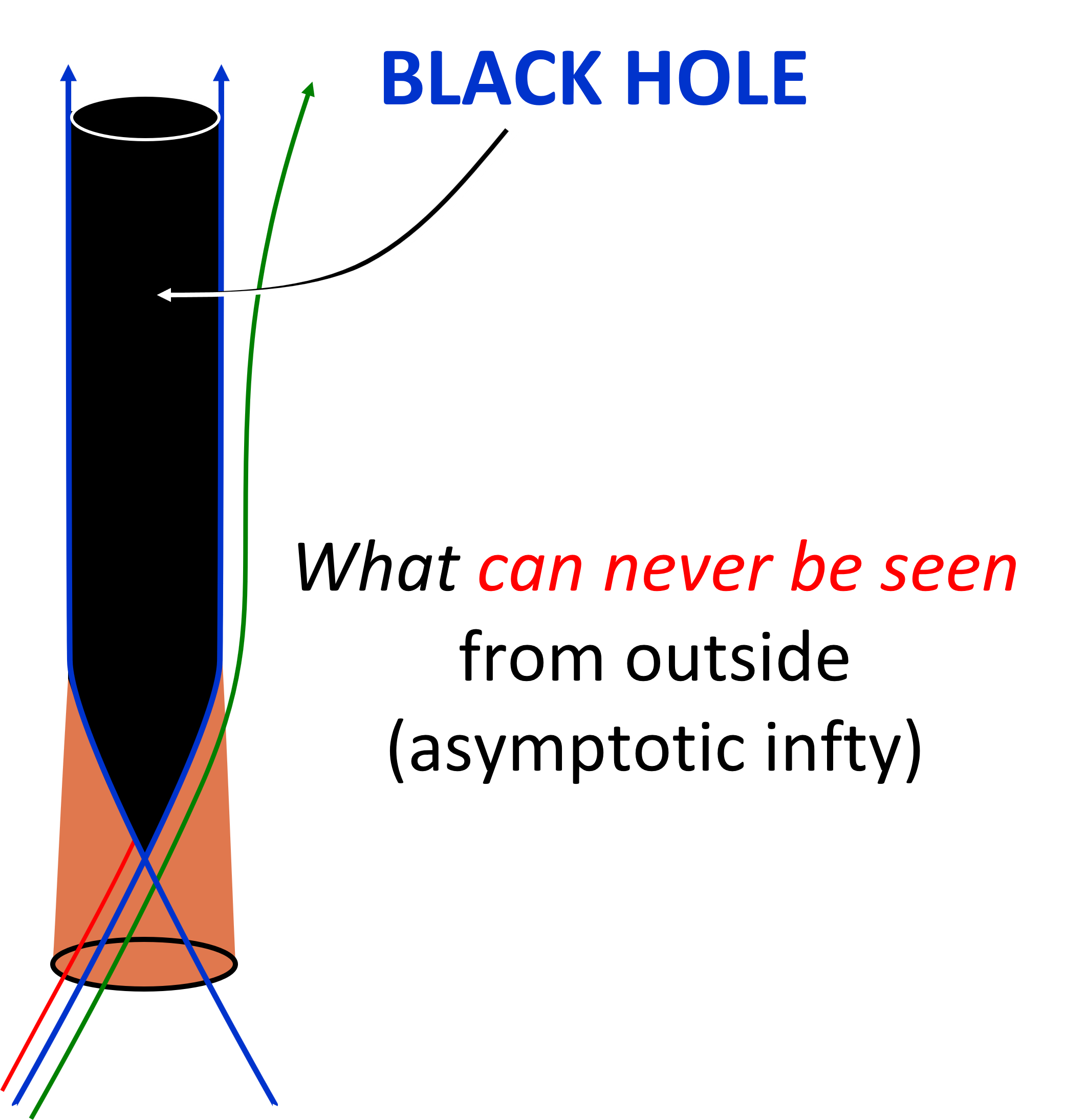}
  \caption{Black hole.}
  \label{fig:bh}
\end{subfigure}\hfill
\caption{Definitions of black hole and event horizon.}
\label{fig:blackhole}
\end{figure}

\paragraph{Event horizon and black hole.}
The event horizon is the surface traced
(generated) by these frozen null geodesics that bipartition the spacetime into causally separated regions, in the sense that nothing beyond it can be seen by any
external observers. Hence its name (figure~\ref{fig:event_horizon}).

The region of spacetime enclosed by the event horizon is what we call the \emph{black hole}. No signals, including any type of light, sent from the black hole can reach the far asymptotic region. Hence its name (figure~\ref{fig:bh}).

The event horizon is a three-dimensional null
hypersurface in four-dimensional spacetime, but it is very common to talk about the horizon as the two-dimensional spatial sections of this hypersurface, which in
this case are spheres. 

In a collapsing situation, the
null geodesics only begin to form the event
horizon at some instant. The points
where this happens are caustics (since light rays cross at
them). They are singular points of the surface,
but not singularities of spacetime.

\begin{figure}
\centering
\begin{subfigure}[t]{0.5\linewidth}
  \centering
\includegraphics[scale=0.4]{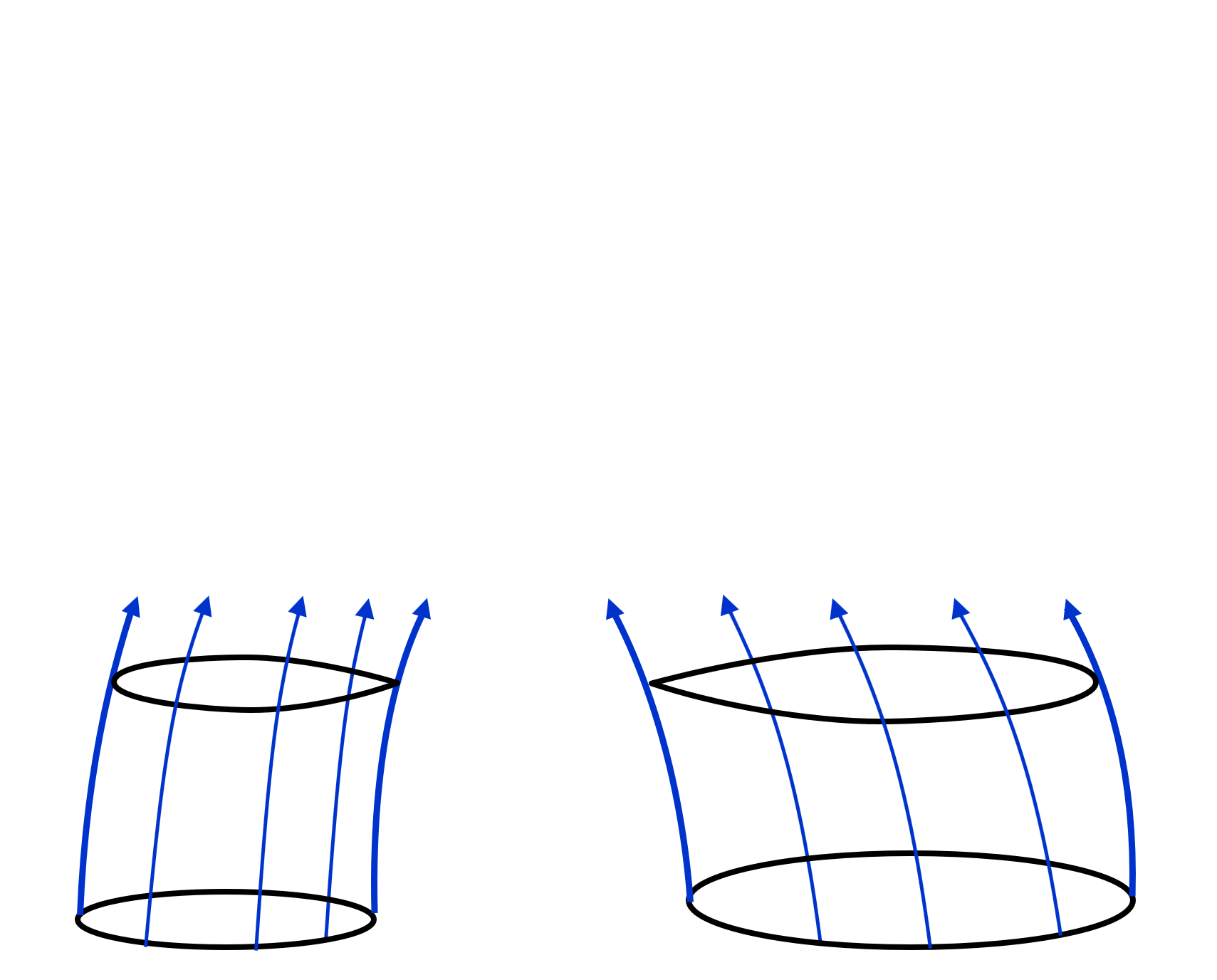}
  \caption{Two black holes approaching each other.}
\label{fig:1}
\end{subfigure}\hfill
\begin{subfigure}[t]{0.5\linewidth}
  \centering
\includegraphics[scale=0.35]{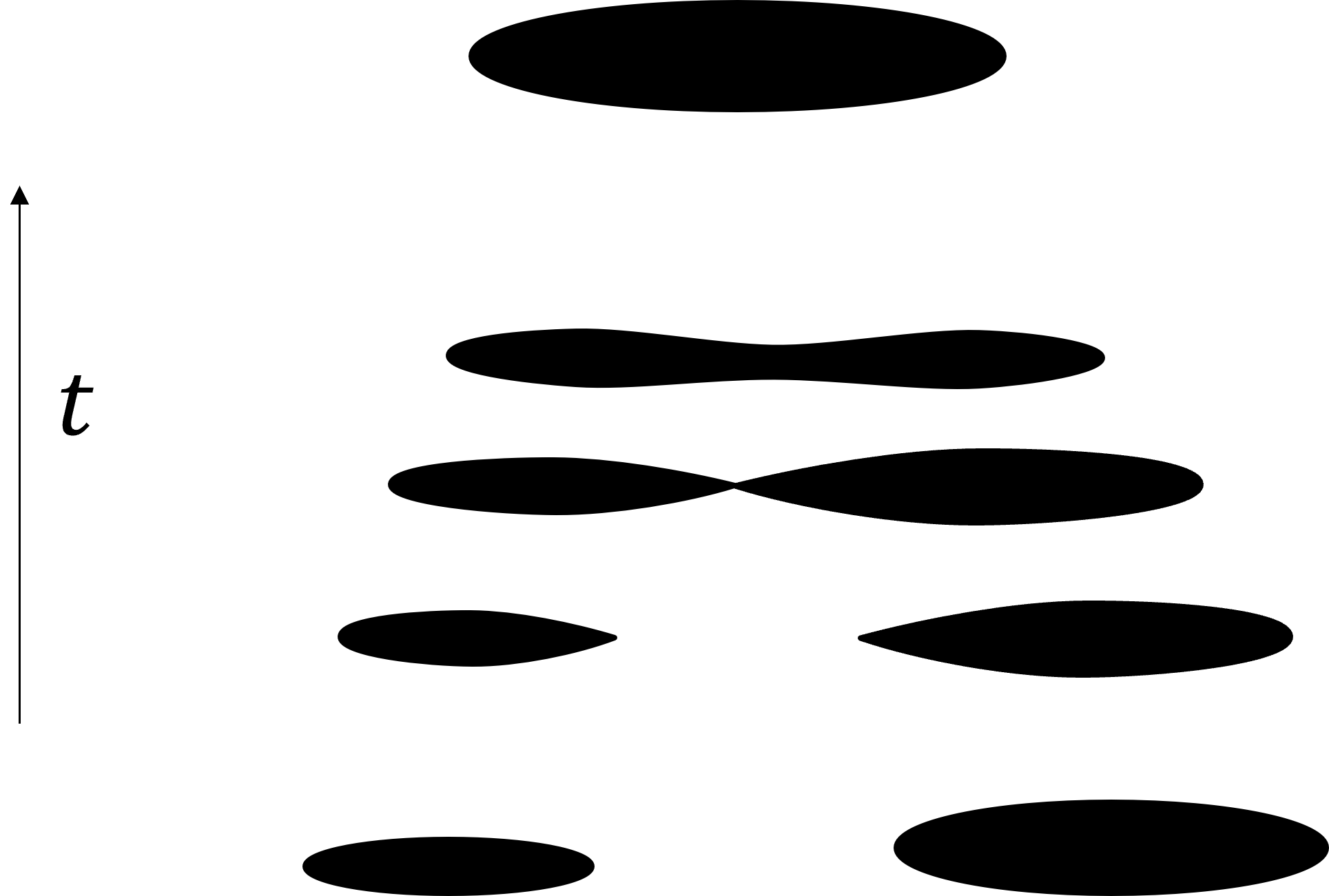}
  \caption{Constant time snapshots: the black holes fuse.}
  \label{fig:2}
\end{subfigure}

\vspace{30pt}

\begin{subfigure}[t]{0.5\linewidth}
  \centering
\includegraphics[scale=0.4]{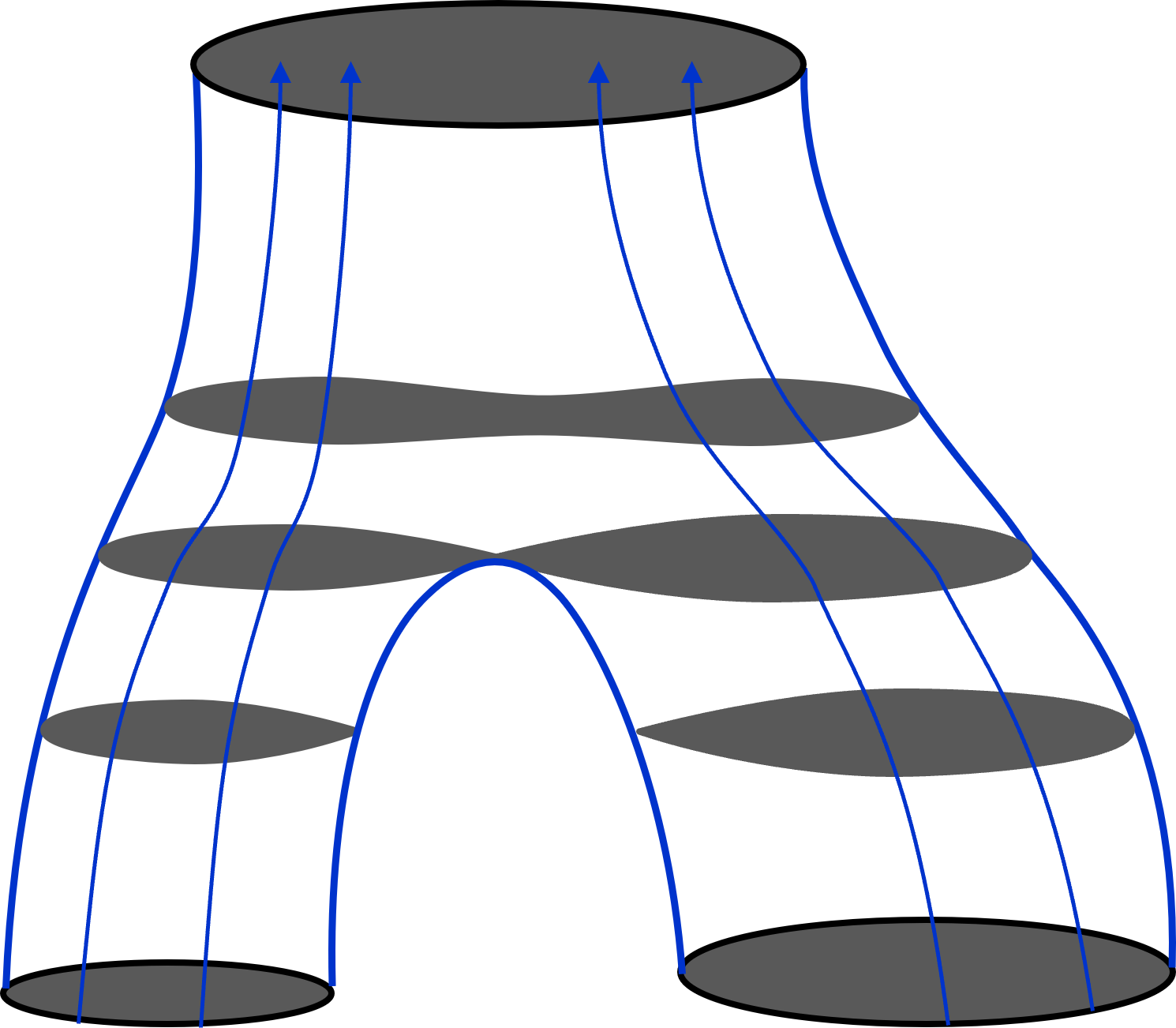}
  \caption{Continuously trace the horizons during the merger.}
\label{fig:3}
\end{subfigure}\hfill
\begin{subfigure}[t]{0.5\linewidth}
  \centering
\includegraphics[scale=0.4]{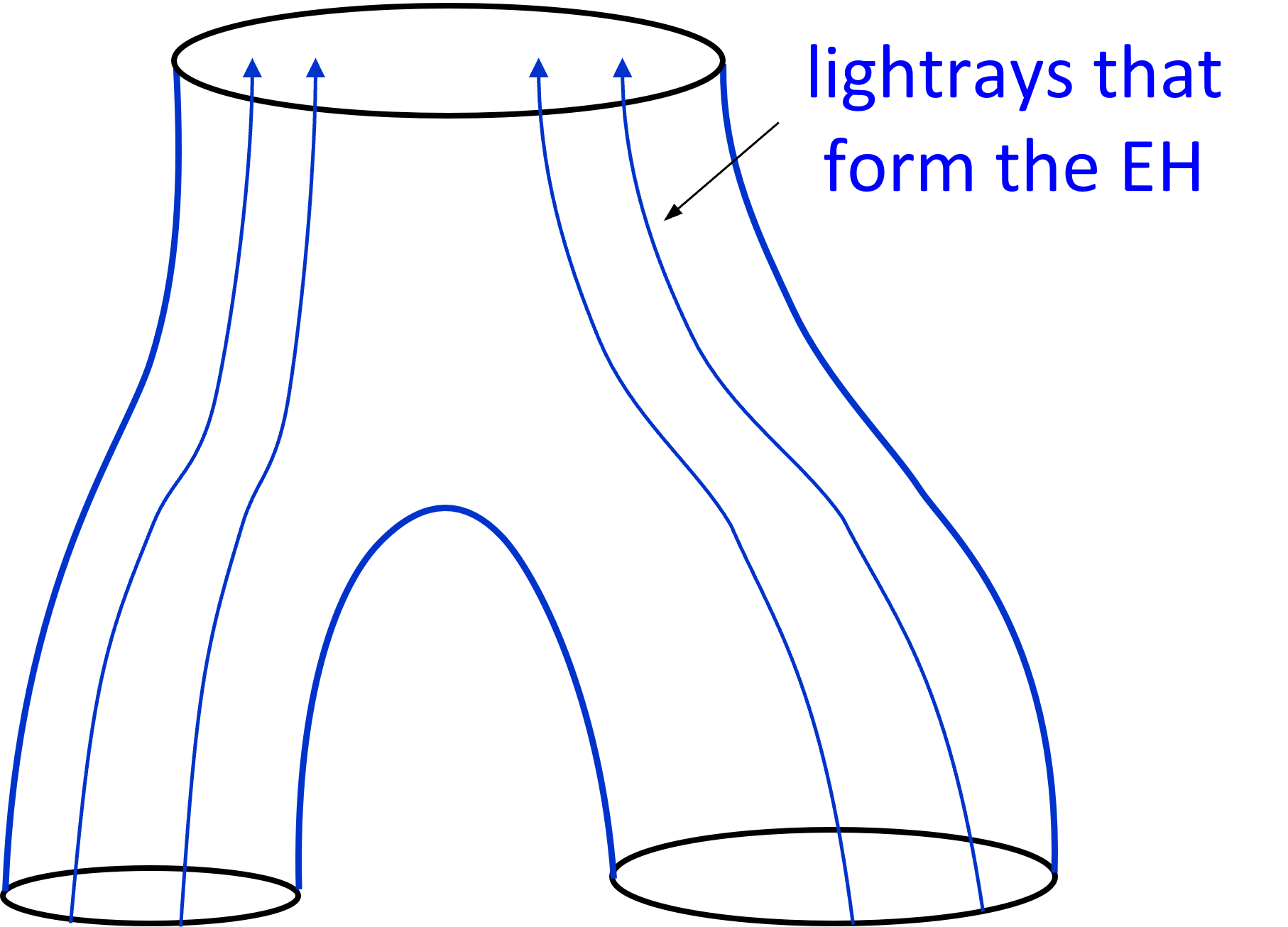}
  \caption{The event horizon is a pair of pants made of light rays.}
  \label{fig:4}
\end{subfigure}
\caption{Event horizon in the fusion of two black holes.}
\label{fig:pants}
\end{figure}

\paragraph{Event horizon in a binary black hole merger.}
Now that we understand what the event horizon is---a family of null geodesics, with the property that they bound a region of spacetime causally amputated from the late asymptotic region---let us consider the event horizon in a
process where two black holes merge to form a
single one (figure~\ref{fig:pants}). 

We begin with two cylindrical null surfaces,
corresponding to the event horizons of the initial
black holes. Viewed in constant time snapshots, we expect
that the two (approximately) spherical black holes
come together, and then fuse into a single one.
We can continuously trace the null surface along the merger to find the
shape of the event horizon: it takes the form of a pair of pants.\footnote{The surface is not completely smooth, since new light rays are added to it at the crotch of the pants, where a crease forms. This is a spacelike set of points which is not shown in figure~\ref{fig:pants}.}

\subsection{General theorems: Singularity (Penrose) and Area (Hawking)}

The central results of the classical theory of black holes are two theorems that apply very broadly and have deep and wide consequences. Penrose's singularity theorem is especially relevant for the collapse that forms a black hole. Hawking's area theorem instead brings out consequences for the merger of two black holes. The two results are indeed so profound that they also point towards directions beyond Einstein's classical theory.

\subsubsection{Trapped surfaces, apparent horizons, and the singularity theorem} 

\begin{figure}
\centering
\begin{subfigure}[t]{0.45\linewidth}
  \centering
\includegraphics[scale=0.32]{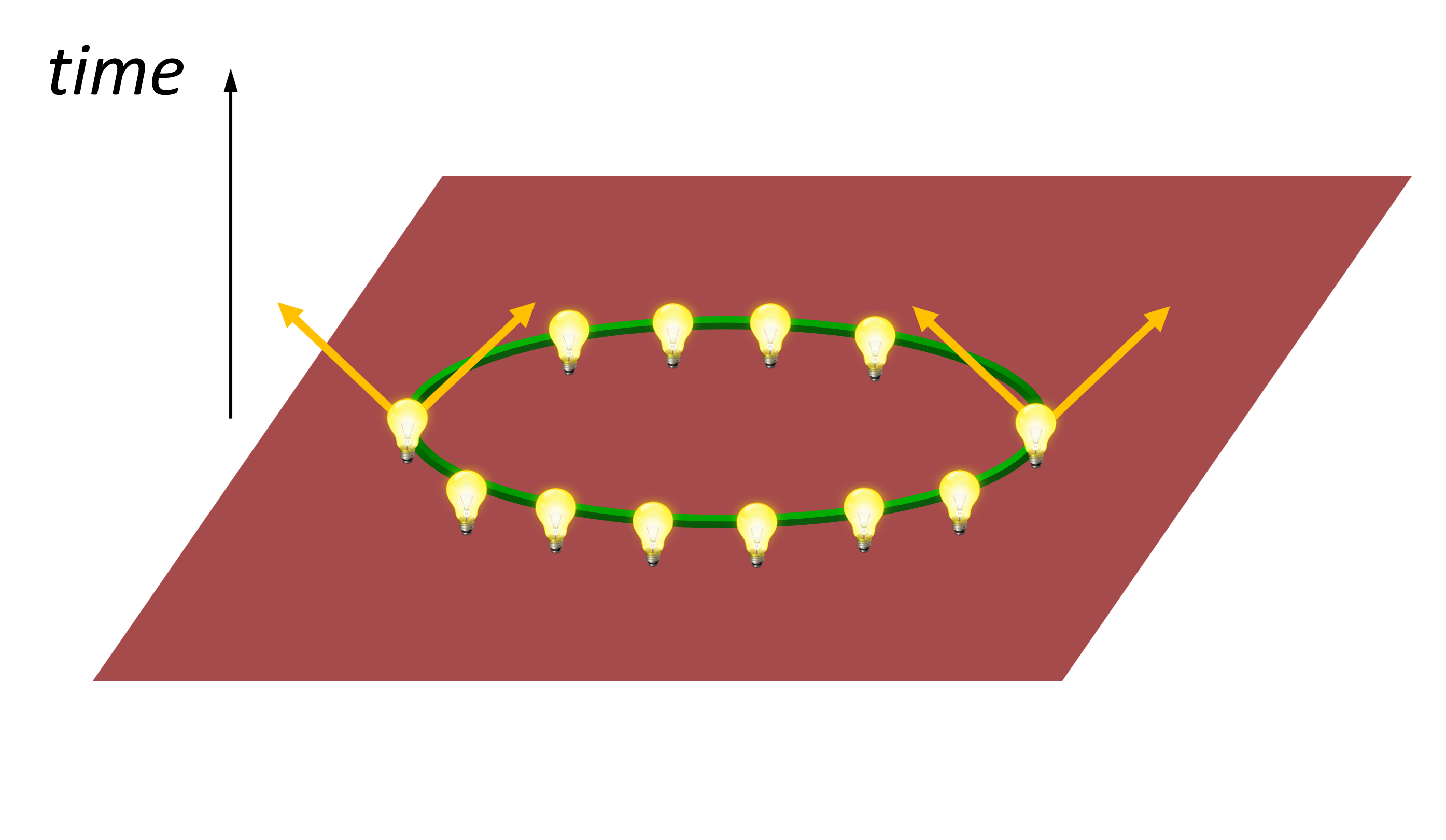}
  \caption{Distribute light bulbs on a closed surface (\eg a sphere) and flash them at a given instant in time. Then, follow the lightfronts that emanate from the surface.}
\label{fig:5}
\end{subfigure}\hfill
\begin{subfigure}[t]{0.45\linewidth}
  \centering
\includegraphics[scale=0.3]{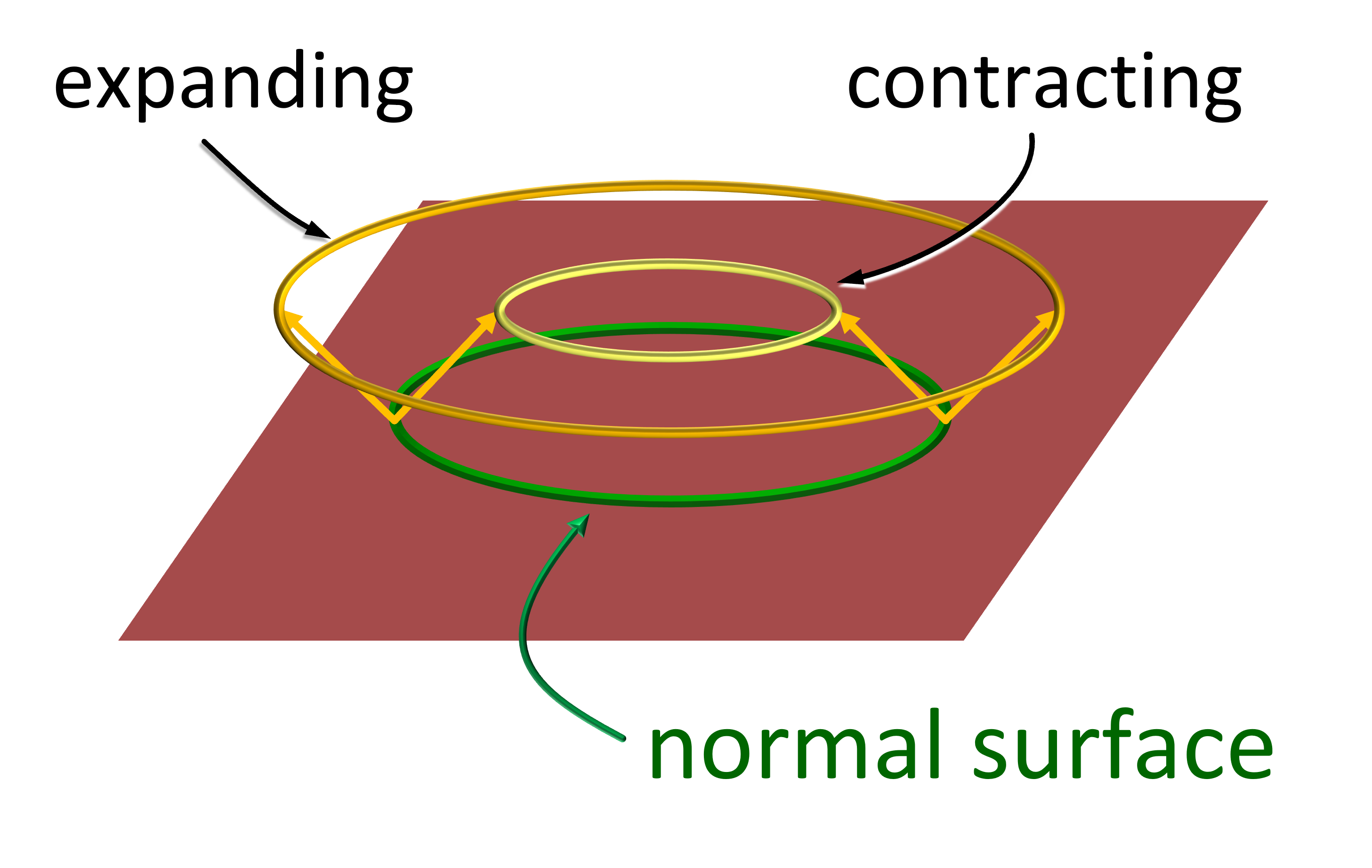}
  \caption{There will be a lightfront that moves outward from the surface and another inward. In a normal situation, the former will expand and the latter will contract.}
  \label{fig:6}
\end{subfigure}

\vspace{30pt}

\begin{subfigure}[t]{0.45\linewidth}
  \centering
\includegraphics[scale=0.3]{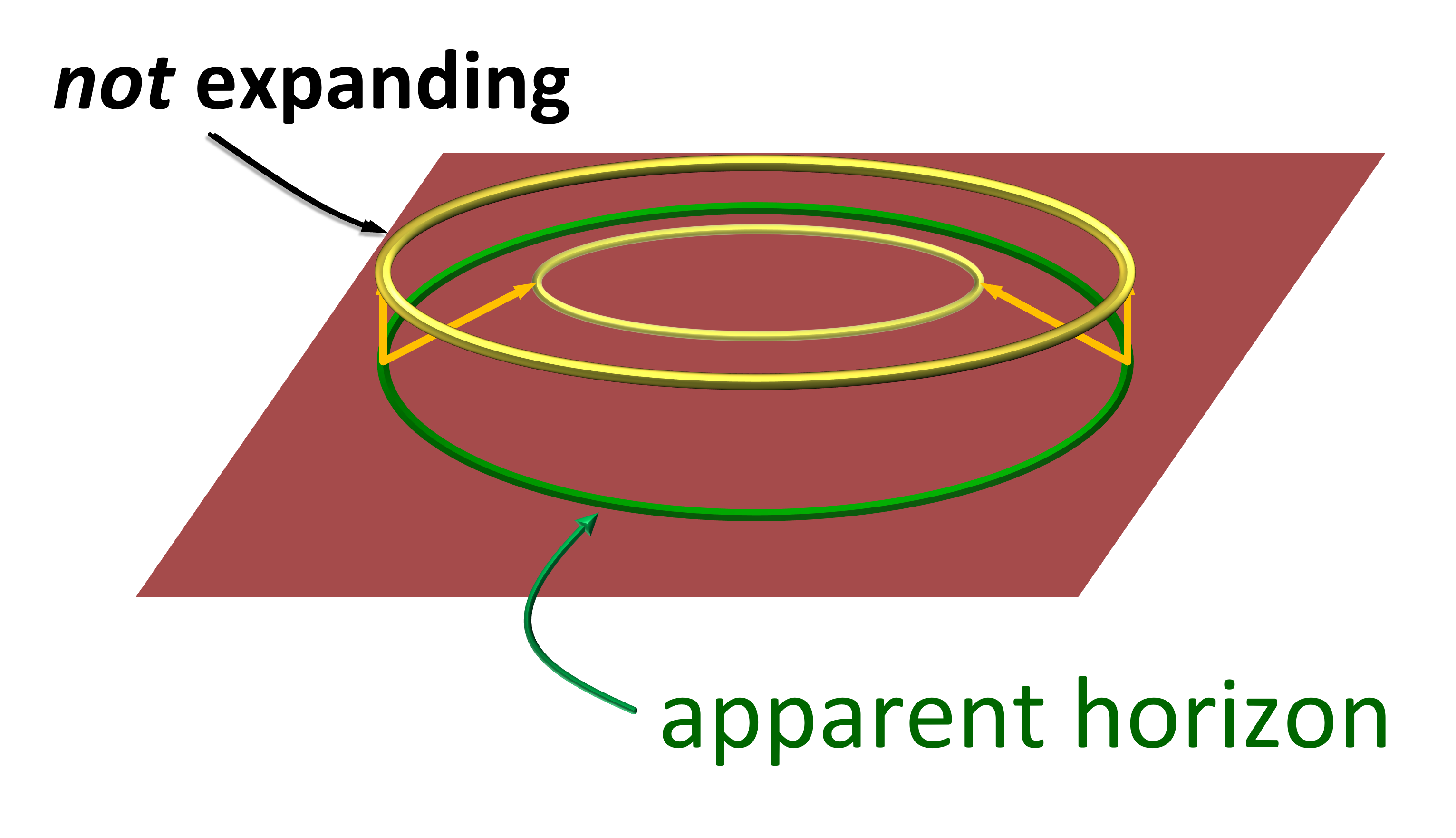}
  \caption{When gravity is very strong, it can happen that lightfronts sent outwards do not expand. We say that the surface is an apparent horizon.}
\label{fig:7}
\end{subfigure}\hfill
\begin{subfigure}[t]{0.45\linewidth}
  \centering
\includegraphics[scale=0.3]{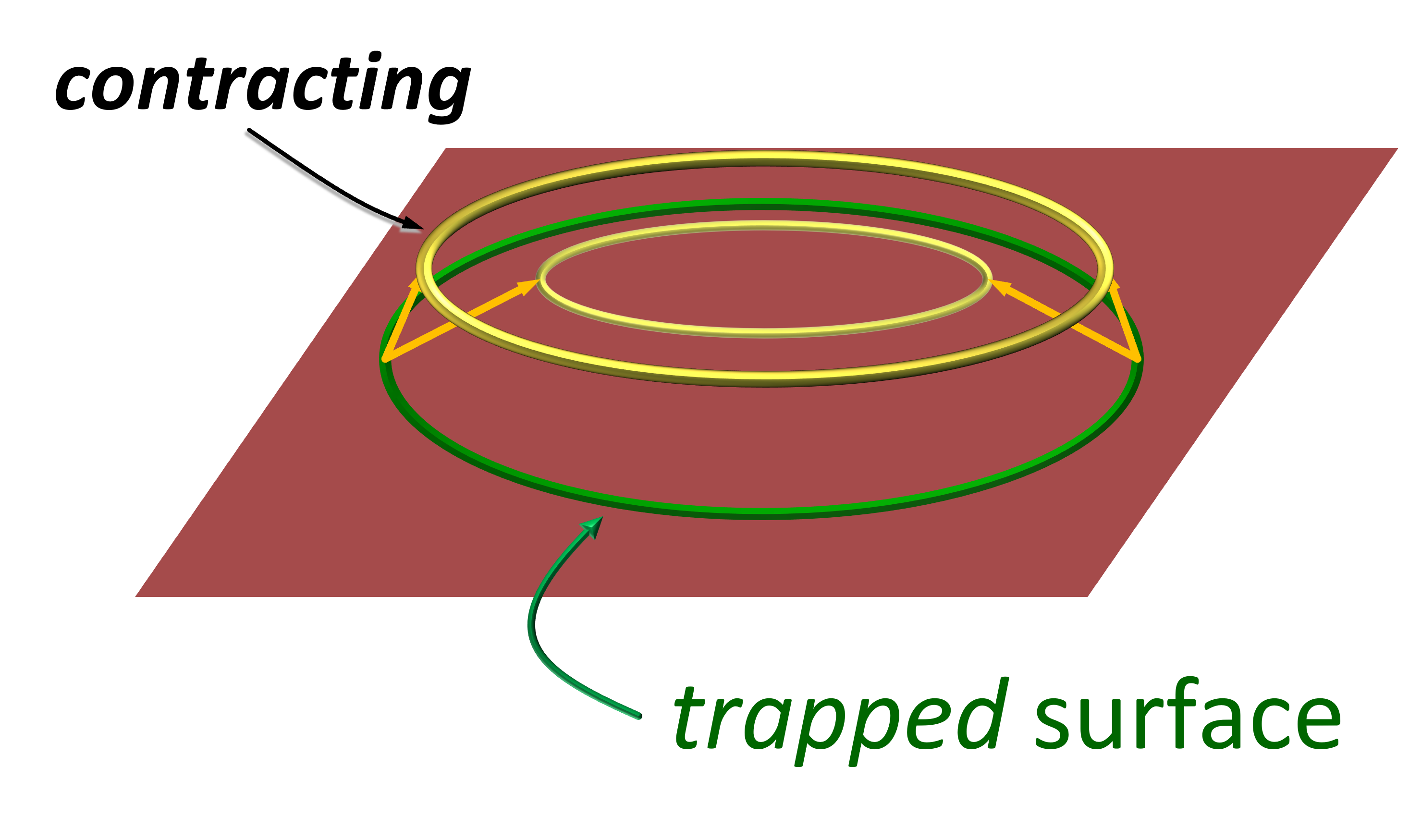}
  \caption{It may even occur that the outgoing lightfronts actually contract. Then the surface is trapped.}
  \label{fig:8}
\end{subfigure}
\caption{Apparent horizon and trapped surfaces. Penrose's theorem predicts that if trapped surfaces appear at some moment, then a singularity will form in a finite time in the future.}
\label{fig:trapped}
\end{figure}

In a revolutionary (and eventually award-winning) article in 1965, Penrose introduced the notion of \emph{trapped surface} and proved that it is an indicator of a collapse so strong that it necessarily leads to a singularity in the future---meaning an \textit{instant} beyond which no predictions can be made using Einstein's theory. 

In order to understand what this means, we begin by considering a closed surface at some instant of time, such as a sphere (see figure~\ref{fig:trapped}). We distribute a set of light bulbs on it and flash them at a given moment. The light rays that emanate from the surface will form lightfronts, some of which will travel outwards from the surface while others will propagate inwards. Normally, the outgoing front will expand and the ingoing will contract.

But then Penrose imagined a situation where the pull of gravity is so strong that the outgoing lightfronts fail to expand but instead, their area remains constant. You may recall that we have seen this before: it happens on the event horizon of the static black hole. But here we are not assuming any specific spacetime geometry, nor are we attempting to follow whether these light rays escape or not to infinity: we are simply watching if the lightfronts leaving from the surface instantly grow or not. A surface with non-expanding outgoing lightfronts is called an \emph{apparent horizon}. In the Schwarzschild static solution, the event horizon\footnote{That is, a constant-time spherical section of it.} is an apparent horizon, but we are allowing for more general time-dependent situations where the two notions need not coincide.

Furthermore, when gravity is so strong as to force the outgoing lightfronts to actually \emph{contract}, then we say that the surface is \emph{trapped}. How is such weird behavior possible at all? We can see from our previous study that surfaces of this kind actually exist inside the Schwarzschild black hole (a weird place indeed), but forget about that for a moment and think about what is going on here: \emph{the geometry of spacetime itself is collapsing so strongly that it drags everything with it, including light}. You may then fear that the universe is catastrophically headed towards a Big Crunch. The collapse, however, need not encompass the entire universe but only a smaller, compact region of it. How will this end?

Badly, said Penrose. His theorem asserts that in this situation a singularity will form within a finite time in the future. 

So, if you find yourself caught between two collapsing lightfronts, then you are trapped and your time will come to an end. So it goes.

Let us be more precise. Assume that:
\begin{itemize}
    \item Energy along null geodesics is non-negative (this is called the \emph{null energy condition}).
    \item In some noncompact constant-time slice of spacetime, there exists a compact trapped surface.
\end{itemize}
Then it follows that
\begin{itemize}
    \item there must exist null geodesics that are incomplete, and their further evolution cannot be predicted using Einstein's equations.
\end{itemize}
The first condition effectively implies that the gravity created by any matter in the geometry has an attractive effect on light (such as we see in the light rays in fig.~\ref{fig:cones}).

The incompleteness of geodesics may sound like a strangely mathematical concept, but Penrose introduced it as a way to express that some sort of singular behavior arises---evolution along a null geodesic stops at some moment. It is a very weak definition of a singularity, since it says nothing about what is happening to spacetime as the singularity is approached. In particular, it does not say that the curvature is \emph{necessarily} diverging, even though that is physically the most natural and interesting possibility. It would mean that the spacetime geometry of Einstein's classical theory is fated to break down and be superseded by a deeper quantum notion. 

Nevertheless, even with these physical limitations, geodesic incompleteness is mathematically a very convenient concept, since it allows to prove theorems for the appearance of singularities.

Finally, observe that the theorem refers to apparent horizons but not anywhere to event horizons---and it is the latter that are boundaries to causal communication. In particular, the theorem does not assert that an event horizon will appear hiding the singularity from the view of faraway observers. That is, it does not predict that a black hole will form as a consequence of the collapse. Still, this seems the most likely possibility, so Penrose was led to hypothesize that indeed it is what will happen: a \emph{cosmic censorship conjecture}.

\subsubsection{The horizon area theorem} In 1971 Hawking used Einstein's theory to show that the total area $\mc A_H$ of sections of the event horizon can never
decrease as time evolves,
\begin{align}
    \Delta \mc A_H\geq 0\,.
\end{align}

His theorem makes two assumptions:
\begin{itemize}
    \item  Energy along null geodesics is nonnegative (null energy condition again).
    \item There are no naked singularities (cosmic censorship).
\end{itemize}
One can then show that the null generators of the horizon can not come close, that is,
the area of a cross-section of a pencil of these null rays cannot shrink. Furthermore, it can also be proven, very generally, that
null generators can be added, but not removed, from the event horizon. These effects can only lead to an \textit{increase} of the total horizon area, and never to a decrease. 

In his original article, Hawking motivated this result by the consequences it has for the amount of gravitational radiation that can be generated during a merger. The energy that is radiated away when two black holes collide must come from the conversion of the mass of the initial black holes into radiation\footnote{If the colliding black holes initially have relativistic velocities, their kinetic energy must be added to the total energy budget.}. The area of the horizon in \eqref{schw2} is 
\begin{align}
   \mc A_H=4\pi r_0^2 = 16\pi M^2\,,
\end{align}
and therefore if a very large fraction of the mass were converted into radiation, there might be a possibility that the final area was less than the total initial black hole area. The area theorem forbids this, and puts an absolute upper bound on the amount of mass-to-radiation conversion in the collision.

This theorem must be revisited for quantum black holes, due to an effect also discovered by Hawking that will be the subject of Section~\ref{sec:bhevap}. Quantum effects can (with some limitations) violate the null energy condition, and make the horizon of the black hole shrink. This
is what happens during the evaporation of a black hole by the emission of Hawking radiation.

\subsection{Non-radial null geodesics: the photon ring}

There are other properties of a black hole, besides its horizon and singularity, which will prove to be important and which are revealed by a study of light rays propagating in its geometry. For this purpose, we turn to the study of non-radial null geodesics.

\begin{figure}
\centering
\includegraphics[width=.55\textwidth]{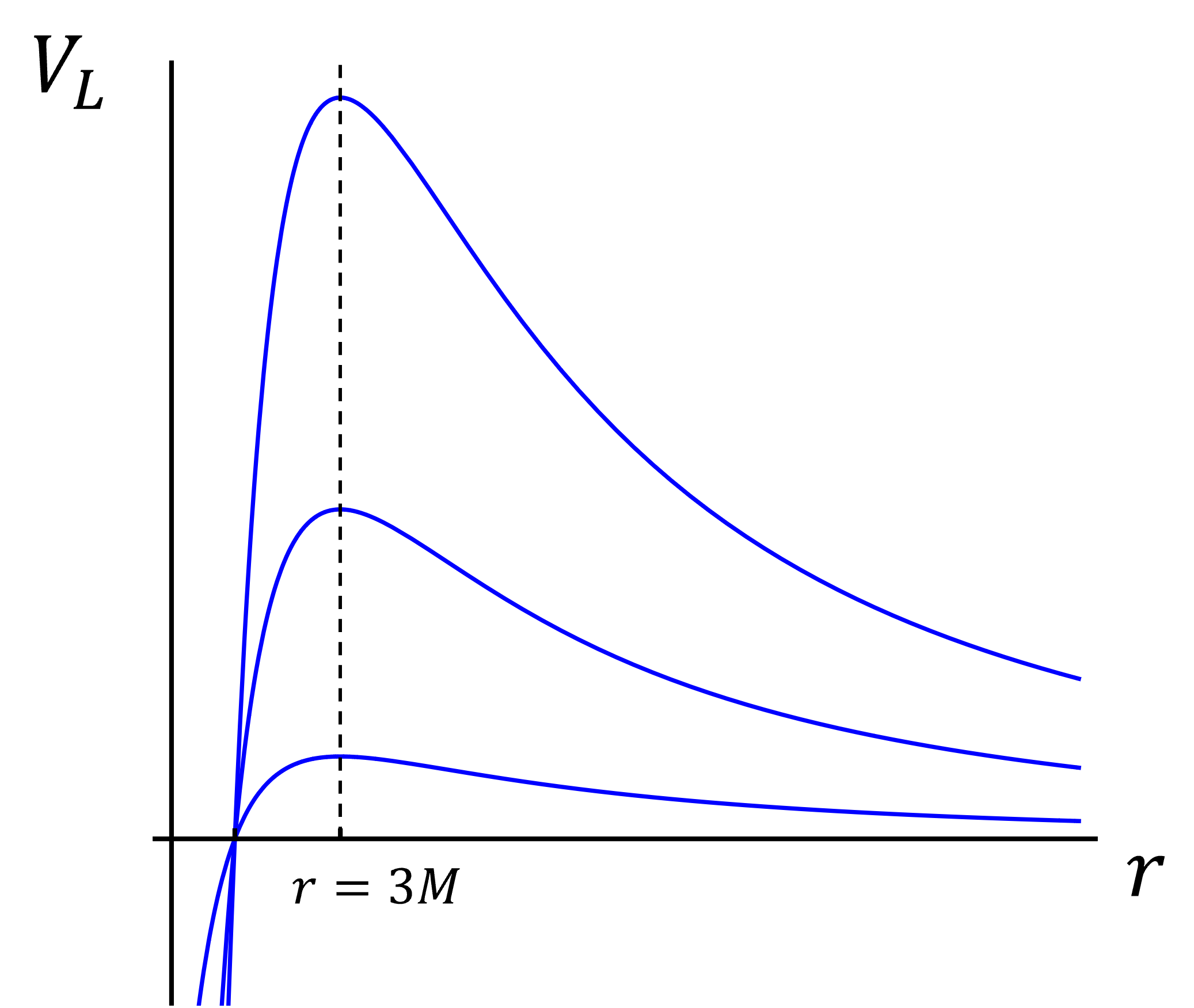}
\caption{Effective radial potential \eqref{v_eff} for circular light rays with angular momenta $L=1,2,3$. The potential vanishes at $r=2M$ and has a maximum at $r=3M$ for all $L$. This maximum defines the radius of the photon ring (or photon sphere). The height of the potential is $\propto L^2$.}
\label{fig:Veff}
\end{figure}

We know that in Newtonian mechanics the conservation of angular momentum fixes the trajectories of planets to lie on a fixed plane. The same is true in GR, not only for particle geodesics but also for null geodesics. Therefore, we choose the trajectory to lie in the equatorial plane $\theta=\pi/2$ of the Schwarzschild geometry. The metric \eqref{schw2} then reduces to 
\beq
ds^2|_{\theta=\pi/2}=-\lp1-\frac{2M}{r}\rp dt^2+\frac{dr^2}{1-\frac{2M}{r}}+r^2d\phi^2\,.
\eeq
This geometry is static and rotationally symmetric, so there will be two constants of motion along the geodesics: energy $E$ and angular momentum $L$.  Introducing an affine parameter\footnote{For massive test particles the proper time is usually taken for the affine parameter. However, for light rays the proper time is zero.} $\lambda$ for the geodesics, these constants of motion are easily found to be
\begin{align}
    E&=\lp 1-\frac{2M}{r}\rp \frac{dt}{d\lambda}\,, \\
    L&=r^2(\lambda)\frac{d\phi}{d\lambda}\,.
\end{align}
Using them we can derive a differential equation for $r(\lambda)$. Since light rays must satisfy $ds^2=0$, we find 
\beq
\label{light_rays_potential}
\lp \frac{dr}{d\lambda} \rp ^2 +\lp 1-\frac{2M}{r} \rp \frac{L^2}{r^2}=E^2\,.
\eeq
Recall now that the energy conservation equation in classical mechanics for a particle in a potential is given by
\beq
\frac{1}{2}\dot{r}^2+V_L(r)=E\,.
\eeq
We notice that the geodesic equation \eqref{light_rays_potential} is of this form if we identify an effective potential
\beq \label{v_eff}
V_{L}(r)=\frac{1}{2}\lp 1-\frac{2M}{r} \rp\frac{L^2}{r^2}\,.
\eeq
Here $L^2/2r^2$ represents a centrifugal barrier term and $-ML^2/r^3$ is a relativistic gravitational attraction term for light rays: observe that the mass $M$ exerts an attractive effect on the rotational (kinetic) energy $\sim L^2/r^2$ of a circular ray at radius $r$. This attraction drastically modifies the behavior of the potential at short enough distances  (see figure~\ref{fig:Veff}).

\begin{figure}
\centering
\includegraphics[width=0.4\textwidth]{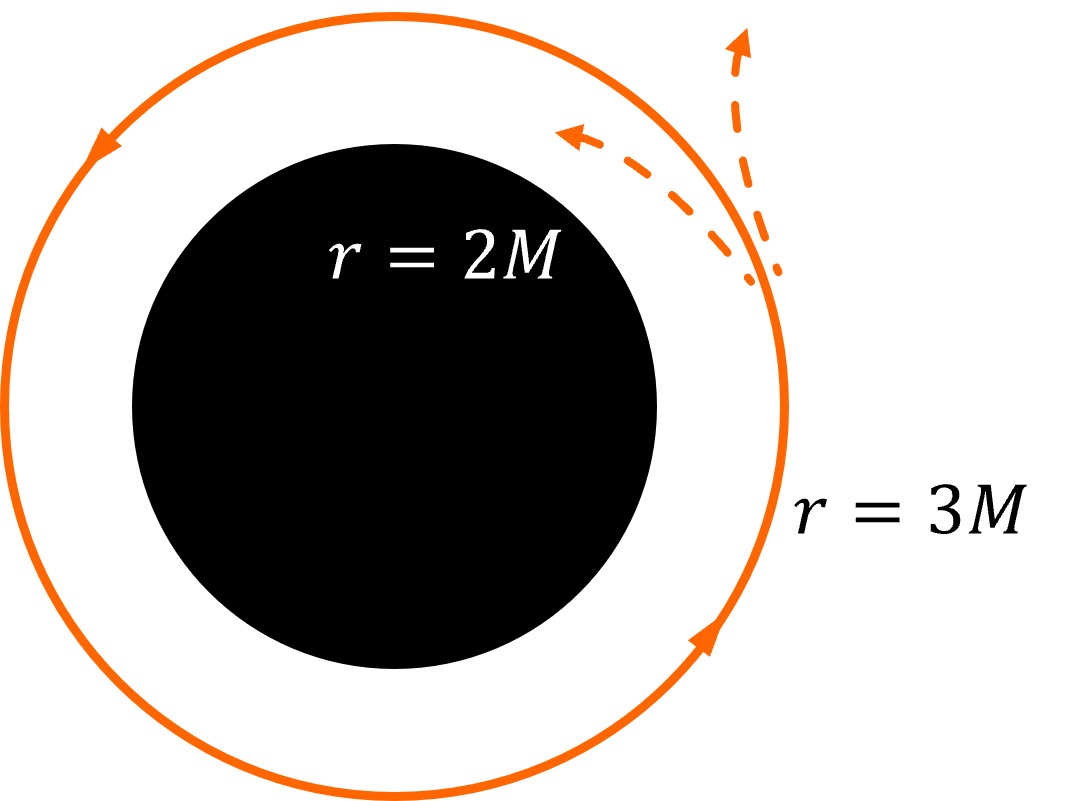}
\caption{Light rays in a circular orbit of radius $3M$ outside a static black hole form the photon ring (or photon sphere). The orbit is unstable.}
\label{fig:lightring}
\end{figure}

 Using the effective potential \eqref{v_eff} we can easily build intuition about circular light rays. First, we notice that $V_{L}(r)$ approaches zero for $r=2M$ (the horizon, so this is expected) and for $r\rightarrow \infty$ (also expected). Since $V_L>0$ for $2M<r<\infty$, there must be a maximum in between.
 We can easily verify that $V'_{L}(r=3M)=0$, and $V''_{L}(r=3M)<0$ so this maximum (a global one) is at $r=3M$. It corresponds to unstable circular trajectories of light rays, or photons, and for this reason this is called the \textit{photon ring} or more generally, the \textit{photon sphere} (see figure~\ref{fig:lightring}). It is a  central feature in black hole imaging, and it will play an important role in the next section.
 
 You can verify with an easy calculation that the time a photon takes to go once around the ring is $\Delta t=6\pi\sqrt{3}M$, and that it has $L/E=3 \sqrt{3} M$. 

\section{The black hole that vibrates}

Now we turn to investigating what happens to the Schwarzschild black hole when we introduce a little perturbation. For instance, we may want to figure out whether it is stable, namely, will it return back to the initial state? If so, as it returns to equilibrium, what kind of signal does the black hole emit, and what information does this signal carry? We refer to this as the \textit{ringdown problem}. 

We will see that the black hole relaxes down to equilibrium like a bell does: with damped oscillations
\begin{align}\label{omega}
    \exp(-i\omega t)=\exp\lp-2\pi i f t-t / \tau\rp\,,
\end{align}
where $f$ is the real part of the oscillation frequency and $\tau$ the damping time. These are quasinormal vibrations---not ``normal'', since the frequencies have an imaginary part owing to the absorption by the black hole, which is a dissipative effect. 

This problem is studied using linear perturbation theory, which in GR is a rather technical subject, but we will sketch its main features. Anticipating the final result, 
 one finds that the black hole does indeed radiate energy into gravitational waves, relaxing back to the initial state through damped oscillations of the form \eqref{omega}
with $\tau>0$. The black hole is therefore `mode-stable', and the amplitude of the perturbations decreases by a factor of $\exp(-1/f\tau)$ with each oscillation.

\subsection{Master equation for black hole perturbations}
The starting point is to introduce a perturbation of the metric,
\beq \label{pertm}
g_{ij}(x)=g_{ij}^{(0)}(x)+\varepsilon h_{ij}(x), \ \lvert \varepsilon\rvert \ll 1\,,
\eeq
where $\varepsilon$ is a parameter considered to be very small and all background (initial) quantities are denoted with the superscript $(0)$. In order to find how this perturbation evolves in time, we insert  \eqref{pertm} into Einstein's vacuum equations,
\beq
R_{ij}\lp g\rp=0 \Rightarrow R_{ij}\lp g^{(0)}+\varepsilon h\rp=\varepsilon\mathcal{L}(h)=0\,.
\eeq
Here we have used that $R_{ij}^{(0)}\lp g^{(0)}\rp=0$, and $\mathcal{L}$ is a fairly complicated differential tensor operator (the Lichnerowicz operator). In this way we find a set of coupled second-order partial differential equations for $h_{ij}$---not so easy to unravel. To further simplify it, we use that the spherical symmetry and time independence of the background metric allow us to decompose $h_{ij}$ as
\beq \label{hdecom}
h_{ij}(x)=e^{-i\omega t} Y_{lm}(\theta,\varphi)h_{ij}^{(\omega,l,m)}(r)\, ,
\eeq
where $Y_{lm}(\theta,\varphi)$ are spherical harmonics\footnote{We are oversimplifying. In the full problem, one needs to consider not only these scalar spherical harmonics but also vector and tensor harmonics. Yes, it is complicated.}. We must bear in mind that there are ambiguities plaguing this problem since some solutions are not physical but are instead pure gauge: they correspond to coordinate transformations such as the ones we saw in \eqref{infdiff}, \ie $h_{ij}=\partial_iv_j(x)+\partial_jv_i(x)$. One possibility is to fully fix the gauge with a specific choice of coordinates. Another option, possibly more common, is to leave at least some gauge freedom unfixed and work with combinations of the metric components that are invariant under the remaining gauge transformations.

With the separation of variables \eqref{hdecom} one hopes to obtain a set of coupled second-order \emph{ordinary} differential equations. After quite some toil, it is indeed possible to not only obtain such a set of ODEs, but, remarkably, also reduce them to a \emph{single} second-order ODE for a gauge-invariant function $\Psi_{\omega l m}(r)$ from which the original metric perturbations, $h_{ij}(r)$, can be recovered up to pure gauge configurations.  This $\Psi_{\omega l m}$ is called a  \textit{master variable}, and the \textit{master equation} that it must solve can be written in the Schr\"odinger form  
\beq \label{mastereq}
\frac{d^2\Psi_{\omega l m}}{d{r_*}^2}+\lp \omega^2-V_{l}(r)\rp\Psi_{\omega l m }(r)=0\, ,
\eeq
with
\beq \label{pot-s2}
V_{l}=\lp 1-\frac{2M}{r}\rp\left[\frac{l(l+1)}{r^2}-\frac{6M}{r^3}\right]\, .
\eeq
The radial variable $r_*$ is the tortoise coordinate introduced in \eqref{tort}, which was useful to study the propagation of light rays, and thus also for massless fields. We regard $r$ as a function of it, $r=r(r_*)$. 

The master equation \eqref{mastereq} with the potential \eqref{pot-s2} describes the propagation of weak gravitational spin-2 fields around a black hole solution. We can also consider the much simpler spin-0 case of a scalar field $\Phi(x)$ and the spin-1 case of the gauge vector field $A_\mu(x)$, which obey the equations 
\beq
\Box \Phi(x)=0, \qquad \nabla_\mu F^{\mu\nu}=0\, ,
\eeq
where $F_{\mu\nu}=\partial_\mu A_\nu-\partial_\nu A_\mu$. Written more explicitly, they are
\begin{align}
    \frac{1}{\sqrt{-g}}\partial_\mu\lp\sqrt{-g}g^{\mu\nu}\partial_\nu\Phi\rp=0\, ,\qquad
    \frac{1}{\sqrt{-g}}\partial_\mu\lp\sqrt{-g}F^{\mu\nu}\rp=0,
\end{align}
where $g=\det(g_{ij})$. These equations are linear, so there is no need to perform any perturbation expansion---we are considering $\Phi(x)$ and $A_\mu(x)$ as `test fields' that do not backreact on the geometry. The equation for the scalar field is easy to work out (see Problem~\ref{prob:scalsch}), while for $A_\mu(x)$ one must, again, bear in mind gauge issues. At the end of the day, the equations can be reduced to a master equation of the form \eqref{mastereq} with
\beq \label{pot_s}
V_{l}^{(s)}=\lp1-\frac{2M}{r}\rp\lp\frac{l(l+1)}{r^2}+(1-s^2)\frac{2M}{r^3}\rp,
\eeq
where 
\beq \label{cases}
s=\begin{cases}
    0, \ \text{scalar field} \\
    1, \ \text{gauge field} \\
    2, \ \text{gravitational field}.
\end{cases}
\eeq

\subsection{No-hair theorem and other basic features} \label{subsec:nohair}

To start with, we study whether there are zero-frequency, static solutions with $\omega=0$ for the different values for $s$. This is not very difficult, and one finds that:
\begin{itemize}
    \item For $s=2$ there exist solutions with $l=0$ and $l=1$. These perturbations add, respectively, a small mass and a small rotation to the black hole. 
    \item For $s=1$ there are again static solutions, in this case just for $l=0$. This perturbation adds charge to the black hole.
    \item For $s=0$ the equation does not have any zero-frequency solutions. 
\end{itemize}
These results provide a linearized version of the \textit{no hair theorem}: Stationary black holes are fully characterized by their mass, angular momentum, and charge, and they cannot have any scalar ``hair".

\begin{figure}
\centering
\includegraphics[width=.8\textwidth]{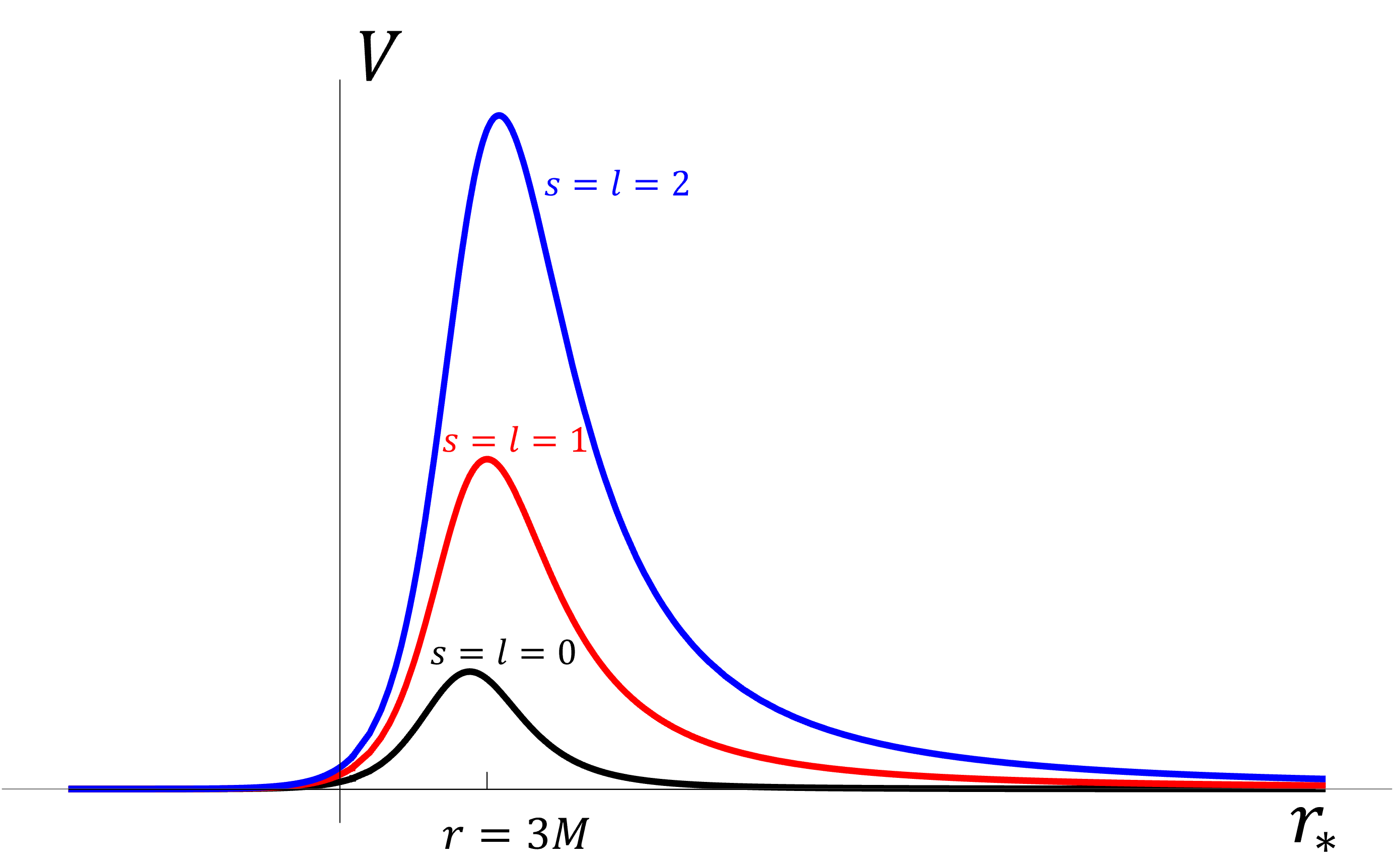}
\caption{Effective radial potential for the propagation of scalar fields (black), gauge fields (red), and gravitational fields (blue) in the Schwarzschild black hole background. We use the coordinate $r_*$, for which the horizon is at $r_*\to-\infty$. In each case, we show the potential for the lowest value of $l$ for which there are propagating modes, $l=s$. We can see that scalar s-waves will be easily absorbed, while gravitational waves have to surmount a large centrifugal barrier. The peak of the potential is close to the photon ring radius $r=3M$.}
\label{fig:Vsl}
\end{figure}

Let us now examine general features of the potential  \eqref{pot_s} (see fig.~\ref{fig:Vsl}). For gravitational wave astrophysics the interesting case is $s=2$, but since the potentials for the other $s$ are similar, the qualitative features do not differ much. For $l\gg1$, which is a limit of large quantum numbers called the eikonal limit, we recover the same potential as for the motion of light rays in \eqref{v_eff},
\beq
V_{L}(r)\approx V_{l}^{(s)}(r), \quad \text{with} \ L\approx l\gg 1\,.
\eeq
In general (and not only in this eikonal limit), the peak of $V_{l}^{(s)}(r)$ occurs at a value of $r$ close to the maximum of $V_{L}(r)$, namely the photon ring radius $r=3M$. We will see that this simple fact is extremely important to understand how and when black holes vibrate. Near this maximum, and for large $l$, the potential is approximately 
\beq\label{Vlmmax}
V_{l,\text{max}}^{(s)}\approx \frac{l^2}{27M^2}, \quad \text{for} \ l\gg 1\, .
\eeq
Its growth $\propto l^2$ is expected: this is a centrifugal barrier.

From the shape of this potential, we can readily infer that incident waves with $\omega^2>V_{\text{max}}\approx V(r_{\text{light-ring}})$ travel ballistically into the black hole, while low-frequency waves with $\omega^2\ll V_{\text{max}}$ are mostly reflected.

\subsection{Quasinormal vibrations of the black hole}\label{subsec:qnms}

We are now interested in the proper vibrational modes of the black hole, which for the gravitational field will have $l\geq 2$. By proper vibrations, we mean that there is no external field exciting the black hole. Moreover, the horizon can only absorb and not emit classical waves. Thus we require, as boundary conditions, that there are no incident waves infinitely far away from the black hole and no waves coming out from the horizon. Recalling that the horizon is at  $r_*\to-\infty$, \eqref{r*hor}, then we must impose that
\begin{align}
    &\Psi\sim e^{-i\omega(t-r_*)}, \quad \text{as} \ r_*\rightarrow \infty\,, \\
    &\Psi\sim e^{-i\omega(t+r_*)}, \quad \text{as} \ r_*\rightarrow -\infty\, .
\end{align}
That is, we have a boundary value problem for the ODE for $\Psi$ which will admit solutions only for a discrete set of complex frequencies $\omega$ of the general form \eqref{omega}. These are the \textit{quasinormal mode spectrum} of the black hole. Mode stability is ensured if $\tau>0$. For observations, the most important modes are the least-damped, longest-lived ones. 

All the scales in the Schwarzschild black hole are set by $M$, so by simple dimensional analysis the mass dependence of the frequency must be
\beq
\omega \propto \frac{c^3}{GM}\,.
\eeq
Solving for the fundamental mode $l=2$ one finds
\beq \label{mode}
\omega = (0.37-0.089i)\frac{c^3 }{GM}\, .
\eeq
In observational units, using that $GM_\odot/c^3\approx 5$ ms,\footnote{Recall that we can measure mass in seconds. This is, approximately, the light-crossing time for a black hole of mass $M_\odot$.}, \ie  $c^3/GM_\odot\approx 200$ Hz, we have
\begin{align}
    &f=1.21 \frac{10M_{\odot}}{M} \ \text{kHz}\, , \\
    &\tau=0.55 \frac{M}{10M_{\odot}} \ \text{ms} ,
\end{align}
where we have taken 10 solar masses as reference because it is the characteristic mass of the black holes we currently observe. Since $f\tau$ is a number of order one, the black hole does not ring for long---its quality factor as an oscillator is low. The amplitude decreases after each oscillation by a factor
\beq
\exp(-1/f\tau)\approx 0.22\,.
\eeq
Black holes do not make good bells.

\begin{figure}
\centering
\includegraphics[width=.7\textwidth]{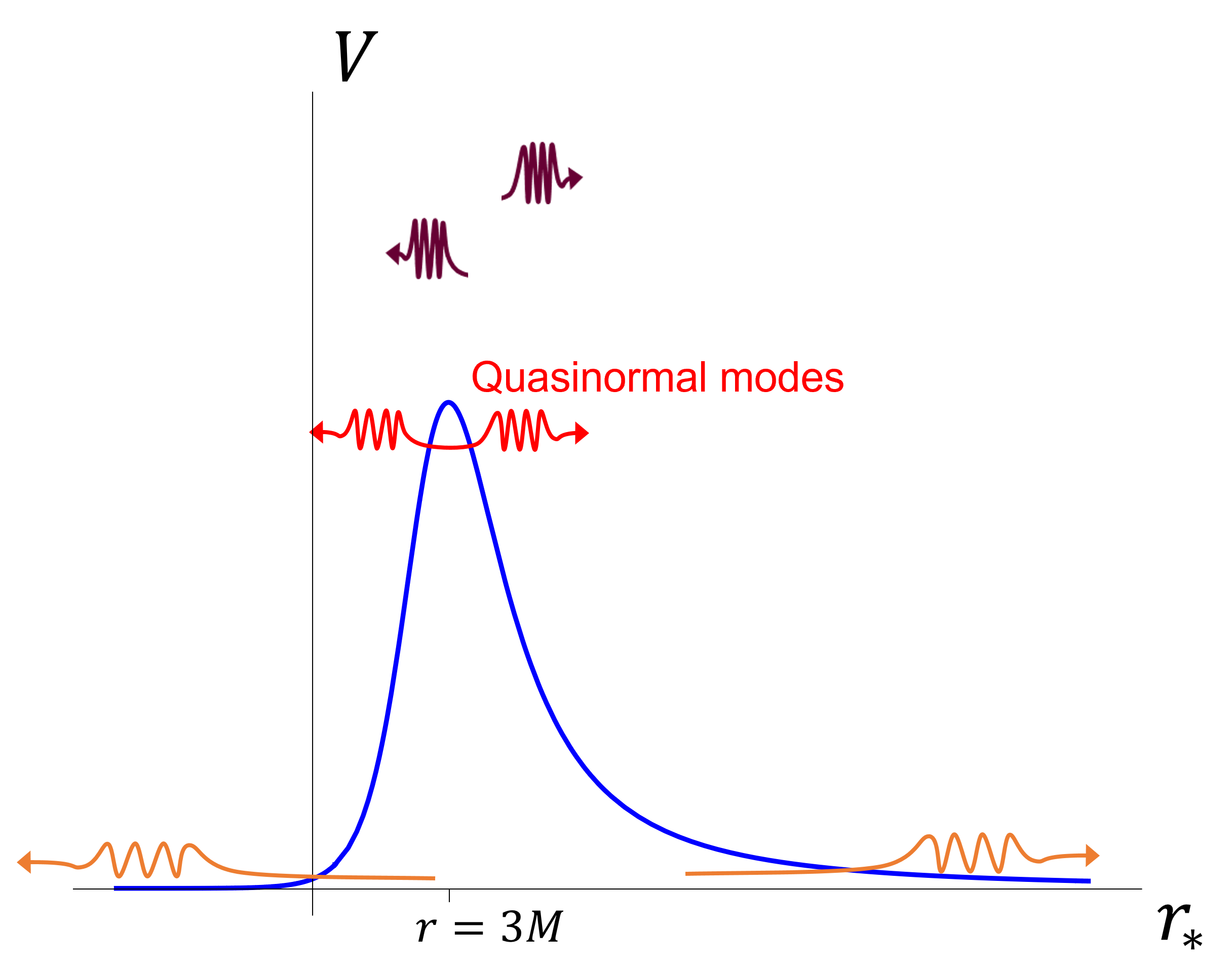}
\caption{Quasinormal modes are waves that are ingoing towards the horizon at $r_*\to-\infty$ and outgoing towards infinity at $r_*\to +\infty$. Waves of very high frequency (purple) cannot satisfy this: they will be either purely ingoing or outgoing, passing ballistically well above the potential barrier. Waves of very low frequency (orange) are damped too strongly below the barrier. Ingoing and outgoing waves (red) with frequency near the peak of the potential at the photon ring $r=3M$ can be matched with only small damping, and thus form a quasinormal mode.}
\label{fig:Vqnm}
\end{figure}

\paragraph{Properties of quasinormal modes.} Let us try to gain some intuition behind these quasinormal modes (see figure~\ref{fig:Vqnm}). We have defined them so that they match an ingoing wave on the horizon and an outgoing wave at large distances. Clearly, no such wave is possible for frequencies  $\omega^2\gg V_{\text{max}}$, since they are so energetic that they pass well above the potential as purely ingoing or outgoing waves. For $\omega^2\ll V_{max}$, on the other hand, the wave encounters a very high barrier, below which it will be extremely damped, so at these low frequencies, the waves remain mostly on one or the other side of the potential. The optimal case is to search for solutions with frequencies just below the peak of the potential, where we can match an outgoing and an ingoing wave that are only slightly damped below the peak. 

From these considerations, we infer that:
\begin{itemize}
    \item The least-damped, dominant QNMs are localized near the photon ring, where the potential reaches a maximum.
    \item They can be computed using the WKB approximation for inverse parabolic potentials.
    \item They have higher frequencies for higher $l$ (see \eqref{Vlmmax}).
    \item Their complex frequencies \eqref{mode} carry information about the mass of the black hole. Measurements of $f$ and $\tau$ for a static black hole provide two independent determinations of $M$: a test of general relativity.
\end{itemize}

From the first point above we learn a perhaps surprising fact: 
\begin{quote}
The proper vibrations of the black hole should not be thought of as vibrations of the horizon. Instead, most of the vibration of the geometry occurs at a distance $\sim 3M$.
\end{quote}
All the vibrations that start out closer than the maximum of the potential at the photon ring get quickly absorbed. Having most of the damped vibrational modes centered around $r=3M$ is not a coincidence, since light rays on the photon ring are a classical (eikonal) limit of the quasinormal perturbations of the QNMs. 
In general, the normal modes of a wave in a box can be regarded as stationary oscillations that bounce around in the box. It is similar for the quasinormal vibrations of a black hole: the circular photon ring is a stationary trajectory for waves that travel around the black hole. But this trajectory is unstable, with a characteristic decay time, which is the eikonal counterpart of the dissipative decay of QNMs. Indeed, the frequency and characteristic decay time of a light ray trajectory in the photon ring provide a good approximation to the $f$ and $\tau$ of the black hole QNMs, even at relatively low wave numbers.

\paragraph{Exciting quasinormal oscillations.} The localization of QNMs around the photon ring has other interesting consequences. An accretion disk around a Schwarzshild black hole cannot be closer than the innermost stable circular orbit (ISCO) for test particles, which is located at $r=6M$. Thus, the particles in the accretion disc are located at a radius quite bigger than the photon ring and do not excite the QNMs. In contrast, particles that fall into the black hole, from the ISCO or farther away, will excite the QNMs as they cross the photon ring. 

For the same reason, extreme-mass-ratio inspirals (EMRIs), where a small object (\eg stellar-mass black hole or neutron star) orbits around a supermassive black hole, excite very little the QNMs. It is only at the end of the EMRI, when the small inspiralling object plummets from the ISCO into the black hole, that it crosses the photon sphere. But being a very small object in a quick drop, the emission from this final plunge is tiny and short. The events that excite QNMs the most are binary mergers between black holes of similar masses. The QNMs are then those of the resulting black hole.

Imagine, on the other hand, that instead of a black hole we have an exotic compact object (ECO), that is, an impostor such that instead of the horizon it has a hard surface slightly above $r=2M$, but otherwise its exterior geometry is much like in the black hole. In this case, the proper oscillations of the spacetime geometry would again occur mainly near the photon ring. These waves, as we saw, will have outgoing and ingoing components. The latter would encounter not a perfectly absorbing horizon but a hard reflecting surface. This would give rise to \emph{echoes} from the reflection of the ingoing part of the QNM wave. Then, the detection of echoes of QNMs in the ringdown would reveal the presence of an object different than the black hole that General Relativity predicts. This would be a harbinger of radically new physics.

\paragraph{Other properties:}
\begin{itemize}
    \item Unlike normal modes, QNMs are not a complete basis of functions. There are ``late time tails" in the radiation, from backscattering in the gravitational potential outside the black hole, that can not be represented by QNMs.
    \item They do not provide good initial data: if $\omega=2\pi f-\frac{i}{\tau}$ with $\tau>0$, then $e^{i\omega (t-r_*)} $ diverges at constant $t$ and $r_*\rightarrow +\infty$.
\item QNMs are also very important in the AdS/CFT correspondence, where they describe the relaxation to equilibrium of the thermal plasma that is holographically dual to a black hole in Anti-deSitter spacetime.
\end{itemize}

\section{The black hole that rotates}
\subsection{Black hole uniqueness}
Now we turn to more realistic black holes. Since the collapse of a massive object typically involves some angular momentum, the geometry in the exterior of such an object will not be described by the Schwarzschild solution. Of course, the collapse need not produce a black hole. Nevertheless, as the \textit{black hole uniqueness theorem} teaches us, if indeed a black hole (and not a star) is born, then there is a unique solution for the geometry. This is the Kerr solution and it is the subject of this section. 

Indeed, we had a preview of this uniqueness in section~\ref{subsec:nohair}, when we discussed the linearized gravitational perturbations of the Schwarzschild black holes. There, we saw that the only deformations of the geometry of the black hole that remain stationary (zero frequency) were modes that added either mass or angular momentum to it (or charge, when a gauge field is present). Any other distortion of the shape will not remain stationary but will instead be absorbed or radiated away.

The much more powerful uniqueness theorem, proven through a sequence of results in the early 1970s by Hawking, Carter, and Robinson, states that the only stationary, asymptotically flat solution of the vacuum Einstein equations that is regular on and outside a (non-degenerate) event horizon is the Kerr black hole with mass $M$ and angular momentum $J$.

Moreover, this solution proves to be dynamically stable. In other words, if we perturb the metric linearly, as we did in \eqref{pertm},
and try to solve the equations for the perturbation $h_{ij}$ (a problem even harder than the static case, as we will see in Sec.~\ref{subsec:teuk}), one finds that the perturbation decays in time\footnote{This can be shown for linearized perturbations beyond the mode analysis.}. Proving nonlinear stability turns out to be much more difficult. However, in numerical simulations, the Kerr solution always appears as the unique, stable final state of the collapse of massive objects to form black holes, or in binary mergers. For this reason, even if the stability of the Kerr black hole, and also its uniqueness (whose proof requires a large degree of differentiability) may not yet have completely satisfactory mathematical proofs behind them, there is strong reason to take them as true. They have an immediate striking consequence: 

{\center 
\boxed{\text{The Kerr solution gives the exact description of all astrophysical black holes in the universe.}}
}

This means that for a black hole such as the photogenic M87*, we only need to determine two parameters, $M$ and $J$, to be able to deduce all its physical behavior---and this for an object that is 15 Mpc away from us. For comparison, just think about how many numbers you would need to begin to parametrize and understand the behavior of the Sun, of your next-door neighbor, or even of your partner. Black holes are, by a very wide margin, the simplest macroscopic objects in the universe.

Before we delve into the Kerr solution and its properties, let us mention briefly what to expect. We know that in linearized gravity, with $g_{ij}=\eta_{ij}+h_{ij}$, the equations of motion are schematically
\beq\label{perteqs}
\Box h_{ij}\sim T_{ij}\,.
\eeq
If there is linear momentum along the $x$ direction then $P_x\sim T_{tx}$. Likewise, a nonzero angular momentum along $\phi$ means $J_\phi\sim T_{t\phi}$. Therefore
\begin{align}
    & P_x\neq 0 \Rightarrow h_{tx}\neq 0\,, \\
    & J_{\phi} \neq 0 \Rightarrow h_{t\phi}\neq 0.
\end{align}
The presence of a non-zero component $h_{t\phi}$ implies that the black hole drags the spacetime around it, and thus also the matter in its vicinity. This has momentous consequences that we will explore below.

\subsection{Kerr's solution}

It took until 1963 to find the rotating extension of Schwarzschild's solution---almost fifty years, which gives an idea of how difficult it is to go from diagonal metric solutions to non-diagonal ones. The exact solution of the vacuum equations that Roy Kerr found is
\beq \label{Kerr}
ds^2=-\lp1-\frac{2M}{\Sigma}\rp dt^2-2a\sin^2{\theta}\frac{2Mr}{\Sigma}dtd\phi+\frac{(r^2+a^2)^2-\Delta a^2\sin^2{\theta}}{\Sigma}\sin^2{\theta}d\phi^2+\Sigma\lp\frac{dr^2}{\Delta}+d\theta^2\rp,
\eeq
where 
\begin{align}
    \Delta=r^2-2Mr+a^2\,, \qquad
    \Sigma=r^2+a^2\cos^2{\theta}\,.
\end{align}
We have now two parameters that (with $G=c=1$) have dimensions of length: $M$ and $a$. One can easily show that by taking $a=0$ we recover the Schwarzschild solution \eqref{schw2}. We can expect (and will verify) that $M$ stands for the mass, and we will presently see what $a$ measures. For now, we note that the metric \eqref{Kerr} is time-independent and axisymmetric, \ie independent of the angular variable $\phi$. Thus, it describes stationary rotation around a fixed axis. 

We can confirm that $M$ is indeed the mass from the expansion of $g_{tt}$ at large distances, as we saw in \eqref{gttPhi},
\beq
-g_{tt}=1-\frac{2M}{r}+\mathcal{O}\lp\frac{1}{r^2}\rp\,.
\eeq
Similarly, we should measure the angular momentum from its effects on the geometry at large distances. In general, the angular momentum for a distribution of rotating matter can be read from the asymptotic behavior of $g_{t\phi}$, 
\beq
g_{t\phi}=-\frac{2J}{r}\sin^2{\theta}+\mathcal{O}\lp\frac{1}{r^2}\rp\,.
\eeq
Comparing to the expansion of the $g_{t\phi}$ component of \eqref{Kerr} we find
\beq\label{JMa}
J=a\,M\,,
\eeq
so the parameter $a$ is the angular momentum, or spin, per unit mass. From now on, without loss of generality, we take $a\geq 0$. 

\subsection{Singularities and horizons}

Looking at the Kerr metric \eqref{Kerr} we notice two instances where the coefficients become singular: $\Delta=0$ and $\Sigma=0$. The latter is reached for $(r=0,\theta=\pi/2)$ and it is a true curvature singularity: one can verify that the Kretschmann scalar $R_{ijkl}R^{ijkl}$ diverges there. We will not deal more with this singularity in these lectures, since it has no relevance to astrophysics (furthermore see footnote~\ref{foot:scc}), but focus instead on the other apparent singularities.

The solutions to $\Delta=0$ are
\beq \label{Khorizon}
r=r_{\pm}=M\pm\sqrt{M^2-a^2}\,.
\eeq
Let us assume that $M\geq a$ so these are real roots. Then, as for the Schwarzschild black hole, we can find Eddington-Finkelstein coordinates (Problem~\ref{prob:EFKerr}) to show that $r=r_{\pm}$ are indeed just coordinate singularities. The largest one, $r_+$, gives the location of the event horizon.\footnote{The inner horizon at $r=r_-$ is widely believed to be unstable to becoming a severe singularity, if not by classical effects then by quantum ones. Effectively, this \emph{strong cosmic censorship} removes all the fun you could have in the Kerr interior---a timelike ring singularity, passages to other universes, time machines---and makes the resulting inner singularity not an `object' you could see, but a terminal event.  So it goes. \label{foot:scc}}

The condition $a\leq M$ for the existence of the horizon is equivalent, temporarily restoring $G$ and $c$ and using \eqref{JMa}, to
\beq \label{Kcondition}
J\leq\frac{G}{c}M^2\,.
\eeq
That is, there is an upper bound on the angular momentum of a black hole of a given mass (see Problem~\ref{prob:rotbhs}). The bound is saturated in the \textit{extremal Kerr limit} for which $J=M^2$, or equivalently $a=M$. One may wonder what happens if one tries to overspin an extremal Kerr black hole past this limit, by throwing particles with a small mass and a high impact parameter towards it. One can solve the geodesic equations in \eqref{Kerr}, and find that these particles miss the black hole precisely when their impact parameter \ie their orbital angular momentum divided by their energy, would be large enough to cause the black hole to have $J>M^2$ if it absorbed the particles.\footnote{If such an absorption were possible, presumably it would not create a naked ring singularity (as is sometimes said), but rather generate a large, violent reaction of the black hole, \ie it would be a sign of a strong instability of the solution.}

\begin{figure}
\centering
\includegraphics[width=.6\textwidth]{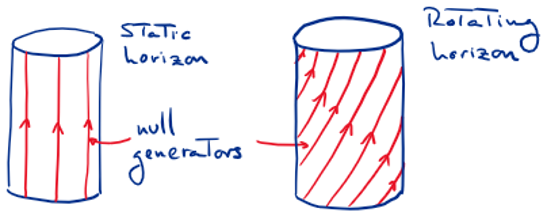}
\caption{Null generators of the event horizon in static and rotating black holes.}
\label{fig:roth}
\end{figure}

By going to Eddington-Finkelstein coordinates one can see that the null rays that remain fixed at $r=r_+$ are not trajectories with tangent vector $\partial/\partial t$---that is, $g_{tt}$ does not vanish at $r=r_+$. Instead, one finds that the vector that vanishes there is 
\beq \label{killing}
\xi^\mu=(\partial/\partial t)^\mu+\Omega_H(\partial/\partial\phi)^{\mu}\,,
\eeq
with 
\begin{align}
    \Omega_H=\frac{a}{2M r_+}\,.
\end{align}
This is a Killing vector field that has a Killing horizon at $r_+$. The difference between $\partial_t$ and the horizon generator $\xi$ allows us to make precise the sense in which the black hole (which is nothing but empty spacetime) is rotating (fig.~\ref{fig:roth}). We can think of the vector $\partial_t$ as the four-velocity vector of observers who remain static at infinity: in the flat asymptotic geometry, this timelike vector is unit-normalized, and generates motion only in time, not in any spatial direction. Then, the null geodesic rays parallel to $\xi$ that generate the event horizon (recall section~\ref{subsec:collmer}) are rotating relative to these static asymptotic observers. 

The four-velocity $u^\mu$ of an observer who is instead very close to the horizon and corotating with it must be a vector almost proportional to $\xi^\mu$. The observer at infinity measures the rotation of this corotating observer, in the limit to the horizon, to be 
\beq \label{angvel}
\omega=\frac{d\phi}{dt}=\frac{\frac{d\phi}{d\tau}}{\frac{dt}{d\tau}}=\frac{u^{\phi}}{u^t}\equiv \Omega_H\,.
\eeq
It is in this sense, relative to the observer at infinity, that the black hole rotates with constant angular velocity $\Omega_H$. 

\subsection{Ergosphere}

\begin{figure}
\centering
\includegraphics[width=.65\textwidth]{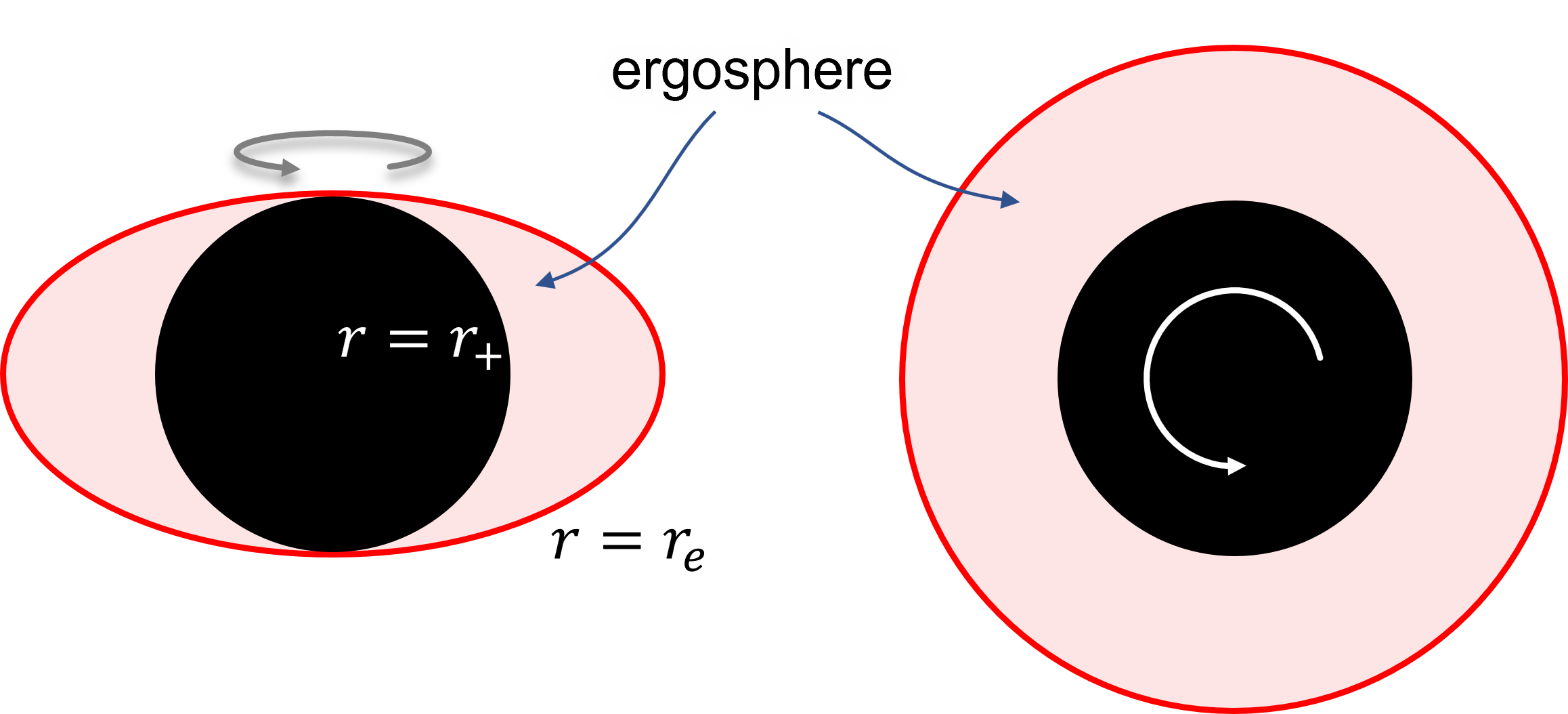}
\caption{Ergosphere of a rotating black hole, viewed from the side (left) and from above (right).}
\label{fig:ergosphere}
\end{figure}

Particles at finite distances that follow trajectories with a four-velocity parallel (i.e., proportional) to $\partial/\partial t$ will remain at rest relative to the asymptotic static observers. However, in order for a particle to have such a four-velocity, the vector $\partial/\partial t$ at the location of the particle must be timelike, \ie we must have $g_{tt}<0$ there. In the Schwarzschild solution, this happens everywhere outside the horizon, but in the Kerr geometry, this is not guaranteed. From \eqref{Kerr} we find that this will happen only if $r>r_e$, where
\beq
r_e=M+\sqrt{M^2-a^2\cos^2{\theta}} \geq r_+\,.
\eeq
The surface $r=r_e$ is topologically spherical (similar to an oblate ellipsoid) and lies strictly outside the horizon, except at the poles $\theta=0,\pi$, where the two surfaces touch. At the equator, $r_e(\theta=\pi/2)=2M>r_+$ (figure~\ref{fig:ergosphere}).

The region bounded by $r_+<r<r_e$ is called the \textit{ergosphere} (sometimes, ergoregion).
The fact that $\partial/\partial t$ becomes a spacelike vector implies that no $u\propto \partial/\partial t$ can be a timelike velocity vector inside the ergosphere, and therefore there cannot be any particle (or light ray) in a trajectory in the ergosphere that remains static relative to the asymptotic observers. Any particles in the ergoregion must necessarily move along $\phi$, since the rotational dragging is so strong that it cannot be counteracted by any acceleration in the opposite direction.

The event horizon is a causal boundary but the ergosphere is not: it is perfectly possible to escape from it towards asymptotic infinity, and particles can flow into and out of the ergosphere. This will be important in the next section.

The main implications of the ergosphere are therefore not related to causality but to energy (hence its name). Inside the ergosphere, a particle that locally has positive energy (as measured in its own reference frame) can have negative energy relative to observers at infinity. To see this, note that for a particle with four-momentum $p^\mu$, the energy conjugate to the time coordinate $t$ (the time of asymptotic observers at rest) is
\beq
E=-p_{\mu}\lp \frac{\partial}{\partial t}\rp^\mu\,.
\eeq
Since $p^\mu$ is a timelike vector, then wherever $\partial/\partial t$ is also timelike (i.e., outside the ergosphere) we will have $E>0$. However, in the ergosphere $\partial/\partial t$ is spacelike, and there, the local interpretation of $E$ is not as the energy of the particle but as a component of its spatial momentum, which can certainly be negative. A particle in the ergosphere with $E<0$ will, from the viewpoint of asymptotic infinity, have negative energy. 

In the Schwarzschild geometry, there is a sense in which these (classical) negative energies can happen too, but only for particles inside the horizon, so they are outside of causal contact with asymptotic infinity. In  Kerr, a particle with negative $E$ can escape outside the ergosphere, but it would need to absorb enough energy to compensate for $E$, since in the asymptotic region only positive energies are allowed.

\subsection{Penrose process}

The possibility of negative energy particles in a region outside the event horizon permits a mechanism for extracting rotational energy out of the black hole. This is the \textit{Penrose process}.
\begin{figure}
    \centering
    \includegraphics[width=.9\textwidth]{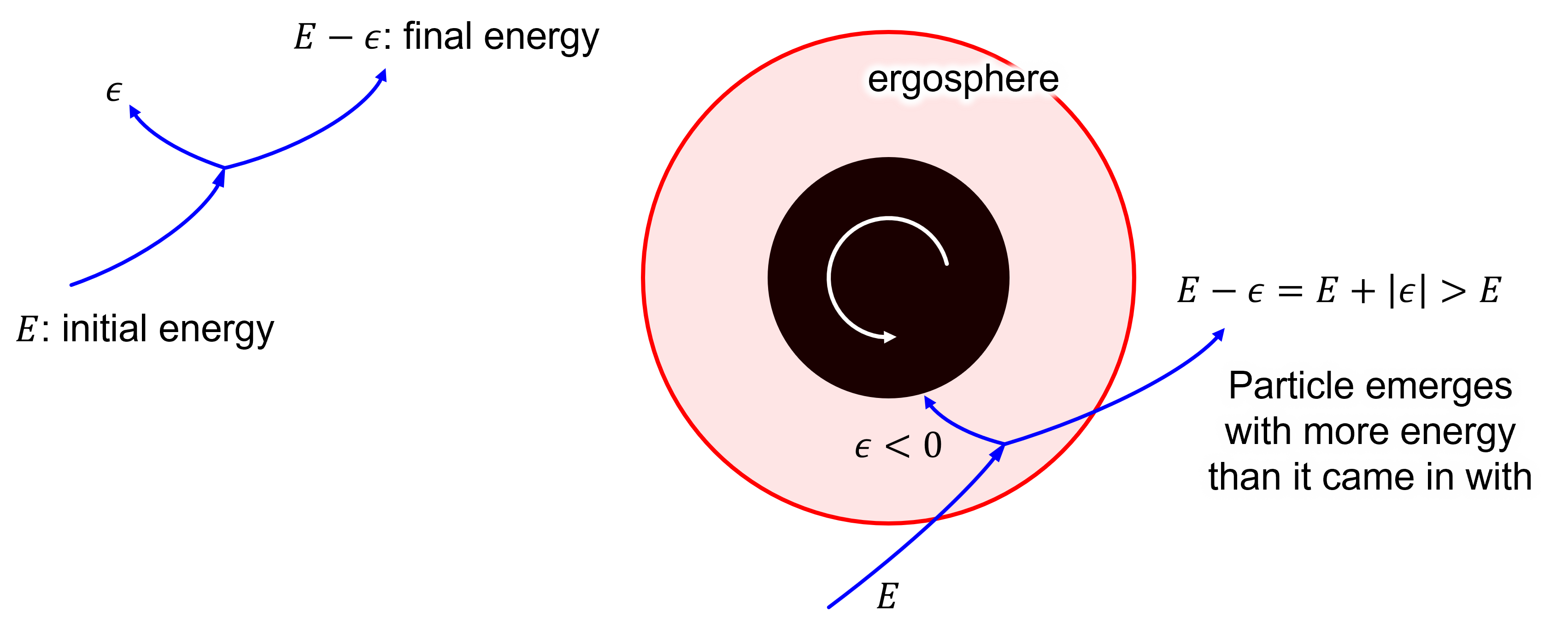}
    \caption{Penrose mechanism for extracting energy from a rotating black hole. A particle with energy $E$ is designed to fission into two fragments with energy $E-\epsilon$ and $\epsilon$. When the fission happens inside the ergosphere, it is possible to have $\epsilon<0$ for one of the fragments, which then falls into the black hole in a counterrotating trajectory. The particle that escapes has more energy than it came in with. In the process, the black hole loses rotational energy and spins down.}
    \label{fig:energy-ergo}
\end{figure}

Avail yourself of a particle with an energy $E$, which can be timed to fission into two fragments with energies $E-\epsilon$ and $\epsilon$. 

Now throw your particle towards the black hole and arrange that it fissions inside the ergosphere, adjusting the trajectory in such a way that one of the fragments has negative energy $\epsilon<0$ conjugate to the asymptotic time $t$. The other fragment escapes the ergosphere and, since the components of the four-momentum along $\partial/\partial t$ are conserved, its energy as it reaches the asymptotic region will be
\begin{align}
    E-\epsilon=E+|\epsilon|>E\,.
\end{align}
The outgoing fragment comes out with more energy than the initial particle!

The extra energy is taken away from the rotational energy of the black hole: it is easy to prove that for the infalling fragment to have energy $\epsilon<0$, it must fall inside the black hole in a counterrotating trajectory. So the black hole will slow down when it absorbs the particle. The balance between the energy and the spin that the black hole loses in the process is such that the ratio $J/M^2$ decreases. Then, the mechanism can continue until the black hole loses all its spin, at which point the ergosphere ceases to exist.

One can use also classical field waves instead of particles. The phenomenon is then called \textit{superradiance} and can be viewed as follows: inside the ergosphere, the field polarizes into two components, and one of them is absorbed by the black hole, in a wave analog of the Penrose process. If the wave is a mode of frequency $\omega$ and angular momentum number $m$, then one can prove that if
\begin{align}
\label{supercond}
0\leq \omega \leq m\Omega_H\,
\end{align}
is satisfied, the outgoing wave has a higher amplitude than the ingoing one (Problem~\ref{prob:srrad}). Eq.~\eqref{supercond} is known as the superradiance condition.

Let us make a remark anticipating aspects that we will discuss in Section~\ref{sec:bhevap}. Superradiance can be interpreted as stimulated emission of radiation. It is related to spontaneous emission of radiation (Hawking radiation) by the usual relation between Einstein's $A$ and $B$ coefficients, \ie by detailed balance. Equivalently, if a mode characterized by $(\omega, l, m)$ has a decay rate $\Gamma_{lm}(\omega)$, detailed balance requires that when there is an incident flux of the field $\mathcal{F}_{in}$, we have 
\begin{align}\label{detbal}
    \mathcal{F}_{in } \sigma_{l m}(\omega)+\Gamma_{l m}(\omega)=0\,,
\end{align}
where $\sigma_{\ell m}(\omega)$ is the absorption cross section for this mode.

Superradiant modes spontaneously decay, so $\Gamma_{lm}(\omega)>0$ and hence $\sigma_{lm}(\omega)<0$.
One can prove using flux conservation that
\beq
\mathcal{F}_{out}-\mathcal{F}_{in}=C_l(\omega) \mathcal{F}_{in}\sigma_{lm}(\omega)\,,
\eeq
where $C_l(\omega)>0$ is a universal coefficient (essentially a conversion factor between plane waves for the fluxes and spherical waves for the absorption $\left.\sigma_{lm}(\omega)\right)$.
Then, since $\sigma_{l m}(\omega)<0$, we have $\mathcal{F}_{out}>\mathcal{F}_{in}$ : the outgoing wave is amplified. For large occupation numbers, stimulated emission is a classical process, whereas spontaneous emission is always quantum.

\subsection{Superradiant instability, or the black hole bomb}

Imagine that outside a black hole there is a potential barrier for the propagation of a field. The barrier could be the mass of the field---so its waves can reach infinity only if their energy is equal or above this mass, otherwise they bounce back inwards---or some geometric effect such as created by the negative cosmological constant in Anti-deSitter space, which acts as a sort of `covariant box' for anything propagating in its interior.
\begin{figure}[ht]
    \centering
    \includegraphics[width=.3\textwidth]{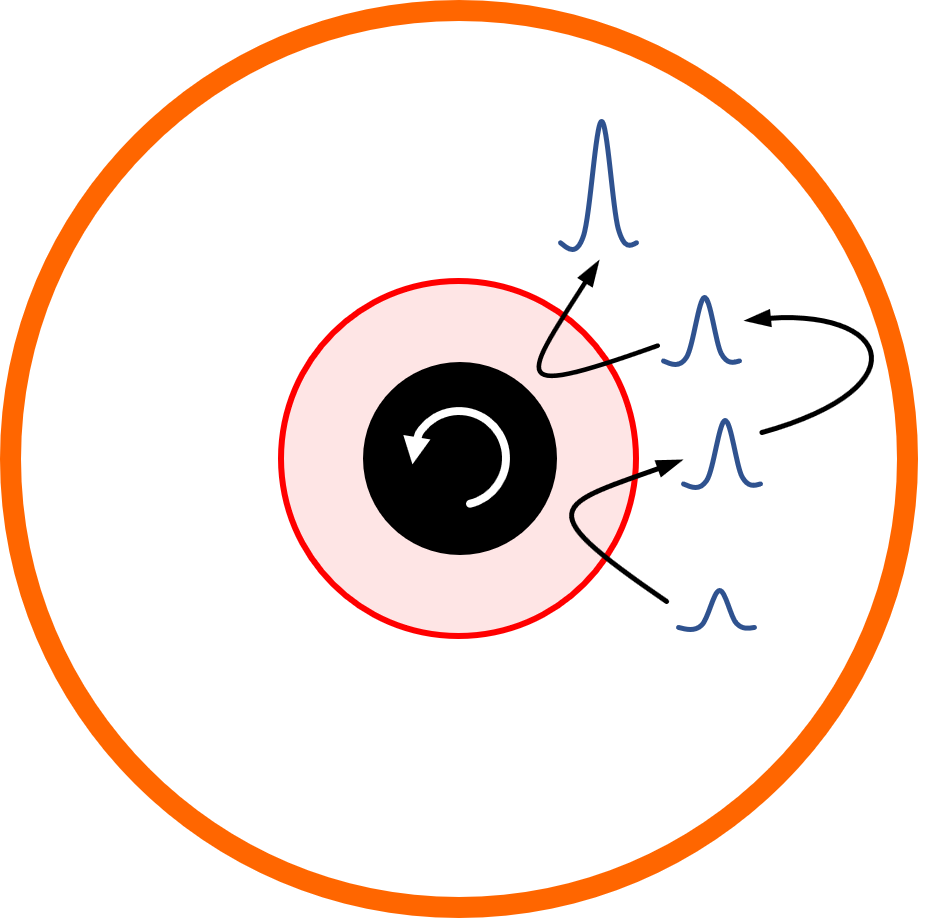}
    \caption{Instability of a rotating black hole surrounded by a barrier. An initial small classical wave that satisfies the superradiant condition \eqref{supercond} can go back and forth between the ergosphere and the barrier, extracting energy and angular momentum from the black hole and eventually forming a rotating cloud around it.}
    \label{fig:bh_bomb}
\end{figure}

Let us look at what would happen to a small scalar perturbation propagating in this scenario.  Superradiant amplification will take place for field modes that satisfy the condition \eqref{supercond} when they propagate inside the ergosphere, so the wave will emerge with higher amplitude. However, as the wave travels further out, it encounters a barrier that acts as a mirror, and it gets reflected back toward the ergosphere, where it will undergo another round of amplification. The process will repeat itself as the wave goes back and forth, giving rise to a self-amplifying field cloud around the black hole, whose angular momentum will be gradually depleted. This process is called the \textit{black hole bomb instability}. Depending on the nature of the barrier, the cloud may not exist as a strictly stationary configuration (which would be a form of hair), but it may nevertheless be long-lived (pseudo-hair).

The simplest possibility for such a mirror to exist in an astrophysical setting is provided by the mass of a scalar field. For the mechanism to work, the Compton wavelength of this scalar field must be bigger than the black hole---otherwise, the field barrier would be inside the ergosphere and the waves could not propagate away from it. Since the Compton wavelength is inverse to the mass of the field, $\lambda_c=\hbar/(mc)$, the scalar field must be extremely light. There are many theoretical scenarios that allow or predict the existence of ultralight scalar fields in the universe---very often they are not really scalars but pseudoscalar axions, but this does not affect their superradiant amplification. The superradiant instability offers the possibility of detecting their presence using their gravitational (\ie universal) coupling to black holes.
 
 If such fields do exist, there must be a range of spins that black holes of a certain mass would not have, since they should have been depleted by superradiant spin extraction. Therefore, by studying the spectrum of black hole spins and masses in the universe it might be possible to detect the existence of an ultralight scalar field in nature. Fields of such low masses would be extremely difficult to detect with the usual particle detection mechanisms, but nothing can escape gravity, and the superradiant instability provides a way to amplify the presence of these fields.

\subsection{Perturbations of the Kerr black hole: Testing Kerr}\label{subsec:teuk}

As in the case of the Schwarzschild black hole, we can slightly perturb the Kerr black hole and observe how the disturbance develops through time. As a warm-up, we may study the simpler case of a scalar field in the Kerr background. We can decompose it as
\begin{align}
    \Phi=\phi(r,\theta)e^{-i\omega t +im\phi}\,.
\end{align}
Now we do not have full spherical symmetry of the background to simplify the analysis. Only axial symmetry is present, and the problem appears considerably more complicated. 

Plugging this field in the scalar field equation we get a PDE for the amplitude function $\phi(r,\theta)$. Fortunately, due to the existence of a hidden symmetry\footnote{The presence of a conserved quantity, the Carter constant, related to the existence of a higher order symmetry of the Kerr metric generated by a Killing tensor field.}, we can separate variables into $R(r)\Theta(\theta)$. The functions $\Theta(\theta)$ are spheroidal harmonics, which are known numerically, and analytically for small $a$. Then we can write down a radial equation for $R(r)$ and study the QNMs of a scalar field. 

Gravitational perturbations are more complicated. One may hope to still be able to separate the variables $r$ and $\theta$, but it is unclear whether one will succeed in the task of decoupling the equations to obtain a single, second-order master equation. Impressively, such a decoupling was achieved by Teukolsky, and the remarkable master equation he obtained bears his name. After decoupling it, the gauge-invariant master variable $\Psi(r,\theta)$ of the Teukolsky equation can be separated into $R(r)\Theta(\theta)$. Explicit inversion formulas to recover the metric perturbations from $\Psi(r,\theta)$ are known, but they take a complicated form.

It is possible to numerically solve the Teukolsky equation to obtain the spectrum of quasinormal frequencies,
\beq
\omega(M,J)=2\pi f(M,J)-\frac{i}{\tau(M,J)}\,.
\eeq
For all of them, $\tau>0$. This proves the mode stability of the Kerr black hole.

The ringdown phase of a black hole merger provides the possibility of measuring both the frequency $f(M,J)$ and decay time $\tau(M,J)$ for the slowest quasinormal mode, from which we can determine $M$ and $J$. If our data are accurate enough to measure \emph{two} QNMs, then we can test the validity of the `Kerr hypothesis', or the `no-hair theorem', namely, whether it is indeed true that all the properties of a Kerr black hole, including all the higher QNMs, are fully determined by just $M$ and $J$. At present, the observations of ringdown are too noisy to allow precise tests, but this will definitely improve in the future. Einstein's theory will then be subjected to unprecedented scrutiny in the strong-field regime where its nonlinear character is in full swing.

By now we have covered a good deal of the behavior of black holes as purely classical systems. Time to enter the quantum world.

\section{The black hole that evaporates}\label{sec:bhevap}

In 1974 Hawking studied quantum fields propagating on a
geometry that collapses to form a black hole. He found out that during the process of collapse, quantum radiation is emitted, as expected in a
time-dependent situation. But more surprisingly, after the black hole has
settled into a stationary configuration and the transient
effects have died out, there remains a steady outflow of radiation, which observers at a large distance detect as a black body
spectrum\footnote{Filtered by frequency-dependent `greybody factors'
from the propagation of the field from the black hole out to infinity.} with
temperature 
\beq 
T_H=\frac{\hbar
\kappa}{2\pi}\,,
\eeq
where $\kappa$ is the surface gravity of the black hole horizon.\footnote{See Problems~\ref{prob:rindler} and \ref{prob:surfg}.} We will not attempt to rigorously derive this result but rather provide a physical, heuristic argument for it.

\subsection{Particle production in an external field} We intend to study
the production of particles in a given background field, due to
fluctuations of a quantum field.
Virtual quantum fluctuations of a field with mass $m$ are described by
\beq
\langle \phi(x)\phi(0)\rangle \sim e^{-x\frac{mc}{\hbar}}\,.
\eeq
This is the probability amplitude that a pair of field-quanta separated by a distance $x$ would spontaneously form. In vacuum, these
fluctuations do not materialize in the production of a real pair since
that would violate energy conservation. But the energy required for this materialization can be provided if there is an
external field to which the quantum field couples. 

Let us denote the field strength (force per unit charge) by  $F$, and the coupling (charge) by $\lambda$. Assuming the field is uniform, in order to
materialize the pair of quanta, we need the equality 
\beq
\lambda F x =2 mc^2\,, 
\eeq
for energy conservation to hold. Then the probability for a pair creation per unit volume and unit time
is given by
\beq
\Gamma\sim |\langle \phi(x)\phi(0)\rangle|^2 \sim e^{-\frac{4m^2
c^3}{\hbar \lambda F}}\,.
\eeq

A proper calculation in quantum field theory involves a tunneling
process (hence the exponential suppression) which can be evaluated in
the WKB approximation, and indeed gives
\beq
\Gamma\sim A\, e^{-\gamma\frac{\pi m^2 c^3}{\hbar \lambda F}}\,,
\eeq
where $A$ is the quantum one-loop determinant factor, which we will ignore in
the following (it is not easy to compute, and yields subdominant corrections), and $\gamma$ is a numerical factor of order one, which
depends on the specific type of particle and its coupling to the field.

This is a process of pair creation by a background field, where the latter is a
semi-classical, coherent state involving a large number of quanta in the
background. The process is described in terms of a non-perturbative
instanton bounce. It is different than the perturbative process of pair
creation, \eg\ $e^+ e^-$ creation by photon-photon collision. The
non-perturbative $e^+ e^-$ production in a background electric field was
studied by Schwinger in a classic paper in 1950. In this case
$\lambda=e$, $F=E$, and $\gamma=1$. Schwinger's leading order
result yields
\beq\label{schwin}
\Gamma_{e^+e^-}\sim A\, e^{-\frac{\pi m^2 c^3}{\hbar e E}}\,.
\eeq
The energy for the creation of the pair is provided by the background
field, which as a result decays gradually.

A black hole creates a strong gravitational field, so we might also
expect particle pairs to form near the horizon. In this case the
coupling is the particle's mass, while for the force we take the surface gravity,
\beq
\lambda=m\,,\qquad F=\kappa\,.
\eeq
With these choices,  we find that
\beq
\Gamma\sim e^{-\gamma\frac{\pi m c^3}{\hbar \kappa}}\,.
\eeq
Observe that the exponent is proportional to $m$, \ie to the energy $E=m c^2$
of the particle. Thus we can write it as
\beq
\Gamma\sim e^{-E/T_H}\,,
\eeq
where
\beq\label{TH}
T_H=\frac{\hbar \kappa}{\gamma \pi c}\,.
\eeq
This is a \textit{thermal spectrum with temperature $T_H\propto
\kappa$.} So, the black hole is expected to radiate like a blackbody. In
contrast, the Schwinger production rate \eqref{schwin} is not thermal.
The reason that it is thermal in the case of a black hole is that
gravity couples to the particle's energy. It is very suggestive that the universal character of gravity appears to be related to a universal
thermal behavior.

Some caveats about this heuristic argument:
\begin{itemize}
\item We have not pinned down the value of $\gamma$.
For this we need Hawking's proper calculation, which yields $\gamma=2$.

\item The argument was made for massive particles. However, Hawking's result applies as well to massless quanta, with $E=\hbar\omega$.

\item The energy required by the creation of the pair is supplied by the
black hole. The energy of the black hole decreases, since one of the members of the pair has negative energy (relative to asymptotic
observers) and falls inside the black hole. In more detail, if the pair have four-
momenta $p_1$ and $p_2$, four-momentum conservation requires
that 
\beq
p_1+p_2=0\,. 
\eeq
If $\partial_t$ is the generator of asymptotic
time-translations, so that $t$ is conjugate to the energy measured by
asymptotic observers, then the energy of a particle with four-momentum
$p$ is 
\beq
E=-p\cdot\partial_t\,.
\eeq 
Thus for the particle pair we must have
$E_1+E_2=0$. Now, if particle 1 is to escape to infinity, it must have
$E_1>0$. If particle 2 goes inside the black hole, then in that region
$\partial_t$ is spacelike, so $E_2$ is actually not an energy but a
component of momentum, which can be negative. Thus it is consistent to
create the pair if one of the particles falls inside the black hole. In
this case, the total mass of the black hole will decrease by an amount
\beq
\delta M=E_2=-E_1\,,
\eeq 
consistent with conservation of the energy as
measured by outside observers.

\end{itemize}

 Let us provide yet another viewpoint on why quantum physics in the presence of a black hole horizon gives rise to what may seem an unexpected consequence. A classical particle outside a stationary black hole perceives that nothing changes. A quantum field can be excited when the geometry that it lives in is time-dependent, a phenomenon of parametric excitation well-known in cosmology. The exterior of the black hole is static, and so naively one might conclude that no field excitation will occur. However, a quantum wave function can have support simultaneously in the exterior and the interior of the black hole. This is a crucial property since the interior geometry of a black hole is dynamically collapsing (a Little Big Crunch) in a time scale $\sim \kappa^{-1}$ (in units $c=1$). A quantum field will be sensitive to the time dependence of the collapsing interior and therefore will be excited. The characteristic time of this collapse implies that the field excitations will have typical frequency $\sim \kappa$, and the energy of the produced quanta will be $\sim \hbar\kappa$, which agrees with \eqref{TH}. 

 \begin{figure}
\centering
\includegraphics[width=.7\textwidth]{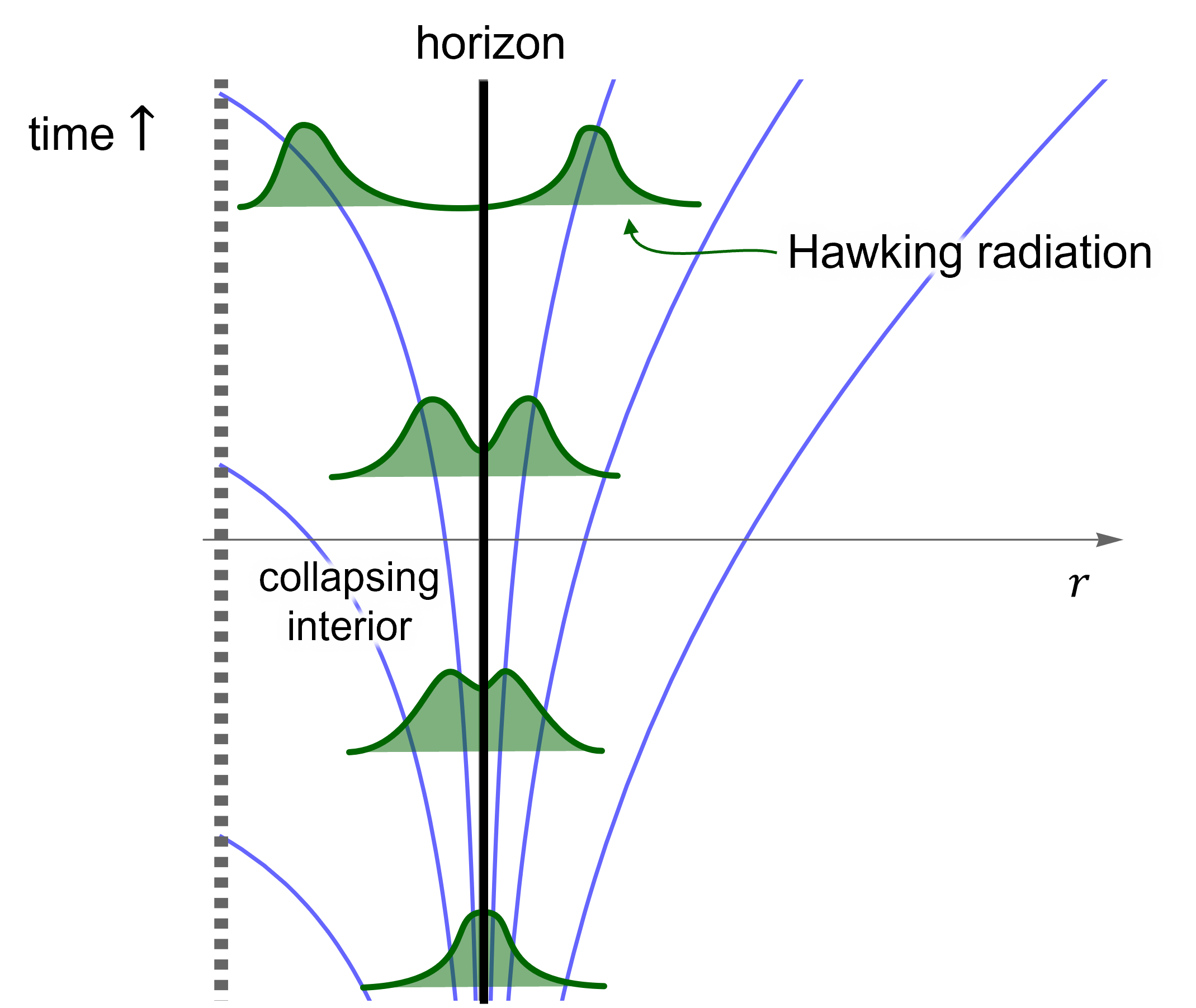}
\caption{Hawking emission as quantum pair production, in an Eddington-Finkelstein diagram (cf.\ fig.~\ref{fig:lightrays}). A quantum wavefunction initially straddling the horizon is stretched by the collapse of the black hole interior, until one of its components escapes away as a Hawking quantum and the other falls to the singularity. The pair is created in an entangled Bell-like state.}
\label{fig:stretch}
\end{figure}
 The part of the wave function that is in the interior will be dragged in by the collapse, and stretched until it becomes a wave packet separated from its exterior partner (see figure~\ref{fig:stretch}). 
 This produces a pair of quanta, with positive and negative energies relative to asymptotic time. The quantum that is radiated away will be entangled with the quantum in the interior since they are part of the same wave function, so they will be in a maximally entangled Bell-like state.

 Observe also that, according to this interpretation, a black hole with a time-independent interior (static or stationary) must not give rise to Hawking radiation. This is indeed what happens in extremal black holes, which have $\kappa=0$ and therefore $T_H=0$ too. Nevertheless, extremal black holes can decay through non-thermal spontaneous emission of superradiant modes \eqref{detbal}.

\subsection{Further aspects of Hawking radiation and black hole evaporation}

We can now draw two major consequences: black holes must have entropy, and they will evaporate. 

\paragraph{Back to the 19th-century, with a black hole.}

Rudolf Clausius explains to us that black holes must be assigned an entropy on very general phenomenological grounds: an object with energy $E$ that radiates at a temperature $T$ has an entropy $S$ given by
\begin{align}\label{clausius}
    S=\int \frac{d E}{T}\,. 
\end{align}

Let us see how this works. As the seasoned experimentalists that we are, we can measure the energy (mass) of the black hole and its temperature from far away without even knowing that the radiating object is a black hole. We proceed to collect data using a dynamometer and a bolometer and thus we obtain $T(M)$.
Our measurements yield
\begin{align}
T=\frac{\hbar}{8 \pi G M}\,,  
\end{align}
which, using that $E=M$, we plug into \eqref{clausius} to find
\begin{align}
d S&=\frac{8 \pi G}{\hbar} M d M\,, \\
\Rightarrow \quad S & =\frac{4 \pi G}{\hbar} M^2 \, ,
\end{align}

When we work out the numbers we are at first surprised to find that this entropy is (as we will see in a moment) enormous. But we attribute it to having a peculiar object that has a very large energy but is radiating at an extremely low temperature.

By now we have also figured out, \eg by scattering particles or waves, that the radiating object is a black hole whose radius we have measured to be $r_H=2GM$. Then we realize that we can write 
\begin{align}
S & =\frac{4 \pi (2GM)^2}{4G\hbar} =\frac{\mc A_H}{4 G \hbar}  \\
& = \frac{\text { Horizon area }}{\text { Planck area }} \, .
\end{align}
Therefore, the entropy of the black hole is nothing but its area measured in Planck units---truly, an enormous number for any macroscopic area! But then Clausius looks puzzled at us: why should entropy---a quantity that he introduced for understanding the efficiency of heat exchanges---bear any relation with that most basic entity, the geometry of space and time?

The identification of the black hole area with an entropy was first proposed---very boldly, and not without controversy---in 1973 by Bekenstein. It was then put on a firm footing by Hawking's discovery that black holes emit thermal radiation with a precise temperature. For this reason,
\begin{align}
    \boxed{S_{BH}=\frac{\mc A_H}{4 G \hbar}}
\end{align}
is called the Bekenstein-Hawking Black Hole entropy formula. It contains $G$ and $\hbar$, and over the last half-century it has provided the deepest and most fruitful guidance towards a quantum theory of gravity.

Now we go to interrogate other 19th-century physicists about the implications of this result. Ludwig Boltzmann informs us that, ultimately, it implies the existence of many microscopic states corresponding to the macroscopic system that we characterize by this mass and temperature \ie the black hole. Indeed, he continues, there must be as many as $e^{S_{BH}}$ states! In other words, a black hole---which is nothing but strongly warped space and time---must somehow be made of a humongous number of microscopic degrees of freedom, even if we do not see them at all. 

Boltzmann raises an eyebrow and asks: if we are to think of these degrees of freedom as somehow related to `atoms of spacetime' (an idea he seems to relish), shouldn't their number scale like the spatial volume, instead of the area? He begins to warn us about the tribulations of applying statistical reasoning to entities that have long been regarded as paradigms of determinism, but it is time that we take leave of him (some of his concerns will reappear later) and continue with other consequences of Hawking's discovery.

\paragraph{Evaporation rate and black hole lifetime.} The black hole
will evaporate by emitting radiation like a blackbody. The
radiating power of a blackbody of area $A$ and temperature $T$ is
\beq
\frac{dE}{dt}=\sigma A T^4\,,
\eeq
where $\sigma$ is the Stefan-Boltzmann factor, which depends on the
specific (effectively massless) fields that are being radiated. For a real scalar
field, $\sigma=\pi^2/60$.
As a first approximation, we can take $A\simeq \mc A_H$. For a
Schwarzschild black hole,
\beq
\mc A_H=16\pi M^2\,,\qquad T=\frac{1}{8\pi M}\,,
\eeq
and since $dE=-dM$ we can obtain the evaporation rate of the black hole as
\beq
\frac{dM}{dt}\simeq-\frac{\sigma}{256\pi^3}\frac{1}{M^2}\propto M^{-2}\,,
\eeq
so that the total evaporation time derived from
\beq
\int_0^{t_{\rm evap}}dt\propto -\int_M^0 dM'\, {M'}^2 \,,
\eeq
is simply
\beq
t_{\rm evap}\propto M^3\,.
\eeq
It is then obvious that the black hole evaporates in a finite time. This calculation is of
course very rough since it neglects the back-reaction effect that the
emission of radiation and loss of mass have on the black hole geometry
and on the radiation process itself. These effects should be small
when the energy of emitted quanta is much smaller than the mass of the black
hole, 
\beq
T_H\ll M\,,
\eeq
\ie as long as $M\gg M_{\rm Planck}$. Therefore, for most of the black hole lifetime the approximation is good, and its mass will reach Planck size in a time
$\propto M^3$.

\paragraph{Astrophysical (ir)relevance of Hawking evaporation. Cosmic dominance of black hole entropy.} Restoring full units, the Hawking temperature takes the form
\beq
T_H=\frac{\hbar c^3}{8\pi k_B G M}=6\times 10^{-
8}\frac{M_\odot}{M}\,\mathrm{K}\,,
\eeq 
so for a solar-mass black hole, $T_H\sim 10^{-7}$ K. This is much colder
than the temperature ($\sim 3$ K) of the CMB! Thus the accretion of CMB
photons alone is a stronger effect for these black holes than Hawking 
evaporation. Of course, it gets even worse for supermassive black holes.

The smallness of the effect should not be surprising: it is a quantum
effect and therefore one expects it to be small for macroscopic objects
(and for a black hole, macroscopic means larger than the Planck scale).

The initial mass of a black hole that started evaporating in the early
universe and ends its evaporation today, so $t_{\rm evap}\sim 10^{10}$ yr,
is $M\sim 10^{15}$ g. This is roughly the mass of a kilometer-high
mountain. Black holes with these masses are necessarily primordial, \ie formed by density fluctuations in the early universe, since astrophysical collapse
cannot yield black holes lighter than a couple of solar masses (Chandrasekhar limit).

On the other hand, the entropy of astrophysical black holes is enormous:
\beq
S_{BH}=\frac{c^3}{\hbar G}\frac{\mc A_H}{4}\simeq 10^{76}\lp\frac{M}{M_{\odot}}\rp^2\,.
\eeq
A single galactic black hole, with $M\sim 10^{6-9}M_\odot$, has more
entropy than all the matter and radiation in the universe ($S_{\rm CMB}\sim
N_{\mathrm{photons}}\sim\mathrm{volume~of~universe~in~mm^3}\sim 10^{87}$). See Problem~\ref{prob:entBHs}.

Black holes are at the same time the \textit{simplest} classical objects in the universe and the most \textit{complex} quantum objects.

With black holes, it is always all or nothing. They are the ultimate drama queens.

\paragraph{Black holes are small radiators.} The wavelength of Hawking
quanta is given by
\beq
\lambda_H\sim \frac{\hbar}{T_H}\sim GM\,,
\eeq
which is comparable to the Schwarzschild radius of the black hole (it had to be, since
this is the only scale in the system). Including numerical factors, one
in fact finds $\lambda_H\gtrsim R_{Schw}$.

The black hole is therefore a small radiator, with a size comparable to or
smaller than the wavelength of the radiation. Thus, Hawking radiation
cannot be traced to any point on the horizon. The image one forms of a
black hole from its Hawking quanta is a blurred one. This is unlike, \eg
the Sun, whose size ($\sim 10^9$ m) is much larger than the wavelength
of the radiation it emits ($10^{2-3}$ nm), and therefore we can use it
to get a detailed image of the star. Another consequence is that black
holes radiate mostly in low-multipole waves.

Furthermore, observe that the typical frequency and
wavelength of Hawking quanta is the same as
that of the classical quasinormal vibrations of
the black hole. The difference is that in quasinormal ringdown
the occupation numbers are very large, while
in Hawking emission they are of order one.

\paragraph{Scanning all the particle spectrum.} During black hole
evaporation, $T_H$ increases as $M$ decreases, all the way until
Hawking's approximations break down. Thus, in its evaporation the black
hole will produce any particle that is permitted by local conservation
laws, with mass possibly all the way up to the Planck energy. So, initially the
black hole will radiate mostly photons, gravitons, and then
neutrinos, and as it reaches different mass thresholds, all other
particles will be produced: electrons and positrons around $T_H\sim
1$~MeV, mesons at $T_H\sim O(100)$~MeV, nucleons at $T_H\sim 1$~GeV,
Higgs
bosons at $\sim 126$~GeV,, then X? at ???ev etc.

However, given that $t_{\rm evap}-t\sim M^3$, we have $T_H\sim
(t_{\rm evap}-t)^{-1/3}$, and therefore the black hole spends most of its
lifetime, and releases most of its energy, emitting low-energy quanta in
copious quantities. By the time it reaches the threshold to produce more
interesting massive stuff, little energy is left and relatively few of these
particles are produced.

\paragraph{Negative specific heat.}

The specific heat of the black hole is negative:
\beq	
C=\frac{dE}{dT}=\frac{dM}{dT_H}=-8\pi M^2<0\,.
\eeq
This means that the black hole is thermodynamically unstable. The black
hole heats up by radiating energy, and cools down by absorbing it. 
So if we try
to keep it in equilibrium with a radiation bath at temperature $T_H$,
then if the black hole absorbs a little more energy than required, it
will become cooler than the bath, and then it will tend to absorb more
and get further away from equilibrium. Conversely, if it absorbs
less energy than demanded by equilibrium, it will heat up and radiate even more.

This is unlike conventional thermodynamic systems in equilibrium, but it
is in fact typical of gravitating systems. For instance, a star that
emits radiation reduces its pressure and contracts, which raises its
temperature\footnote{For instance, at the core of a
newly-formed neutron star $T\sim 10^{11-12}$K, while at the center of
the Sun $T\sim 10^{7}$K.}. In fact, this property of gravitating
systems to increase
their entropy by becoming more concentrated is \textit{absolutely
crucial} for the universe to evolve structure from an initial thermal,
almost homogeneous state.

In these arguments we have been considering the Schwarzschild black hole. Indeed
the thermodynamic instability is typical of vacuum black holes. But
there do exist other black hole solutions with positive specific heat,
which are thermodynamically stable. One way to achieve this is to
add charge to the black hole: the Reissner-Nordstrom solution near
its extremal limit has positive specific heat. 

Another possibility is to
put the black hole in a box, say by limiting the radial coordinate to be
smaller than some $r_\mathrm{box}>r_\mathrm{Schw}$. For a given temperature of
the box, there are two possible black holes with that temperature, a
small and a large one. The small one hardly feels the presence of the
box and is qualitatively similar to the one in asymptotically flat
space. The large one feels, so to speak, that radiation cannot easily
fit in the box and comes to a stable equilibrium with it. A covariant version of
this box is provided by Anti-deSitter spacetime, which has a confining
effect on radiation. Black holes in AdS with a size larger than the
cosmological radius have positive specific heat and play an important
role in the AdS/CFT correspondence.

\paragraph{Generalized Second Law (Bekenstein 1972).}

The second law of thermodynamics is extended to include the
contributions from black hole
entropy so that
\beq
\Delta S_{\rm total}=\Delta S_{BH}+\Delta S_{\rm matter}\geq 0\,.
\eeq
Each of the two contributions can separately be negative, but only as
long as their total sum is
positive.

In the process of collapse to a black hole, $\Delta S_{\rm matter}<0$ but
the entropy $S_{BH}$ that is produced overwhelms this decrease by far. A
simple estimate for the collapse of a weakly-gravitating ball of radiation gives
\beq
\frac{\Delta S_{BH}}{|\Delta S_{\rm matter}|}\sim
\lp\frac{M}{M_\mathrm{Planck}}\rp^{1/2}>1\,,
\eeq
which is in fact $\gg 1$ for any astrophysical black hole. In order to
understand the huge increase in the entropy, observe that the ordinary
matter before the collapse naturally probes only a small, low-energy subset of all the degrees of freedom of the fundamental theory (\eg string theory), so it
has a low entropy. The black hole, instead, samples all the
stringy (fundamental) states available.\footnote{That it manages to do so even though its temperature is very low is surprising and you should pause to ponder it.}

On the other hand, in the process of black hole evaporation we have $\Delta
S_{BH}<0$, but the radiation carries sufficient entropy
to compensate for it. Again a simple calculation for the entropy production when massless radiation is emitted  can be seen to yield 
\beq
\frac{\Delta S_{\rm rad}}{|\Delta
S_{BH}|}\simeq \frac{4}{3}>1\,.
\eeq
Observe that in this case the entropy produced is of the same order as
the entropy of the black hole. The reason is that the evaporation
produces a number of quanta $N\sim S_{BH}$. The factor $4/3$ comes from
the relation $s_{\rm rad}=\frac{4}{3}\epsilon_{\rm rad} T$ between the
entropy and energy density of massless
radiation at temperature $T$ (in
$D$ spacetime dimensions, $s_{\rm rad}=\frac{D}{D-1}\epsilon_{\rm rad} T$). The
number is guaranteed to be $>1$ for any form of radiation or matter that
is thermodynamically stable.

\subsection{The black hole information loss problem}

Shortly after his discovery that black holes emit quantum thermal radiation, Hawking realized a surprising consequence: black hole evaporation implies that classical general relativity and quantum mechanics are incompatible. 

It was already known that the merger of the two theories was marred by the non-renormalizability of the quantum theory of the massless spin-two field (the graviton). But the latter is a problem of the ultraviolet, high-energy structure of the theory and becomes manifest only when trying to probe the nature of spacetime close to the Planck scale, where the curvatures are so large that the very notion of spacetime geometry becomes dubious. It is therefore a situation where we naturally expect that general relativity, regarded as an effective theory, breaks down.

The problem that Hawking identified is much more surprising, since it appears in situations where the geometry of spacetime is expected to be smooth. The curvature near the horizon of a black hole scales inversely to its size, so for a very large black hole the region around the horizon can be as weakly curved as we wish. Hawking's problem arises for semiclassical black holes with sizes arbitrarily larger than the Planck length. Since there is no diff-invariant quantity that diverges in these regions, why should the effective theory break down?

This has turned out to be a surprisingly deep and subtle question---and a very confusing one too. We will not be able to do proper justice to all its intricacy, but we will attempt to make the reader aware of its gist, without going into the resolutions that have been proposed.

\subsubsection{Classical prelude: Black holes have no (or too little) hair}

A main property of black holes, mentioned in the previous sections, is
that they are characterized by very few parameters. The black hole phase
space is therefore of low dimensionality.

We referred to this by saying that `black holes have no hair': the space of 
black hole solutions is completely characterized by the conserved
charges of the black hole, namely its mass, angular momentum, and
charges associated to local conservation laws such as electric charges. 
Nowadays we know of the existence of many kinds of black hole hair. Nevertheless, the phase space of black hole solutions is
still characterized by a number of parameters that, for all we
know, is much smaller than the
huge number $\mc A_H/\ell_{\rm Planck}^2$ that the Bekenstein-Hawking
formula suggests.

Therefore, when a system collapses to form a black hole, it appears that
almost all of the initial information of the configuration of collapsing
matter is lost in the process. The final state only knows about a few
aggregate, macroscopic quantities. Classically, there is no way that this information
can be retrieved. Therefore the black hole appears to be a sink where
phase-space volume is destroyed. 

\subsubsection{Basic formulation of the information loss problem}

One may still regard this as not truly problematic, because the
information about the state of the matter that formed the black hole may
somehow be preserved inside the black hole. However, if the black
hole evaporates and disappears then the problem comes back\footnote{There is
a version of the information loss problem that does not involve the
collapse and evaporation of a black hole, but instead deals with the
very-long-time behavior of fluctuations in the system of a black hole in
equilibrium with its radiation. Significant progress has recently been made in this direction, but we will not discuss it in these
notes.}. We will formulate the puzzle in different manners, each at a different
level of sophistication and detail. The first formulation contains the
core of the problem. The next one makes it more precise. Further formulations are possible that highlight other subtleties of the problem, but we will not discuss them here. 

\paragraph{Fundamental irreversible evolution.} The original configuration of matter
may have been arbitrarily complicated, needing a huge number of
parameters to describe it. All but a handful of these data are lost to
the outside observers when the black hole forms. The black hole
evaporates by emitting
Hawking radiation, but this radiation, having a Boltzmann spectrum, only
carries information about the total mass of the black hole, and about
the conserved charges conjugate to `chemical potentials' such as angular
velocities, electric potentials etc. Thus, when the black hole finally
disappears, the initial information is lost. 

This is a problem of fundamental irreversibility at a deeper level than statistical thermodynamic irreversibility: from the final radiation state we cannot reconstruct, \textit{even in principle}, the initial matter state.

\paragraph{Non-unitary quantum evolution.} In the previous paragraph we
used `information' in a rather loose sense. Let us now make it more
precise by phrasing the previous
problem in quantum mechanical parlance. States in quantum mechanics can
be generically described using density matrices $\rho$. For pure states,
which correspond to vectors $|\Psi\rangle$ (properly, rays) in a
Hilbert space, the matrix $\rho_\textrm{pure}=|\Psi\rangle\langle \Psi|$ has von
Neumann entropy $-\mathrm{Tr}\rho_\textrm{pure}\log\rho_\textrm{pure}=0$. Mixed states
are characterized by having instead $-
\mathrm{Tr}\rho_\textrm{mixed}\log\rho_\textrm{mixed}>0$. Thermal ensembles
$\rho_\textrm{th}=Z^{-1}\sum_i e^{-E_i/T}|i\rangle\langle i|$ are of
this kind and in fact maximize the entropy. Unitary evolution in quantum
mechanics takes a pure state into another pure state
\beq
|\Psi\rangle \to U(t)|\Psi\rangle\,,\qquad \rho_\textrm{pure}\to \rho_\textrm{pure}\,,
\eeq
($U^{-1}=U^\dagger$) and therefore no (fine-grained) entropy can be
generated. This is the quantum-mechanical counterpart of Liouville's
theorem that Hamiltonian evolution conserves phase space volume, \ie
there are no sinks in phase space. Unitary evolution is also the
quantum-mechanical statement that the initial state can be reconstructed
out of the final state. So unitarity can be regarded as implementing the notion of fundamental reversibility of quantum-mechanical evolution.

We may take the initial state of the matter that will collapse into a
black hole to be a pure state. During most of the evaporation the
radiation is thermal to a very good approximation. When the evaporation is completed, only this radiation remains. Thus
the evolution of the system violates unitarity, which is one of the
basic assumptions of quantum mechanics. The `black hole information
problem' is now stated as the `black hole unitarity problem'.

Imagine that we manage to collect all the radiation and all the final products of the evaporation of the black hole, isolating them so well as to prevent any losses of coherence to external systems. We pass these black hole remains through a quantum super-computer (still under construction) which attempts to reconstruct the initial state, that is, it tries to find a unitary transformation that, acting on the final state, would yield back the initial one. Hawking affirmed that the quantum computer will fail.

\subsection{Cut the knot?}
The problem seems to be a quintessential theorist's riddle, one which may be far from having any consequences for real-world observations. This is not only because, as we saw, the Hawking radiation from black holes of stellar mass or larger is a negligible effect; also, the task of verifying whether information is fundamentally lost or not appears to require exquisite control over the system in order to monitor its degree of coherence, and involves computational resources that are virtually unimaginable and very likely of exponential complexity. 

Still, desperate times call for desperate measures, and instead of trying to carefully untie the Gordian knot, one may be tempted to brutally cut it and say that there never was a problem since black holes do not actually exist---that is, there are objects in the universe that can mimic black holes very well, but which do not have smooth, perfectly absorbing horizons. These are what we referred to in Sec.~\ref{subsec:qnms} as extremely compact objects. Might there be observational signatures of their presence, in the ringdown signal as we discussed, or possibly in EHT-type radio imaging? Perhaps. But, what if the mimicker is so good that, even if it does not have a horizon, it behaves, for most external probes and to high accuracy, as if it did? This could thwart any conceivable effort at telling it apart from a true black hole. 

At any rate, even if the ECO hypothesis may seem far-fetched, given what is at stake---nothing less than our fundamental theory of space and time---it is worth looking for any telltale signs of them.

\subsection{Creative confusion}

The black hole information paradox, still unresolved almost fifty years after its initial formulation, is an excellent example of \emph{creative confusion} in science. For all the difficulties that it poses when trying to state it and investigate it in a precise way (and here we have only scratched the surface), this perplexing conundrum has proven to be a remarkably fertile source of ideas for the quantum theory of gravity.

In recent years, the problem has been reformulated in sharper ways and addressed in contexts that allow great technical control. This has been done by considering a class of low-dimensional gravity models that afford solvability for the quantum matter and its backreaction on the geometry, while still preserving a version of the paradoxes. A host of new, deep, and intriguing ideas have emerged, emphasizing the role of quantum entanglement, quantum information, chaotic behavior, and the relevance of semiclassical wormhole geometries to resolve the unitarity puzzle. 

A completely satisfactory solution has not been achieved yet---\eg it remains to clarify the experience of an observer falling into an `old' black hole (one that has evaporated more than half its entropy)---, but a consensus has grown that quantum mechanical unitarity can be upheld in the presence of black holes in a very subtle manner. We are also getting glimpses of how the geometry of spacetime emerges out of the quantum entanglement between non-gravitational, fundamental degrees of freedom. But a proper discussion of these fascinating topics will have to wait for another school. Vidimo se uskoro!

\acknowledgments We are very grateful to the Local Organizing Committee for the excellent organization of the school and warm hospitality in Belgrade; also to them and the Global Organizing Committee, for the invitation to deliver and then write up these lectures; and, not least, to the students, whose enthusiastic participation made for a truly enjoyable experience.

RE is supported by MICINN grant PID2019-105614GB-C22, AGAUR grant 2017-SGR 754, and State
Research Agency of MICINN through the ``Unit of Excellence María de Maeztu 2020-2023'' award to the Institute of Cosmos Sciences (CEX2019-000918-M).

\appendix

\section{Guided problems}\label{problems}

\renewcommand\thesubsection{Problem~\arabic{subsection}}
\renewcommand\thesubsubsection{\arabic{subsection}.\alph{subsubsection}}

\newcommand{\nocontentsline}[3]{}
\newcommand{\tocless}[2]{\bgroup\let\addcontentsline=\nocontentsline#1{#2}\egroup}

\tocless\subsection{Coordinates and surface gravity}\label{prob:coords}

The Schwarzschild metric, given by
\begin{equation}
ds^2=-\lp 1-\frac{2M}{r}\rp dt^2+\frac{dr^2}{1-\frac{2M}{r}}+r^2d\Omega_2\,,
\end{equation}
becomes singular at $r=2M$. Since a photon of frequency $\omega$ at radius $r$ is redshifted to
frequency $\sqrt{-g_{tt}(r)}\,\omega$ as it travels to infinity (where
$-g_{tt}=1$), then $r=2M$ is a surface of infinite redshift. We have nevertheless seen that this surface can be reached in finite affine parameter by null (light-ray) trajectories. Indeed, using Eddington-Finkelstein coordinates adapted to ingoing light rays, we found that the geometry is smooth there\footnote{It is not difficult to see that $r=2M$ is also reached in finite proper time by timelike (particle) trajectories. The so-called Painlev\'e-Gullstrand coordinates are adapted to radially in-falling particles and yield a metric that is manifestly regular at $r=2M$. }.
In order to further illuminate the geometry near  $r=2M$ we will explore it in other ways. 

\subsubsection{Rindler spacetime near the horizon and surface gravity}\label{prob:rindler}

Let us get closer to $r=2M$ by taking
\beq	
r-2M\simeq \frac{\xi^2}{8M}\,,
\eeq
with $\xi\ll\sqrt{M}$.

$\bullet$ Prove that the metric becomes
\beq
ds^2\simeq -\frac{\xi^2}{16 M^2}dt^2+d\xi^2+4M^2 d\Omega_2\,.
\eeq
The term for the 2-sphere with constant radius $2M$ is not important in
what follows\footnote{Strictly, since we are focusing on length scales much smaller than the sphere radius $2M$, this sphere should be approximated by a plane.\label{footrind}}. It is the $(t,\xi)$ part of the metric that matters to us
here: it is reminiscent of the plane in polar coordinates
\beq
ds^2=\rho^2d\phi^2+d\rho^2\,,
\eeq
and in fact it becomes of this form if we make the transformation $\phi\to it/(4M)$, $\rho\to\xi$. We know that in this case $\rho=0$ is just a coordinate singularity, and we can remove it by changing to Cartesian coordinates $x=\rho \cos\phi$, $y=\rho\sin \phi$. 
Following this lead, we perform the change of coordinates as
\beq
X=\xi\cosh (t/(4M))\,,\qquad T=\xi\sinh (t/(4M))\,.
\eeq

$\bullet$ Show that the metric becomes
\beq
-\frac1{16 M^2}\xi^2dt^2+d\xi^2=-dT^2+dX^2\,.
\eeq
Thus we see that the metric on the left, which is called Rindler spacetime, is locally equivalent to 2D Minkowski space. So even if the Rindler metric is singular at the horizon ($\xi=0$), a simple change of coordinates allows to extend it across the horizon as smoothly as in Minkowski space.

Notice that $X^2-T^2=\xi^2$. Thus the horizon $\xi=0$ is manifestly a null surface in Minkowski (actually two surfaces: $T\pm X=0$), and trajectories of constant $\xi\neq 0$ are hyperbolas, which we know are trajectories of uniform acceleration equal to $1/\xi$. So Rindler spacetime is the geometry of observers following trajectories of uniform acceleration. Of course we know that in order to hover at fixed $r$ above a black hole and not fall in it, you must accelerate away from it. We have found that, close to the horizon, you are approximately a Rindler observer.

More generally, one can prove (see \ref{prob:surfg}) that the geometry near a black hole horizon takes the form 
\beq
ds^2\approx -\kappa^2 x^2d\tau^2+dx^2+r_0^2d\Omega\,,
\eeq
where $\kappa$ is a constant called the \emph{surface gravity} of the horizon. For the Schwarzschild black hole we have
\beq
\kappa=\frac1{4M}\,.
\eeq
$\kappa$ has dimensions of inverse time, which is like acceleration (in natural units $c=1$). Its operational meaning is as follows: Imagine an observer at a large distance $r\gg 2M$, who is slowly lowering towards the black hole a unit mass that is attached to the endpoint of a rope. As the mass is lowered, the tension of the rope increases, and the observer must exert a stronger force to keep it in place. When the horizon is approached, the tension of the rope near the mass diverges, but this tension is redshifted (as can be seen from stress-energy conservation) upwards along the rope. As a result,  to hold the unit mass hovering right above the horizon, the observer at infinity must pull the rope with a finite force equal to $\kappa$.

\subsubsection{Surface gravity for static spherical black holes}\label{prob:surfg}

Consider a static, spherically symmetric metric of the generic form
\beq\label{statmet}
ds^2=-f(r)dt^2+\frac{dr^2}{g(r)}+r^2 d\Omega\,.
\eeq
Assume that $f(r)$ and $g(r)$ vanish linearly at $r=r_0$, \ie\ $f(r)=(r-r_0)f'_0 +\mathcal{O}(r-r_0)^2$ and similarly for $g(r)$.

$\bullet$ Show that near $r=r_0$ they take the form of Rindler spacetime\footnote{See footnote~\ref{footrind} again. The transverse space does not play any role in this analysis.}
\beq
ds^2\approx -\kappa^2 x^2d\tau^2+dx^2+r_0^2d\Omega\,,
\eeq
with
\beq
\kappa=\frac12\sqrt{f'(r_0)g'(r_0)}\,.
\eeq

\bigskip

\noindent NB: horizons with $\kappa\neq 0$ are called non-extremal, non-degenerate, or bifurcate horizons. Horizons with $\kappa=0$ (such as when $f$ and $g$ have double zeroes) are called extremal or degenerate, and they require separate treatment.

\subsubsection{\emph{Outgoing} Eddington-Finkelstein coordinates}\label{prob:outEF}

In the Schwarzschild spacetime, argue that radially outgoing light rays are given by
\beq
t=r_* + \textrm{const}
\eeq
where the tortoise coordinate $r_*$ is defined by
\beq
dr_*=\frac{dr}{\left|1-\frac{2M}{r}\right|}\,.
\eeq

$\bullet$ Show that if we now introduce a new coordinate $u=t-r_*$, such that outgoing light rays are $u= \textrm{const}$, and change coordinates $(t,r)\to (u,r)$, then the metric in these coordinates is regular at $r=2M$.

$\bullet$ Find that the equation for radial light rays is also solved by: $r=2M$ (which are the light rays that generate the horizon); and $u=-2r_* +\textrm{const}$. Verify that the latter agrees with the solution that we found using the coordinate $v=t+r_*$, i.e, $v=\textrm{const}$.

$\bullet$ Draw the trajectories of these light rays in a diagram where the vertical axis is $u+r$ (which at large $r$ and $t$ approaches $t$) and the horizontal axis is $r$. Observe that, now, the outgoing light rays $u=\textrm{const}$ cross the horizon \emph{outwards}, while the ingoing light rays $u=-2r_* +\textrm{const}$ \emph{never} cross the horizon, but only approach it asymptotically to the future. That is, light (and henceforth particles) can escape from inside the horizon, but never enter it --- precisely the opposite of what we found earlier! What is going on here? (You are encouraged to try exercise 1.d for a fuller understanding).

\subsubsection{Kruskal coordinates}
\label{prob:kruskal}

Change to null ingoing and outgoing coordinates $(t,r)\to (u=t-r_*,v=t+r_*)$ and examine whether in $(u,v)$ coordinates the metric at $r=2M$ is regular or not (Answer: it is not).

Try instead a related set of null coordinates $(U,V)$, defined by
\beq
U=-2M e^{-u/4M}\,,\qquad V=2M e^{v/4M}\,,
\eeq
and, from them, introduce new time and space coordinates
\beq
T=\frac12(U+V)\,,\qquad X=\frac12(U-V)\,.
\eeq

$\bullet$ Verify that, near $r=2M$, these coordinates become, up to constant factors, the same as the $(T,X)$ coordinates we introduced above in the Rindler limit of the solution. Therefore, in these coordinates, the horizon will be manifestly smooth.

$\bullet$ Write the Schwarzschild solution in terms of them, to find
\beq
ds^2=8M\frac{e^{-r/2M}}{r}\lp -dT^2+dX^2\rp +r^2 d\Omega_2
\eeq
where $r(T,X)$ is given implicitly by the relation
\beq\label{rTX}
2M(r-2M)e^{r/2M}=-T^2+X^2\,.
\eeq
The only singularity in these coordinates is at $r=0$. The geometry can then be smoothly (analytically) extended along all the horizons (future and past), resulting in the Kruskal maximal analytic extension of the Schwarzschild solution.
You can try to piece together all of the information we have obtained above to try to draw a picture of the Kruskal geometry in the $(T,X)$ plane. (This takes some work, but the result is rewarding).

\bigskip

\noindent NB: Although Kruskal coordinates help clarify the global nature of the maximal Schwarzschild geometry, the implicit nature of \eqref{rTX} often makes them impractical. Eddington-Finkelstein coordinates are usually the most efficient way of verifying horizon regularity.

\tocless\subsection{Wave propagation in Schwarzschild spacetime}\label{prob:scalsch}

Consider the propagation of a massless scalar field in a spacetime of the form\footnote{This is not the most general static spherical metric. Doing this exercise for the general case \eqref{statmet} is only a little more involved.}
\beq
ds^2=-f(r)dt^2+\frac{dr^2}{f(r)}+r^2 (d\theta^2+\sin^2\theta d\varphi^2)\,.
\eeq

$\bullet$ Write down the form of the wave equation 
\beq
\Box\Phi\equiv \nabla_i \nabla^i \Phi= \frac1{\sqrt{-g}}\partial_i\left(\sqrt{-g}g^{ij}\partial_j \Phi\right)=0\,,
\eeq
(keep it in a compact form). 

$\bullet$ Observe that the
angular part is the same as in the wave equation in flat space, so the
angular dependence can be separated and solved by introducing the spherical
harmonics $Y_{lm}(\theta,\varphi)$. Then, write the wave equation for the
radial field modes
$\phi_{\omega lm}(r)$ in the decomposition
\beq
\Phi(x^\mu)=e^{-i\omega t}Y_{lm}(\theta,\varphi)\phi_{\omega lm}(r)\,.
\eeq

$\bullet$ We know that in flat space (which is approached as $f\to 1$)
it is convenient to introduce a new radial field variable $\psi_{\omega lm}$ defined as
\beq
\phi_{\omega lm}=\frac{\psi_{\omega lm}}{r}\,.
\eeq
In addition, for the propagation of massless excitations in the black
hole background it is convenient to introduce
the tortoise coordinate $r_*$ defined as
\beq
dr_*=\frac{dr}{f(r)}\,.
\eeq
With these changes, you must find an equation of the form
\beq
-\frac{\partial^2 \psi_{\omega lm}(r_*)}{\partial r_*^2}=\bigl(\omega^2-V_{l}(r)\bigr)\psi_{\omega lm}(r_*)\,,
\eeq
in terms of an effective potential $V_{l}(r)$ (which you can leave
expressed in terms of $r$, with the understanding that $r$ is a function of $r_*$).

\medskip

$\bullet$ For the Schwarzschild spacetime, with 
\beq
f=1-\frac{2M}{r}\,,
\eeq
sketch the shape of potential $V_{l}$ vs.\ $r_*/M$ for different values of $l$. 

\medskip 
$V_{l}$ is the effective radial potential
that a massless scalar wave feels when propagating in this background. We have mapped
this problem to one of waves in a one-dimensional potential that extends
in the range $r_*\in(-\infty,+\infty)$. There are many questions that
can be answered qualitatively from the shape of this potential. For instance: argue that there are not any bound states of the scalar field in the
black hole background. This shows (at the perturbative level) that the
black hole does not admit `scalar hair'. 

\tocless\subsection{Rotating black holes}

\subsubsection{Extension across the Kerr horizon }\label{prob:EFKerr}

In the Kerr solution, change to Eddington-Finkelstein ingoing
coordinates $(t,\phi)\to (v, \tilde\phi)$ as
\beq
dv=dt+(r^2+a^2)\frac{dr}{\Delta}\,,\qquad d\tilde\phi=d\phi+a
\frac{dr}{\Delta}\,.
\eeq

$\bullet$ Write the metric in coordinates $(v,r,\theta,\tilde\phi)$ and
show that it is regular at the points $r_\pm$ where $\Delta=0$.

$\bullet$ Find the change to outgoing E-F coordinates $(t,\phi)\to (u, \hat\phi)$
that allow to extend the
metric across the past horizon.

\subsubsection{Rotation parameters for stars, planets, and other objects }\label{prob:rotbhs}

$\bullet$ Estimate, assuming rigid rotation, the dimensionless rotation parameter
$a_*=a/M$, which in conventional units is $a_*= cJ/(GM^2)$, for: (a) the Sun; (b) the Earth; (c) a rapidly rotating
neutron star, of mass $\simeq 1.5 M_\odot$, radius $\simeq 10~\mathrm{km}$ and
rotation period $\simeq 1.5~\mathrm{ms}$; (d) a
ball with radius 1~cm, weight 1~g, spinning at 1~Hz.

$\bullet$ Comment on the results, in particular how/why they differ so much from the maximum value for a Kerr black hole $a_*=1$.

\subsubsection{Superradiance}\label{prob:srrad}

Consider a complex massless scalar field $\Phi$, which satisfies the Klein-Gordon equation
\beq
\nabla_\mu\nabla^\mu\Phi=0\,.
\eeq
For this field we can construct a current
\beq
J_\mu=i(\Phi\nabla_\mu \Phi^* -\Phi^*\nabla_\mu \Phi)\,,
\eeq
which gives the flux of the field (\eg the flux of particles associated to the field, when it is quantized).

$\bullet$ Show that this current is conserved when the Klein-Gordon equation is satisfied. 

\medskip

Now scatter this field off a rotating black hole. Due to the symmetries of the Kerr metric and the linearity of the Klein-Gordon equation, we can expand the field into modes and then consider them individually, \ie\ 
\beq
\Phi=\Phi_{\omega m}(r,\theta)e^{-i\omega t}e^{i m\phi}\,.  
\eeq

$\bullet$ Show that the flux $F$ across the horizon
\beq
F=-J_\mu \xi^\mu\,,
\eeq
where $\xi$ is the horizon generator
\beq
\xi=\partial_t+\Omega_H\partial_\phi\,,
\eeq
is negative for modes satisfying $\omega<m\Omega_H$. In other words, there is a positive flux of these modes out of the horizon: they will reflect off the black hole with larger amplitude than they came in with. This is called superradiance.

\tocless\subsection{Entropy of the black holes in the Universe}

The Bekenstein-Hawking entropy
\beq
S_{BH}=\frac{c^3}{\hbar G}\frac{\mathcal{A}_H}{4}
\eeq
is enormous for astrophysical black holes. Here we have taken
Boltzmann's constant $k_B=1$, so temperatures are measured in units of
energy and the entropy is dimensionless. Ignoring rotation (which would only
introduce corrections by factors of order one) we can write
\beq
S_{BH}\simeq 10^{77}\left(\frac{M}{M_\odot}\right)^2\,,
\eeq
where $M_\odot=2\times 10^{33}~\mathrm{g}$ is the mass of the Sun. In the
following, we will try to obtain order-of-magnitude estimates of the
entropy of astrophysical black holes and compare it to the entropies of
other relevant systems. Our assumptions will be rather crude and may be
off by one or even two orders of magnitude, but the comparisons will
still be significant.

\subsubsection{Entropy of a star} A very crude estimate (but sufficient
for our purposes) of the entropy of the Sun is the following. For an
ideal gas of $n$ particles, the entropy is $S\sim n$. Regard the Sun as
a ball of a gas of particles of mass equal to the proton
mass, $m_p\sim 10^{-24}\mathrm{g}$. What is then the entropy of the Sun?
What is the entropy of a black hole of the same mass?

\subsubsection{Entropy of a galaxy} Estimate in the same manner the
entropy of the galaxy from the following sources:
\begin{enumerate}
\item Luminous matter (stars), if the luminous mass in the
galaxy is $M_\mathrm{galaxy}\sim 10^{11}M_\odot$. 

\item Central black hole. Our galaxy contains a central black hole with
mass $M_\bullet\simeq 4\times 10^6 M_\odot$. 

\item Stellar-mass black holes. The
number of black holes of stellar mass $\sim M_\odot$ in the galaxy is
estimated to be $\sim 10^8$. 
\end{enumerate}
Which of these three contributions to
the entropy of the galaxy is the largest?

\subsubsection{Entropy of the Universe}\label{prob:entBHs}
Estimate the following
contributions to the entropy of the Universe:
\begin{enumerate}
\item Luminous matter (stars in galaxies): take the radius of the Universe to be  $\sim
10^{10}~\mathrm{ly}$ ($\mathrm{ly}=$ light year), and consider that each
galaxy occupies a sphere of radius  $\sim 10^5-10^6~\mathrm{ly}$,
with our galaxy being a typical galaxy.

\item Cosmic microwave background radiation at $T\sim 3 \mathrm{K}\sim
10^{-4} \mathrm{eV}$. The entropy of a gas of photons at temperature $T$
in a volume $V$ is $S\sim V T^3$ (with $\hbar=1=c$). ({\it Hint:} write
$T$ in terms of the wavelength of the radiation). 

\item Entropy in black holes at galactic centers. Assume that each
galaxy has a central black hole of mass $\sim 10^6-10^9
M_\odot$.

\end{enumerate}

You must have found that the total entropy of the Universe is overwhelmingly dominated by black hole entropy. The only known entropy that is larger is that of the cosmological horizon.

\newpage
\bibliographystyle{JHEP.bst}
\bibliography{refs.bib}
\end{document}